\documentclass{aa} 
\usepackage{txfonts,graphicx,longtable,lscape,epsfig,color}

\begin{document}

   \title{The XMM-Newton Serendipitous Survey. V. The
Second XMM-Newton Serendipitous Source Catalogue\thanks{Based on observations
obtained with XMM-Newton, an ESA science mission with instruments and
contributions directly funded by ESA Member States and NASA.
Tables D.1 and D.2 are only available in electronic form at the CDS via anonymous ftp to cdsarc.u-strasbg.fr (130.79.128.5) or via http://cdsweb.u-strasbg.fr/cgi-bin/qcat?J/A+A/.}}

   \titlerunning{The 2XMM Serendipitous Source Catalogue}

   \author{M. G. Watson$^1$, A. C. Schr\"oder$^1$, D. Fyfe$^1$,
C. G. Page$^{1}$, G. Lamer$^2$, S. Mateos$^1$, J. Pye$^1$, M. Sakano$^1$,
S. Rosen$^1$, J.  Ballet$^{3}$, X. Barcons$^4$, D. Barret$^{5}$,
T. Boller$^6$, H. Brunner$^6$, M. Brusa$^6$, A. Caccianiga$^7$,
F. J. Carrera$^4$, M. Ceballos$^4$, R. Della Ceca$^7$, M. Denby$^{1}$,
G. Denkinson$^1$, S. Dupuy$^{5}$, S. Farrell$^{5}$, F. Fraschetti$^{3}$,
M. J. Freyberg$^6$, P.  Guillout$^9$, V. Hambaryan$^{2,16}$, T. Maccacaro$^{15}$,
B. Mathiesen$^{3}$, R. McMahon$^8$, L. Michel$^9$, C.  Motch$^9$,
J. P. Osborne$^1$, M. Page$^{10}$, M.W. Pakull$^9$, W. Pietsch$^6$,
R. Saxton$^{11}$, A. Schwope$^2$, P.  Severgnini$^7$, M. Simpson$^1$,
G. Sironi$^{1,7}$, G. Stewart$^1$, I. M. Stewart$^{1,13}$,
A-M. Stobbart$^{1}$, J. Tedds$^1$, R.  Warwick$^1$, N. Webb$^{5}$,
R. West$^1$, D. Worrall$^{12}$, W. Yuan$^{8,14}$ }

   \authorrunning{M.G. Watson et al.}

   \offprints{M.G. Watson}

   \institute{$^1$~Department of Physics \& Astronomy, University of Leicester, Leicester, LE1 7RH. UK\\
$^2$~Astrophysikalisches Institut Potsdam (AIP), An der Sternwarte 16, 14482 Potsdam, Germany\\
$^3$~AIM, DSM/IRFU/SAp, CEA Saclay, 91191 Gif-sur-Yvette, France\\
$^4$~Instituto de Fisica de Cantabria (CSIC-UC), Avenida de los Castros, 39005 Santander, Spain\\
$^5$~CNRS, Universit\'e Paul Sabatier \& Observatoire Midi-Pyr\'en\'ees, 9 avenue du Colonel Roche, 31400 Toulouse, France\\
$^6$~Max-Planck-Institut f\"ur Extraterrestrische Physik,
     Giessenbachstra{\ss}e 1, 85748 Garching, Germany\\
$^7$~INAF - Osservatorio Astronomico di Brera, via Brera 28, 20121 Milan, Italy\\$^8$~Institute of Astronomy, Madingley Road,
     Cambridge CB3 0HA, UK\\
$^9$~Observatoire Astronomique, UMR 7550 CNRS, Universit\'e Louis Pasteur, 11 rue de l'Universit\'e, 67000 Strasbourg, France\\
$^{10}$~Mullard Space Science Laboratory, University College London, Holmbury St. Mary, Dorking, Surrey RH5 6NT, UK\\
$^{11}$~ESA/ESAC, Apartado 78, 28691 Villanueva de la Ca\~nada, Madrid, Spain\\
$^{12}$~H.H. Wills Physics Laboratory, University of Bristol, Tyndall Avenue, Bristol BS8 1TL, UK\\
$^{13}$~Jodrell Bank Centre for Astrophysics, University of Manchester, Oxford Road, Manchester M13 9PL, UK \\
$^{14}$~National Astronomical Observatories of China/Yunnan Observatory, Phoenix Hill, PO Box 110, Kunming, Yunnan, PR China \\
$^{15}$~INAF - Headquarters, via del Parco Mellini 84, 00136 Rome, Italy \\
$^{16}$~Astrophysikalisches Institut und Universit\"ats-Sternwarte, Friedrich-Schiller-Universit\"at Jena,
        Schillerg\"asschen 3, D-07745 Jena, Germany
}

   \date{Received 7 July 2008; accepted 14 October 2008}

 
  \abstract
   {}
   {Pointed observations with XMM-Newton provide the basis for creating
   catalogues of X-ray sources detected serendipitously in each field. This
   paper describes the creation and characteristics of the 2XMM catalogue.}
   {The 2XMM catalogue has been compiled from a new processing of the
   XMM-Newton EPIC camera data. The main features of the processing
   pipeline are described in detail. }
   {The catalogue, the largest ever made at X-ray wavelengths, contains
   246,897 detections drawn from 3491 public XMM-Newton
   observations over a 7-year interval, which relate to 191,870
   unique sources. The catalogue fields cover a sky area of more than
   500\,deg$^2$. The non-overlapping sky area is $\sim\!360$\,deg$^2$
   ($\sim\!1$\% of the sky) as many regions of the sky are observed more
   than once by XMM-Newton. The catalogue probes a large sky area at the
   flux limit where the bulk of the objects that contribute to the X-ray
   background lie and provides a major resource for generating large,
   well-defined X-ray selected source samples, studying the X-ray source
   population and identifying rare object types. The main characteristics
   of the catalogue are presented, including its photometric and astrometric properties
   }
   {}

        
   \maketitle

\section{Introduction } \label{intro}
 
Surveys play a key role in X-ray astronomy, as they do in other wavebands,
providing the basic observational data  characterising the
underlying source populations. {\sl Serendipitous} X-ray sky surveys, based
on the field data from individual pointed observations, take advantage of
the relatively wide field of view afforded by typical X-ray
instrumentation. Such surveys have been pursued with most X-ray astronomy
satellites since the Einstein Observatory. The resulting serendipitous
source catalogues (e.g., EMSS: Gioia et al. 1990, Stocke et al. 1991;
WGACAT: White et al. 1994; ROSAT 2RXP: Voges et al.\ 1999; ROSAT 1RXH: ROSAT Team 2000;
 ASCA AMSS: Ueda et al., 2005) have been the basis for numerous
studies and have made a significant contribution to our knowledge of the
X-ray sky and our understanding of the nature of the various Galactic and
extragalactic source populations.

The XMM-Newton observatory provides unrivalled capabilities for
serendipitous X-ray surveys by virtue of the large field of view of the
EPIC cameras and the high throughput afforded by the heavily nested
telescope modules. This capability guarantees that each XMM-Newton
observation provides a significant harvest of serendipitous X-ray sources
in addition to data on the original target. In addition, the extended
energy range of XMM-Newton ($\sim 0.2-12$ keV) means that XMM-Newton
detects significant numbers of obscured and hard-spectrum objects that are
absent in many earlier soft X-ray surveys.

This paper describes the Second XMM-Newton Serendipitous Source Catalogue
(2XMM) which has been created from the serendipitous EPIC data from from
3491 XMM-Newton pointed observations made over a $\sim\!7$-year interval
since launch in 1999. The XMM-Newton serendipitous source catalogues are
produced by the XMM-Newton Survey Science Centre (SSC), an international
consortium of ten European institutions, led by the University of
Leicester, as a formal project activity performed on behalf of ESA. The
catalogues are based on the EPIC source lists produced by the scientific
pipe-line used by the SSC for the processing of all the XMM-Newton data.
The first serendipitous source catalogue, 1XMM, was released in 2003
(Watson et al. 2003a; XMM-SSC 2003). The current 2XMM catalogue
incorporates a wide range of improvements to the data processing, uses the
most up-to-date instrument calibrations and includes a large number of new
parameters. In parallel, the 2XMM catalogue processing also produces a
number of additional data products, for example time-series and spectra for
the brighter individual X-ray sources. A pre-release version of the current
catalogue, 2XMMp (XMM-SSC 2006), was made public in 2006. This includes
$\sim\!65$\% of the fields and $\sim\!75$\% of the sky area covered by
2XMM, while $\sim\!88$\% of all 2XMMp sources appear in the 2XMM catalogue.
Around 56\% of all 2XMM sources  already have an entry in the 2XMMp
catalogue.

The 2XMM catalogue provides an unsurpassed sky area for serendipitous
science and reaches a flux limit corresponding to the dominant
extragalactic source contribution to the cosmic X-ray background. The
catalogue is part of a wider project to explore the source populations in
the XMM-Newton serendipitous survey (the XID project; Watson et al. 2001;
Watson et al. 2003b) through optical identification of well-defined samples
of serendipitous sources (e.g., Barcons et al. 2002, 2007; Della Ceca et
al. 2004, Caccianiga et al. 2008, Motch et al. 2002; Schwope et al. 2004;
Page et al. 2007; Yuan et al. 2003; Dietrich et al. 2006). Indeed these identification programs were effectively
based on less mature versions of the XMM-Newton catalogue data processing.
XMM-Newton serendipitous survey results have also been used to study
various statistical properties of the populations such as X-ray spectral
characteristics, source counts, angular clustering, and luminosity
functions (Severgnini et al. 2003; Mateos et al. 2005; Carrera et al. 2007;
Caccianiga et al. 2007; Mateos et al. 2008; Della Ceca et al. 2008; Ebrero
et al. 2008). Other projects based on XMM-Newton serendipitous
data include the HELLAS2XMM survey (Baldi et al. 2002; Cocchia et
al. 2007).

The 2XMM serendipitous catalogue described here is complementary to
``planned" XMM-Newton surveys which provide coverage of much smaller sky
areas, but often with higher sensitivity, thus exploring the fainter end of
the X-ray source population. The deepest such surveys, such as the Lockman
Hole (Hasinger et al. 2001; Brunner et al. 2008) and the CDF-S
(Streblyanska et al. 2004), cover essentially only a single XMM-Newton
field of view, have total integration times $\sim\!300-1000$\,ks and reach
fluxes $\sim$ few $\times 10^{-16}\rm\,erg\,cm^{-2}\,s^{-1}$, close to the
confusion limit. XMM-Newton has also carried out contiguous surveys of
various depths covering much larger sky areas utilising mosaics
of overlapping pointed observations to achieve the required sensitivity and
sky coverage. Currently the largest contiguous XMM-Newton survey is the
XMM-LSS (Pierre at al. 2007) covering $\sim\!5$\,deg$^2$ with typical
exposure time 10\,--\,20\,ks per observation. Other medium-deep surveys of
1\,--\,2\,deg$^2$ regions include the SXDS ($\sim\!1.1$\,deg$^2$,
50\,--\,100\,ks exposures; Ueda et al. 2008), the COSMOS surveys
($\sim\!2$\,deg$^2$, $\sim\!80$\,ks exposures; e.g., Cappelluti et
al. 2007; Hasinger et al.\ 2007), and the Marano field survey (Krumpe et
al. 2007). These larger area surveys typically reach limiting fluxes of
$10^{-14}$ to $<10^{-15}\rm\,erg\,cm^{-2}\,s^{-1}$.

We also note that Chandra observations have been used to compile a
serendipitous catalogue including $\sim\!7000$ point sources (the ChaMP
catalogue; Kim et al. 2007) and plans are underway to compile a
serendipitous catalogue from all suitable Chandra observations (Fabbiano et
al. 2007).

The paper is organised as follows. Section~\ref{instr} introduces the
XMM-Newton observatory. Section~\ref{catobs} presents the XMM-Newton observations used to
create the catalogue and the characteristics of the fields.
Section~\ref{processing} outlines the XMM-Newton data processing framework
and provides a more detailed account of the EPIC data processing, focusing
in particular on source detection and parameterisation, astrometric
corrections and flux computation. Section~\ref{ssp} provides an account of
the automatic extraction of time-series and spectra for the brighter
sources, while Sect.~\ref{extcatcross} outlines the external catalogue
cross-correlation undertaken. Section~\ref{qual_eval} describes the quality
evaluation undertaken and some recommendations on how to extract useful
sub-samples from the catalogue. Section~\ref{catcompile} describes
additional processing and other steps taken to compile the catalogue
including the identification of unique sources.  The main properties and
characterisation of the catalogue is presented in
Sect.~\ref{catchar}. Section~\ref{cataccess} summarises access to the
catalogue and plans for future updates to 2XMM, and Sect.~\ref{conclusion}
gives a summary.

\section{XMM-Newton observatory} \label{instr}

To provide the essential context for this paper, the main features of the
XMM-Newton observatory are summarised here, with particular emphasis on the
EPIC X-ray cameras from which the catalogue is derived.

The XMM-Newton observatory (Jansen et al. 2001), launched in December 1999,
carries three co-aligned grazing-incidence X-ray telescopes, each
comprising 58 nested Wolter-I mirror shells with a focal length of
7.5\,m. One of these telescopes focuses X-rays directly on to an EPIC
(European Photon Imaging Camera) pn CCD imaging camera (Str\"uder et
al. 2001). The other two feed two EPIC MOS CCD imaging cameras (Turner et
al. 2001) but in these telescopes about half the X-rays are diverted, by
reflection grating arrays (RGA), to the reflection grating spectrometers
(RGS; den Herder et al. 2001) which provide high resolution
($\lambda/\Delta\lambda$ $\approx 100-800$) X-ray spectroscopy in the
0.33\,--\,2.5~keV range. The EPIC cameras acquire data in the
0.1\,--\,15~keV range with a field of view (FOV) $\sim$ 30~arcminutes
diameter and an on-axis spatial resolution $\sim\!5$~arcseconds FWHM (MOS
being slightly better than pn). The physical pixel sizes for the pn and MOS
cameras is equivalent to $\sim\!1$ and $\sim\!4$~arcseconds,
respectively. The on-axis effective area for the pn camera is approximately
1400~cm$^{2}$ at 1.5~keV and 600~cm$^{2}$ at 8~keV while corresponding
MOS effective areas are about 550~cm$^{2}$ and 100~cm$^{2}$,
respectively. The energy resolution for the pn camera is $\sim\!120$\,eV at
1.5\,keV and $\sim\!160$\,ev at 6\,keV (FWHM), while for the MOS camera it
is $\sim\!90$\,eV and $\sim\!135$\,eV, respectively. The EPIC cameras can
be used in a variety of different modes and with several filters (see
Sect.~\ref{obssel}). In addition to the X-ray telescopes, XMM-Newton
carries a co-aligned, 30\,cm diameter Optical Monitor (OM) telescope (Mason
et al. 2001) which provides an imaging capability in three broad-band
ultra-violet filters and three optical filters, spanning 1800\,\AA\ to
6000\,\AA; two additional grism filters permit low dispersion ultra-violet
and optical-band spectroscopy. The construction of a separate catalogue of
OM sources is in preparation.

A number of specific features of XMM-Newton and the EPIC cameras which are  
referred to repeatedly in this paper are collected together and
summarised in Appendix~\ref{newapp} together with the relevant nomenclature.

\section{Catalogue observations} \label{catobs}

\subsection{Data selection} \label{obssel}

\begin{figure}[t]
\resizebox{\hsize}{!}{\includegraphics[angle=270]{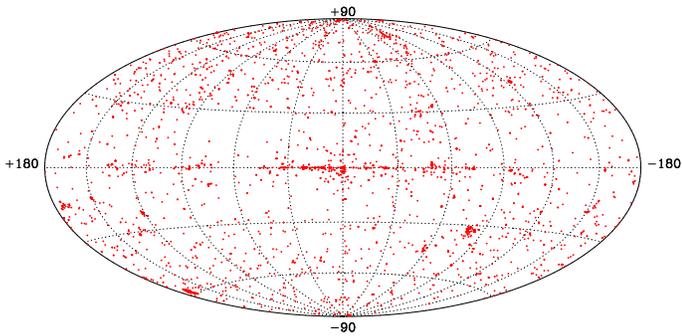}} 
\caption[]{Hammer-Aitoff equal area projection in Galactic coordinates of
  the 3491 2XMM fields.}
\label{skymapfig}
\end{figure}

XMM-Newton observations\footnote{An observation is defined as a single
science pointing at a fixed celestial target which may consist of several
exposures with the XMM-Newton instruments.}  were selected for inclusion in
the 2XMM catalogue pipeline simply on the basis of their public
availability and their suitability for serendipitous science. In practice
this meant that all observations that had a public release date prior to
2007 May 01 were eligible. A total of 3491 XMM-Newton observations (listed
in Appendix~\ref{osc}) were included in the catalogue; their sky
distribution is shown in Fig.~\ref{skymapfig}. Only a few observations(83) 
were omitted, typically because a valid ODF\footnote{The
Observation Data File is a collection of standard FITS format raw data
files created from the satellite telemetry.} was not available or because
of a fewunresolved processing problems. The field of view (FOV) of an
XMM-Newton observation (the three EPIC cameras combined) has a radius
$\sim\!15$ arcminutes. The XMM-Newton observations selected for the 2XMM
catalogue cover only $\sim\!1$\% of the sky (see Sect.~\ref{skycov} for a
more detailed discussion). Certain sky regions have contiguous multi-FOV
spatial coverage, but the largest such region is currently $<10$ deg$^2$.

\begin{figure}[t]
\resizebox{\hsize}{!}{\includegraphics[angle=0]{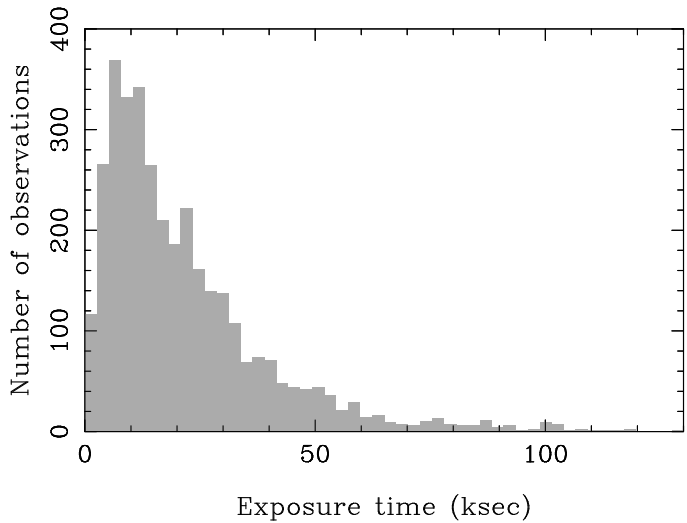}} 
\caption[]{Distribution of total good exposure time (after event filtering)
for the observations included in the 2XMM catalogue (for each observation
the maximum time of all three cameras per observation was used).}
\label{exptimefig}
\end{figure}

By definition the catalogue observations do not form a homogeneous set of
data. The observations selected have, for example, a wide sky distribution
(see Fig.~\ref{skymapfig}, where $\sim\!65$\% are at Galactic latitude $|b|
> 20\degr$), a broad range of integration times (Fig.~\ref{exptimefig}) and
astrophysical content (Sect.~\ref{targetid}), as well as a mixture of
EPIC observing modes and filters, as follows.

The EPIC cameras are operated in several modes of data acquisition. In
full-frame and extended full-frame modes the full detector area is exposed,
while for the EPIC pn large window mode only half of the detector is read
out. A single CCD is used for small window, timing and burst mode (not used
for source detection). In the case of MOS the outer ring of 6 CCDs 
{\it always} remain in standard imaging mode while the central MOS CCD can be
operated separately: in partial window modes only part of the central CCD
is read out, and in fast uncompressed and compressed modes the central CCD
is in timing mode and produces no imaging data. In the MOS refreshed frame
store mode the central CCD has a different frame time and the CCD is not
used for source detection. Table~\ref{modestab} lists all the EPIC camera
modes of observations incorporated in the catalogue, while
Fig.~\ref{examplesmodefig} shows their sky footprints.

\begin{table}[t]
\normalsize
\caption{Data modes of XMM-Newton exposures included in the 2XMM catalogue.} 
\label{modestab}
\small
\centering
\begin{tabular}{lll}
\hline \hline
Abbr. & Designation & Description \\
\hline
\multicolumn{3}{l}{\it \ \ MOS cameras:} \\
PFW  &	Prime Full Window & covering full FOV \\
PPW2 &	Prime Partial W2 & small central window \\
PPW3 &	Prime Partial W3 & large central window \\
PPW4 &	Prime Partial W4 & small central window \\
PPW5 &	Prime Partial W5 & large central window \\
FU &	Fast Uncompressed & central CCD in timing mode \\
RFS &   Prime Partial RFS  & central CCD with different frame \\
    &                      & \ \ time (`Refreshed Frame Store') \\
\multicolumn{3}{l}{\it \ \ pn camera:} \\
PFWE  &	Prime Full Window & covering full FOV \\
      &	 \ \ \ Extended & \\
PFW &	Prime Full Window & covering full FOV  \\
PLW &	Prime Large Window & half the height of PFW/PFWE \\[0.2cm]
\hline 
\end{tabular}
\normalsize		
\end{table}

\begin{figure}[ht!]
\resizebox{\hsize}{!}{\includegraphics{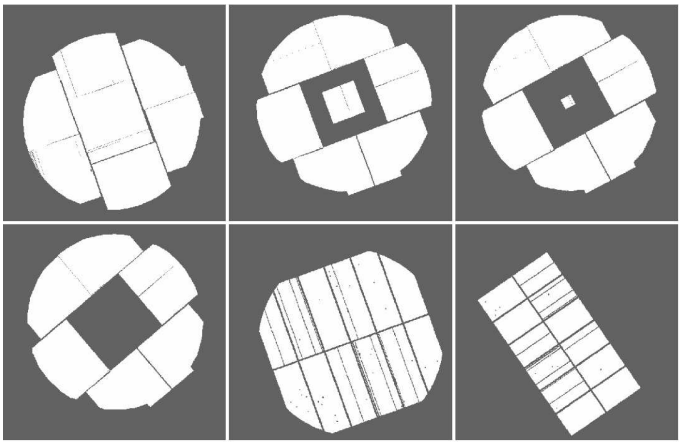}} 
\caption[]{Typical sky footprints of the different observing modes (the FOV
  is $\sim\!30\arcmin$). Noticeable are the CCD gaps as well as columns and
  rows excluded in the filtering process. The effects of vignetting and
  exclusion of CCDs due to much lower exposure times are not shown. Top
  row: MOS full window mode; MOS partial window W3 or W5 mode; MOS partial
  window W2 or W4 mode. Bottom row: MOS fast uncompressed, fast compressed,
  or RFS mode; pn full window mode; pn large window mode.
}
\label{examplesmodefig}
\end{figure}

Each XMM-Newton camera can be used with a different filter: Thick, Medium,
Thin, and Open, the choice depending on the degree of optical blocking\footnote{see Appendix~\ref{newapp}}
desired.  Table~\ref{obssumtab} gives an overview of the data modes and
filter settings used for the 2XMM observations. No Open
filter  exposures passed the
selection criteria  (cf.\ Sect.~\ref{expsel}), while about
20\% of pn observations are taken in timing, burst, or small window mode.

\begin{table}[ht]
\begin{minipage}[ht]{\columnwidth}
\normalsize
\caption{Characteristics of the 3491 XMM-Newton observations included in
 the 2XMM catalogue.}
\label{obssumtab}
\small
\centering
\renewcommand{\footnoterule}{} 
\tabcolsep 0mm
\begin{tabular}{
l @{\extracolsep{7mm}} r @{\extracolsep{1mm}} r @{\extracolsep{1mm}} 
r @{\extracolsep{7mm}} r @{\extracolsep{1.5mm}} r @{\extracolsep{1.5mm}} 
r @{\extracolsep{7mm}} r}
\hline \hline
Camera &\multicolumn{3}{c}{Modes} & \multicolumn{3}{c}{Filters} & Total \\
 & full\footnote{PFWE and PFW modes} 
& window\footnote{pn PLW mode and any of the various MOS PPW modes}
& other\footnote{other MOS modes (FU, RFS)}
& thin & medium & thick \\
\hline
pn	&2441&	233&	--	&1233&	1259&	182&	2674\\
MOS1	&2560&	605&	219	&1314&	1772&	298&	3384\\
MOS2	&2612&	655&	127	&1314&	1777&	303&	3394\\
\hline
\end{tabular}
\end{minipage}
\normalsize		
\end{table}

\subsection{Target classification and field characteristics } \label{targetid}

The 2XMM catalogue is intended to be a catalogue of serendipitous
sources. The observations from which it has been compiled, however, are
pointed observations which typically contain one or more target objects
chosen by the original observers, so the catalogue contains a small
fraction of targets which are by definition not serendipitous. More
generally, the fields from which the 2XMM catalogue is compiled may also
not be representative of the overall X-ray sky.

To avoid potential selection bias in the use of the catalogue, an analysis
to identify the target or targets of each XMM-Newton observation has been
carried out. Additionally, an attempt has been made to classify each target
or the nature of the field observed; this provides additional information
which can be important in characterising their usefulness (or otherwise)
for serendipitous science. In practice the task of identifying and
classifying the observation target is to some extent subjective and likely
to be incomplete (only the investigators of that observation know all the
details). Here, the main results of the exercise are summarised. A more
detailed description is given in Appendix~\ref{newtargbit}.
\begin{itemize}

\item Of the total 3491 observations included in 2XMM, the target could be
unambiguously resolved in terms of its coordinates and classification in
the vast majority of cases ($\sim\!98$\%)

\item In the full set of targets, $\sim 50$\% are classified as spatially
unresolved objects, $\sim\!10$\% as extended objects with small angular
extent ($<3\arcmin$), $\sim\!22$\% as larger extended objects, and around
15\% can be considered to have no discrete target leaving only $\sim\!2$\%
of unknown or problematic cases (see Table~\ref{pcategorytab}).

\item Around 10\% of observations were obtained for calibration purposes;
around 3\% of targets are ``targets of opportunity".

\item Anticipating the discussion in Sect.~\ref{properties}, around 2/3 of
the intended targets are unambiguously identified in their XMM-Newton
observations.

\end{itemize}

\begin{figure*}[ht!]
\resizebox{\hsize}{!}{\includegraphics{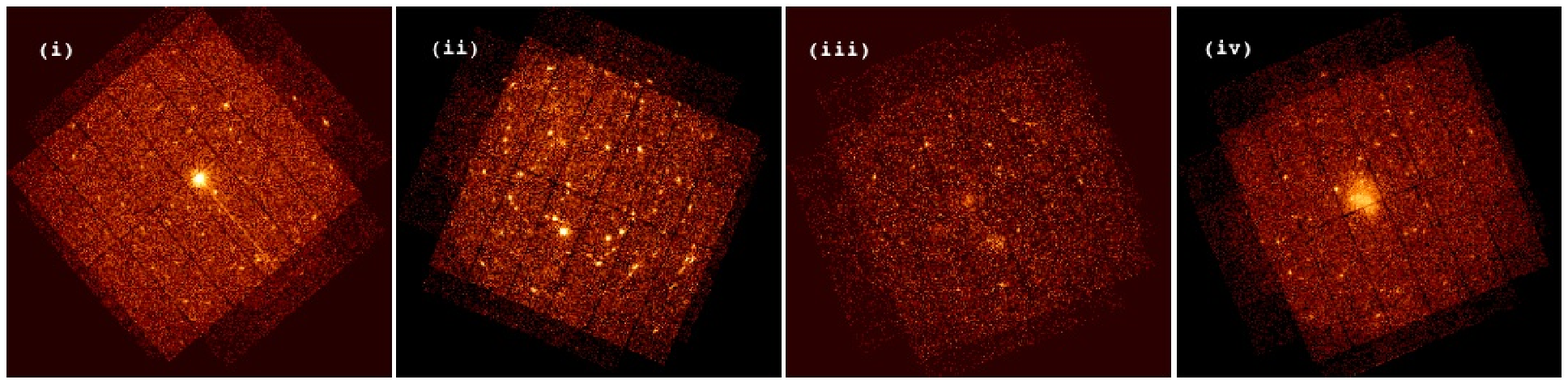}} 
\caption[]{a) Examples of typical 2XMM EPIC images (north is up). From left
  to right: (i) medium bright point source; (ii) deep field observation;
  (iii) shallow field observation with small extended sources; (iv) distant
  galaxy cluster.
  }
\vspace{0.03cm}
\addtocounter{figure}{-1} 
\resizebox{\hsize}{!}{\includegraphics{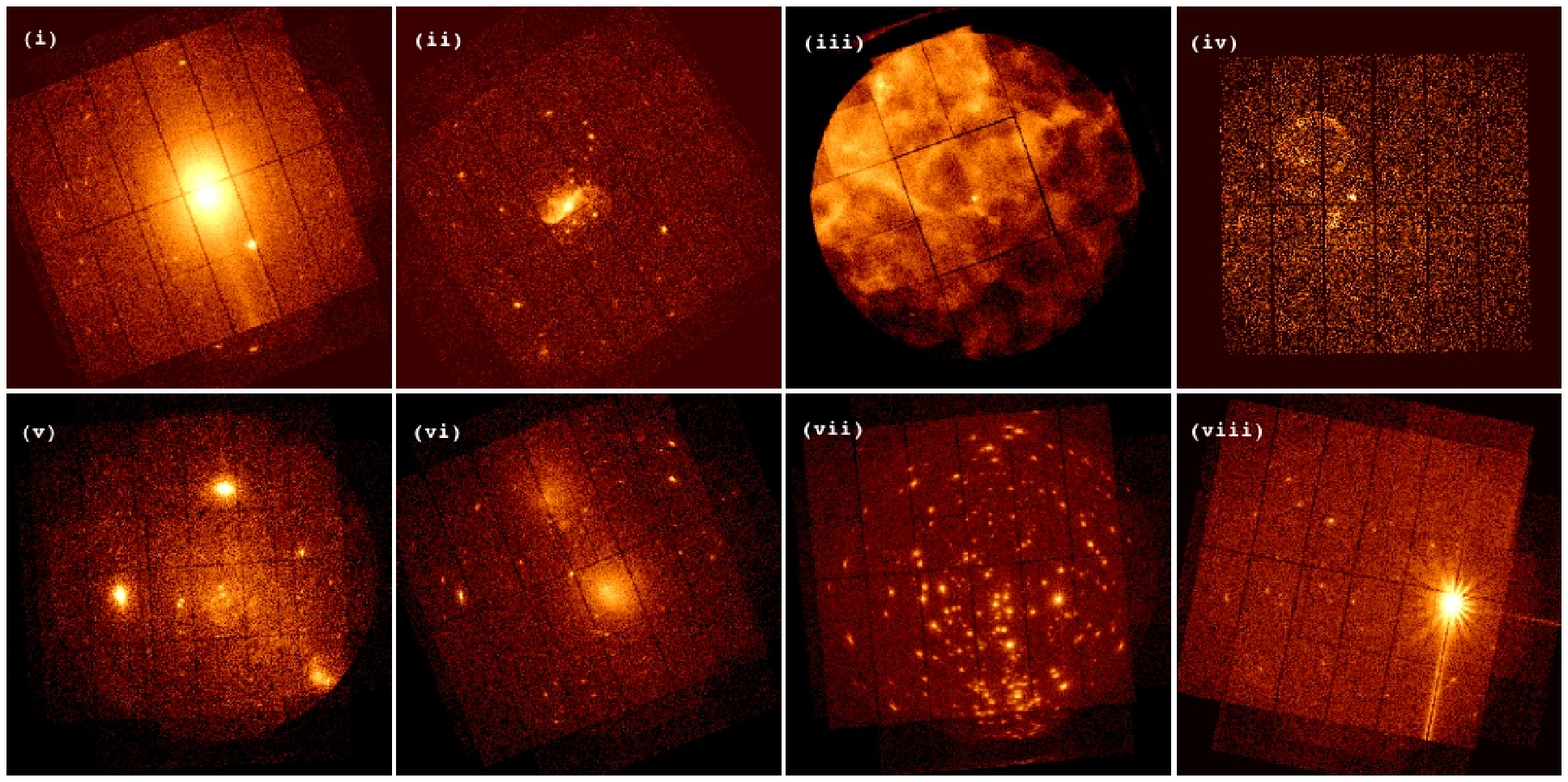}} 
\caption[]{b) Examples of variation in astrophysical content of 2XMM
  observations (north is up); in most of these extreme cases the source
  detection is problematic. Top row, from left to right: (i) bright
  extended emission from a galaxy cluster; (ii) emission from a spiral
  galaxy which includes point sources and extended emission; (iii) very
  bright extended emission from a SNR; (iv) filamentary diffuse emission.
  Second row: (v) complex field near the Galactic Centre with diffuse and
  compact extended emission; (vi) two medium-sized galaxy clusters; (vii)
  complex field of a star cluster; (viii) bright point source, off-centre.
  }
\vspace{0.03cm}
\addtocounter{figure}{-1} 
\resizebox{\hsize}{!}{\includegraphics{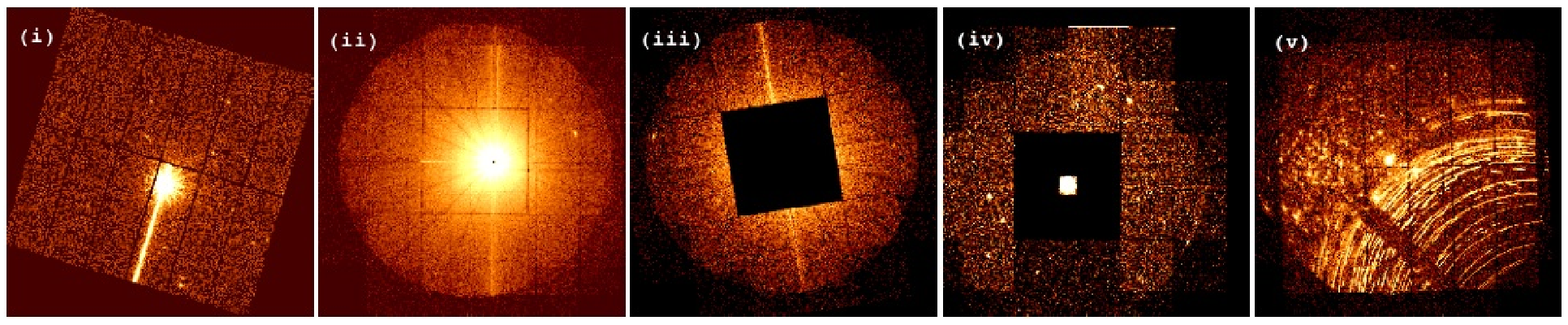}} 
\caption[]{c) Examples of instrumental artefacts causing spurious source
  detection (north is up). From left to right: (i) bright source with
  pileup and OOT events; (ii) very bright point source showing obvious
  pileup, shadows from the mirror spider, and scattered light from the RGA;
  (iii) the PSF wings of a bright source spread beyond the unused central
  CCD causing a brightening of the edges on the surrounding CCDs (which may
  not be well represented in the background map); (iv) obvious noisy CCDs
  for MOS1 (CCD\#4) and for MOS2 (CCD\#5) to the top right; (v) numerous
  and bright single reflections from a bright point source outside the FOV,
  with a star cluster to the left. See Appendix~\ref{newapp} for terminology.
  }
\label{examplesfieldfig}
\end{figure*}

Figure~\ref{examplesfieldfig} illustrates the wide variety in field
content (images are usually combinations of pn and MOS total-band images
that include out-of-FOV areas). Panel (a) shows typical XMM-Newton
observations which may be considered representative of most of the
observations used for the catalogue. Panel (b) shows the variety of
astrophysical content; in many of these cases the source detection is
affected by a dominant bright point or extended source, or by crowding
in high density regions.
Lastly panel (c) illustrates various instrumental or detector artefacts 
which, although relatively rare, cause
significant source detection issues. The most common of these, affecting $\sim\!6$\% of the
observations each, are the OOT events and
X-ray scattering off the RGA (see Appendix~\ref{newapp} for terminology). Both effects occur for all sources but only become
significant for the brightest objects where they may
cause spurious detections and background subtraction problems (as OOT events
of piled-up sources are not represented properly in the background
maps). The rarer problems (also illustrated in panel (c)) are:
\begin{itemize}
\item Pileup$^{\ref{appfoot}}$, which can make the
centroiding of a source difficult, resulting in off-centre detections as
well as spurious extended source detection.
\item The shadows from the mirror spider can
be visible in the PSF\footnote{see Appendix~\ref{newapp} \label{appfoot}} wings of the very brightest sources and affect the
background maps, that is, the source parameters in these areas are
uncertain.
\item Due to the nature of the background maps (spline maps, see
Sect.~\ref{boxmapmode}), sharp edges, caused, for example, by noisy CCDs,
can not be represented well and cause spurious detections. Note that this
problem can affect the parameters of real sources as well. 
\item Finally, the
telescope baffles allow photons from a narrow annular region of sky outside
the nominal FOV to reach the detectors via a single reflection, instead of
the two reflections required for correct focusing. Bright X-ray  objects in
this annular region can give rise to bright arcs in the image, as shown in panel
(v), which typically produce numerous spurious detections.

\end{itemize}

\section{Data Processing} \label{processing}

The SSC operates a data-processing system on behalf of ESA for the
processing of XMM-Newton pointed observations. The system, which can be
considered as a `pipeline', uses the XMM-Newton Science Analysis Software
(SAS\footnote{The description and documentation are available on-line at
the ESAC web site http://xmm2.esac.esa.int/sas/}) to generate high-level
science products from ODFs. These science products are made available to
the principal investigator and ultimately the astronomy community through
the XMM Science Archive (XSA; Arviset et al.\ 2007). In October 2006, the
SSC began to reprocess every available pointed-observation data-set from
the start of the mission. The aim was to create a uniform set of science
products using an up-to-date SAS and a constant set of XMM-Newton
calibration files\footnote{As available on 2006 July 02 plus three
additional calibration files for MOS2 and RGS1.} (the appropriate subset of
calibration files for any given observation was selected based on the
observation date). Of 5628 available observations, 5484 were successfully
processed. These included public as well as (at that time) proprietary
datasets (the data selection for 2XMM observations is discussed in
Sect.~\ref{obssel}).  The complete results of the processing have been made
available through the XSA. The new system incorporated significant
processing improvements in terms of the quality and number of products, as
described below. The remainder of this section details those aspects of the EPIC processing
system which are pertinent to the creation of the 2XMM catalogue.

\begin{figure}[t]
\centering
\includegraphics[width=8cm]{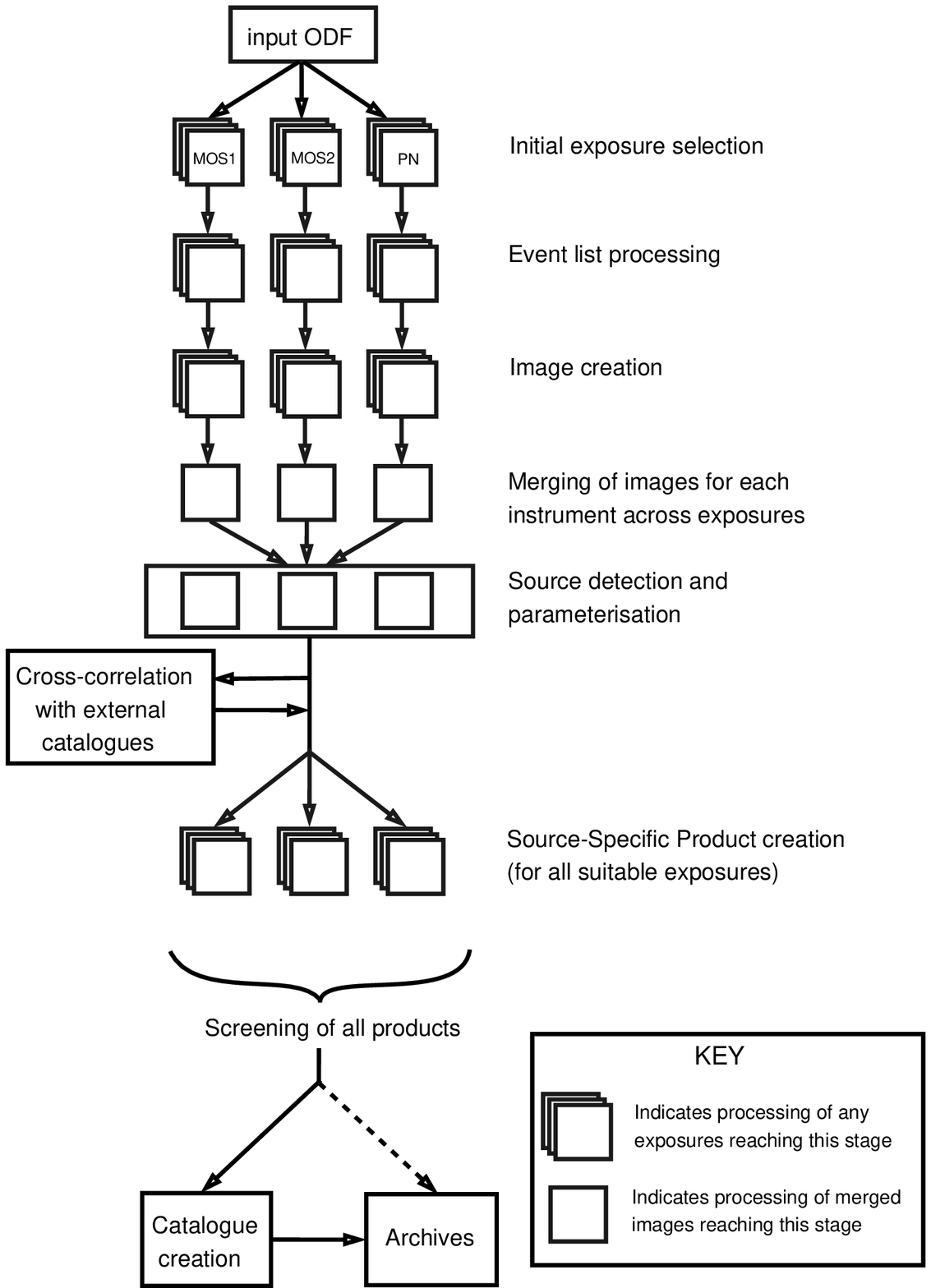}
\caption[]{A simplified schematic of the processing flow for EPIC image
data. Early processing steps treat the data from each instrument and
exposure separately. Source detection and parameterisation are performed
simultaneously on one image from each energy band from each instrument.
Source-specific products can be made, subsequently, from any suitable
exposures in the observation. Observation-level, exposure-level and
source-specific products are screened before archiving and use in making
the catalogue.
}
\label{fig:flowchart}
\end{figure}

The main steps in the data-processing sequence are: production of
calibrated detector events from the ODF science frames; identification of
the appropriate low-background time intervals using a threshold optimised
for point-source detection; identification of `useful' exposures (taking
account of exposure time, instrument mode, etc); generation of
multi-energy-band X-ray images and exposure maps from the calibrated
events; source detection and parameterisation; cross-correlation of the
source list with a variety of archival catalogues, image databases and
other archival resources; creation of binned data products; application of
automatic and visual screening procedures to check for any problems in the
data products. This description and the schematic flowchart in
Fig.~\ref{fig:flowchart} provide a rather simplified view of the actual
data-processing system. They, and the further detail that follows, are
focused on those aspects that are important for an insight into the
analysis processes that the EPIC data have undergone to generate the data
products. A complete description of the data-processing system and its
implementation are outside the scope of this paper.

The criteria employed to select exposures for initial processing and those
to be used for subsequent source detection and source-product generation
are explained further in Sect.~\ref{expsel} but are briefly introduced
here. Several suitability tests were applied during processing to limit
source detection and source-specific product creation to imaging exposures
of suitable quality, mainly by (a) restricting the merging of exposures
(and hence source detection) to imaging exposures with a minimum of
good-quality exposure time, and (b) limiting the extraction of
source-specific products to suitably bright sources.

\subsection{Selection of exposures }  \label{expsel}

Most XMM-Newton observations comprise a single exposure with each of the
cameras, although a significant number of observations are missing
exposures in one or more of the three cameras for a variety of operational
and observational reasons. To avoid generating data products of little or
no scientific use, exposures for each observation were initially selected
for pipeline processing when:
\begin{enumerate}

\item the exposure duration was $> 1000$ seconds;

\item the exposure was taken through a scientifically useful filter. In
practice this requirement rejected all exposures for which the filter
position was closed, calibration, or undefined. The possible filters are
Medium, Thick, Thin1, Thin2 (pn only), and Open.

\end{enumerate}
After event-list processing (Sect.~\ref{exposure_maps}), exposures were
selected for image creation according to the following criteria:
\begin{enumerate}
\addtocounter{enumi}{+2}

\item The quality checks during the event-list processing had been
  successful.

\item The exposure had been taken in a mode which could usefully be
processed by the source detection stage, cf.\ Table~\ref{modestab}. The pn
burst, timing, and small window modes were rejected (the effective FOV in
the latter mode is small, i.e., $258\arcsec \times 262\arcsec$, making the
background fitting stage of the source detection problematic). For the MOS,
all modes, including the outer CCD imaging component of modes where the
central CCD was windowed, missing (non-imaging modes), or modified
(Refreshed Frame Store mode), were included.

\end{enumerate}
A further set of criteria selected the appropriate images for the detection
stage (cf.\ Sect.~\ref{sourcesearch}) which ensured that only high quality
images were used.
\begin{enumerate}
\addtocounter{enumi}{+4} 

\item Background filtering (see Sect.~\ref{images}) must have been
successfully applied. Cases where the sum of high background GTIs\footnote{see Appendix~\ref{newapp}} was less
than 1000 seconds were rejected as unusable. Without background filtering
the source detection is usually of limited value due to the much higher net
background.

\item Each of the five images of an exposure (in the energy bands 1\,--\,5,
see Table~\ref{EnergyBands}) had to contain at least one pixel per image
with more than one event. This further avoided low exposure images being
used.

\item The image must have been in a data mode useful for source detection
(this excluded modes only used for engineering test purposes).

\item Where more than one exposure with a particular camera passed the
above selection criteria, those exposures with the same filter and data
mode were merged and then only the exposure group with the maximum net
exposure time was chosen for use in the source detection stage.

\end{enumerate}

\subsection{Event-List Processing}  \label{EventListProc}

Event-list processing was performed on all initially selected exposures. A
number of checks and corrections were applied to the event lists of the
individual CCDs before they were merged into a single event list per
exposure. Once merged, a further set of checks and corrections was
performed. At each stage of the processing, a quality assessment of the
event lists decided whether to continue the processing. The main steps in
processing the event lists were as follows.
\begin{itemize}

\item The CCD event lists were first examined separately on a frame by
frame basis: corrections were applied to account for telemetry dropouts;
gain and charge transfer inefficiency (CTI) corrections were made; a GTI
list for each CCD was created; frames identified as bad and events
belonging to them were flagged; event patterns\footnote{see Appendix~\ref{newapp}} were identified; events were
flagged if they met criteria such as being close to a bad pixel or edge of
the CCD, which were important to later processing (standard \#XMMEA\_EM for
MOS and \#XMMEA\_EP for pn); invalid events were identified and discarded;
events caused by CCD bad pixels were identified and removed; the fraction
of the detector area in which events could not have been detected due to
cosmic-ray events was recorded for each frame; events caused by CCD bad
pixels as well as cosmic-ray events were identified and removed; EPIC MOS
CCDs operating in low-gain mode were discarded from the event lists.

\item At the point where the event lists from individual CCDs were merged
into exposure event lists, the event positions were converted from CCD
pixel coordinates to the detector (\texttt{CAMCOORD2}) and sky coordinate
systems. This step includes a randomisation within each CCD pixel to
eliminate Moir\'{e} effects. The MOS camera event times were also
randomised within the frame time, to avoid a strong Fourier peak at the
frame period and to avoid possible beat effects with other instrumental
frequencies. Time randomisation was not performed on pn event lists as the
frame time is much shorter than for the MOS.

\end{itemize}

In addition, the spacecraft attitude file was examined for periods of the
observation when the spacecraft pointing direction varied by less than 3
arcminutes from the median of the pointing measurements for the
observation. The 3-arcminute limit was imposed to avoid degradation of the
effective PSF \footnote{see Appendix~\ref{newapp}} which could arise from co-adding data with different off-axis
angles and to avoid a potentially large (but probably low exposure)
extension of the observed sky field.  These attitude GTIs were then further
restricted for each camera to cover only that part of the observation when
the camera was active.

\subsection{Creation of multi-band images and exposure maps } \label{images}\label{exposure_maps}

Periods of high background (mostly due to so-called `soft proton' flares)
can significantly reduce the sensitivity of source detection. Since events
caused by such flares are usually much harder than events arising from
typical X-ray sources, background variation can be disentangled from
possible time variation of the sources in the field by monitoring events at
energies higher than the 12\,keV upper boundary to the `science band',
beyond which point contributions from cosmic X-ray sources are very rare. A
time series of such events, including most of the FOV, was constructed for
each exposure. This event rate was used as a proxy for the science-band
background rate.

The generation of the background time-series differed in detail between pn
and MOS cameras, in particular in terms of the events used to form the
time-series. The MOS high-energy background time-series were produced from
single-pixel events with energies above 14\,keV from the imaging CCDs. The
background GTIs were taken to be those time intervals of more than 100\,s
in duration with a count rate of less than
2~ct~ks$^{-1}$~arcmin$^{-2}$. The pn high-energy background time-series
were produced in the 7.0\,--\,15\,keV energy range. The background GTIs
were taken to be those time intervals of more than 100\,s in duration with
a count rate of less than 10~ct~ks$^{-1}$~arcmin$^{-2}$.

These threshold count rates were chosen as a good compromise 
between reducing background and preserving exposure for detecting point 
sources in the relatively short exposures which make up the bulk of the 
XMM-Newton observations. For comparison, the average quiet level in the MOS cameras,
for example, is $\sim 0.5$ ~ct~ks$^{-1}$~arcmin$^{-2}$.  

For all exposures in imaging mode, images were created for energy bands
1\,--\,5, as listed in Table~\ref{EnergyBands}, from selected events
filtered by event-list, attitude, and high background GTIs (except where
the sum of all high background GTIs was less than 1000 seconds in which
case no background filtering was applied). Note that the event-list GTIs
are CCD dependent and the resulting image can have a different exposure
time in each CCD. The events for pn images were selected by pattern $\le 4$
(for band~1 a stricter requirement of pattern = 0 was adopted) and a cut in
CCD coordinates (Y $> 12$) to reduce bright low-energy edges. Events on CCD
columns suffering a particularly large energy scale offset as well as
events outside the FOV were excluded. For MOS images events with pattern
$\le 12$ were selected and events outside the FOV were excluded. The images
are tangent-plane projections of celestial coordinates and have dimensions
of $ 648 \times 648$ image pixels, with a pixel size of $ 4\arcsec \times
4\arcsec$.

\begin{table}[ht]
\normalsize
\caption{Energy bands used in 2XMM processing}
\label{EnergyBands}
\small
\centering
\begin{tabular}{crl}
\hline \hline
Band   & Energy band & Notes \\
number & (keV)       &       \\  
\hline       
1      & 0.2 -- \phantom{0}0.5  &       \\
2      & 0.5 -- \phantom{0}1.0  &       \\
3      & 1.0 -- \phantom{0}2.0  &       \\
4      & 2.0 -- \phantom{0}4.5  &       \\
\smallskip 
5      & 4.5 -- 12.0 &       \\
6      & 0.2 -- \phantom{0}2.0 & `soft band' \\
7      & 2.0 -- 12.0 & `hard band' \\
8      & 0.2 -- 12.0 & `total band' \\
9      & 0.5 -- \phantom{0}4.5 & `XID band' \\
\hline
\end{tabular}
\normalsize		
\end{table}

Exposure maps represent the GTI-filtered on-time multiplied by the
(spatially dependent) vignetting function, adjusted to reflect telscope and
instrumental throughput efficiency. They were created for each EPIC
exposure in imaging mode in energy bands 1\,--\,5 using the calibration
information on mirror vignetting, detector quantum efficiency, and filter
transmission. The exposure maps were corrected for bad pixels, bad columns
and CCD gaps (cf.\ Fig.~\ref{examplesmodefig}) as well as being multiplied
by an OOT factor which is 0.9411 for pn full frame modes, 0.97815 for pn
extended full frame modes, and 1.0 for all other pn and MOS modes.

\subsection{Source detection \& parameterisation }
\label{sourcesearch}

The fundamental inputs to the 2XMM catalogue are the measured source
parameters which were extracted from the EPIC image data by the multi-step
source detection procedure outlined below. Each step was carried out
simultaneously on each image of the five individual bands, 1\,--\,5, and of
the three cameras. Note that the source counts and rates derived here refer
to the fully integrated PSF.

As a first step, a detection mask was made for each camera. This defines
the area of the detector which is suitable for source detection. Only those
CCDs where the unvignetted exposure map values were at least 50\% of the
maximum exposure map value were used for source detection.

\subsubsection{Sliding-box source detection -- local mode} \label{boxlocalmode}

An initial source list was made using a `box detection' algorithm. This
slides a search box ($20\arcsec \times 20\arcsec$) across the image defined
by the detection mask. The size of the box comprises $\sim\!50$\% of the
encircled energy fraction of the on-axis\footnote{The encircled energy
fraction does not strongly depend on off-axis angle.} PSF. In its first
application (`local mode') the algorithm derived a local background from a
frame ($8\arcsec$ wide) immediately surrounding the search box. In each of
the five bands from each of the three cameras, the probability,
$P_\Gamma(k,x)$, and corresponding likelihood, $L_i$, were computed from the
null hypothesis that the measured counts $k$ or more in the search box
result from a Poissonian fluctuation in the estimated background level,
$x$, i.e.:
\[ L = -\ln P_\Gamma(k,x) , \]
where $P_\Gamma$ is the incomplete Gamma function:
\[ P_\Gamma(k,x) = {1\over \Gamma(k)} \int_0^x {\rm e}^{-t} t^{k-1} dt , \]
and 
\[ \Gamma(k) = \int_0^\infty {\rm e}^{-t} t^{k-1} dt .  \]
The sum of $N$ independent likelihoods, after multiplication by 2, is
expected to have, in the limit of large $N$, the same probability
distribution as $\chi^2$  for $N$ degrees of freedom (Cash 1979). For this reason the total-band EPIC
box-detect likelihood was calculated by summing the band-specific
likelihoods in this way and inserting the result in the standard formula
for the probability for $\chi^2$ to equal or exceed the measured value in
the null hypothesis, i.e.,
\begin{equation}
L \approx -\ln (1-P_{\Gamma}(N,L')) \;\;\; 
     {\rm with} \;\;\;  L' = \sum_{i=1}^{N} L_i \;,
\label{sumlikelihood}
\end{equation}

where $N$ is the number of energy bands and cameras involved. All sources
with a total-band EPIC likelihood above 5 were included in the output list.

\subsubsection{Sliding-box source detection -- map mode} \label{boxmapmode}

After the first pass to detect sources, a background map was created for
each camera and energy band. Using a cut-out radius dependent on source
brightness in each band (specifically the radius where the source counts
per unit area fell below 0.002 ct~arcsec$^{-2}$), areas of the image where
sources had been detected were blanked out. A $12 \times 12$-node spline
surface was fitted to the resulting source-free image to calculate a
smoothed background map for the entire image. For the pn images the
contribution of OOT events was also modelled into the background maps.

A second box-source-detection pass was carried out, creating a new source
list, this time using the spline background maps (`map mode') which
increased the source detection sensitivity compared to the local-mode
detection step. The box size was again set to $20\arcsec\times
20\arcsec$. Source counts were corrected for the part of the  PSF falling
outside the detection box. Only sources with a total-band EPIC likelihood,
cf.\ eq.~(\ref{sumlikelihood}), above 5 were included in this map-mode
source list.

\subsubsection{Source parameter estimation by maximum likelihood fitting}
\label{det_emldetect}

A maximum likelihood fitting procedure was then applied to the sources
emerging from the map-mode detection stage to calculate source parameters
in each input image. This was accomplished by fitting a model to the distribution of counts over 
a circular area of radius $60\arcsec$. The energy-dependent model value, 
$e_i$, in pixel, $i$, is given by 
\begin{equation}
e_i = b_i + \alpha S_i
\label{modelvalue}
\end{equation}
where $b_i$ is the background, derived from the background map, $S_i$ is the 
source profile (i.e. the PSF, convolved with the source extent model 
(Sect.~\ref{extdsrc})) and $\alpha$ is a scalar multiplier of 
the source profile.

For each source, the fitting procedure minimised the C-statistic (Cash
1979)
 \[ C = 2 \sum_{i=1}^{N} (e_i - n_i \ln e_i) \]
to find the best set of model parameters, where $e_i$ is the expected model
value in pixel $i$ (eqn.~(\ref{modelvalue})), $n_i$ the measured number of counts in pixel $i$, and
$N$ is the total number of pixels over all images used.

Free parameters of the fit were source position, extent, and source count
rate. Positions and extent were constrained to be the same in all energy
bands and for all cameras while the count rates were fitted separately for
each camera and energy band.  The fitting process used the multi-band
exposure maps to take account of various instrumental effects (cf.\
Sect.~\ref{exposure_maps}) in deriving the source counts $c_{\rm s}$:
\[ c_{\rm s}(x,y) = R_{\rm s}(x,y) \ t_{\rm map}(x,y) \, , \]
where $R_{\rm s}(x,y)$ is the source count rate in each image pixel as
predicted by the instrumental PSF and source extent model and $t_{\rm
map}(x,y)$ is the corresponding value of the exposure map.

After arriving at those values of the source parameters which minimize $C$,
the detection likelihood (formally, the probability of the null hypothesis)
for those optimum values is then calculated. Cash's prescription for this
is to form the difference
\[
\Delta C = C_\mathrm{null} - C_\mathrm{best} \, ,
\]
where $C_\mathrm{null}$ is the C-statistic of the null hypothesis model
(i.e., with zero source flux) and $C_\mathrm{best}$ is the minimum result
returned by the fitting routine. According to Cash, $\Delta C$ is
distributed approximately as $\chi^2$ for $\nu$ degrees of freedom, where
$\nu$ is the number of fitted parameters. The probability $P(\chi^2 \ge
\Delta C)$ of obtaining the calculated value of $\Delta C$ or greater by
chance fluctuations of the detected background can therefore be obtained in
terms of the  incomplete Gamma function $P_\Gamma$ as follows:
\[
P(\chi^2 \ge \Delta C) = 1-P_\Gamma(\frac{\nu}{2}, \frac{\Delta C}{2}) \, .
\]
Note that the values $L$ which are stored in the source lists are
log-likelihoods, formed from $L = -\ln(P)$. \footnote{Protassov et al. (2002) have highlighted the dangers of using  the probabilities derived from likelihood ratio tests when the null hypothesis is close to the boundary of parameter space. In this regard it is clear that it is inappropriate to interpret the detection likelihoods, $L$, literally in terms of detection probabilities. Instead the relation between the likelihood and the detection probability requires calibration via simulations, as is discussed in Sect.~\ref{simulations}.}

Since the $C$ values are simple sums over all image pixels included in the
fit, one may calculate $\Delta C_i$ for each band $i$ then add the results
together to generate a total-band $\Delta C_\mathrm{total}$ without
destroying the $\chi^2$ equivalence: only the number of degrees of freedom
changes. The source detection procedure thus calculates $\Delta C_i$ and
hence $L_i$ for each $i$th band, using $\nu = 3$ ($=4$ if source extent is
also fitted), then sums the $\Delta C_i$ and calculates $L_\mathrm{total}$
using $\nu = N+2 \ (= N+3)$, where $N$ is the number of bands.

The fitting of the input sources was performed in the order of descending
box(map)-detect detection likelihood. After each fit the resulting source
model was added to an internally maintained background map used for the
fitting of subsequent sources. With this method the background caused by
the PSF wings of brighter sources is taken into account when fitting the
fainter sources.  All sources (as detected by the sliding-box in map mode)
with a total-band detection likelihood $>6$, as determined by the fitting
process, were included in the output source list. Note that for individual
cameras and energy bands, the fitted likelihood values can be as low as
zero.

The calculation of the parameter errors made use of the fact that $\Delta
C$ follows the $\chi^2$-distribution. The $68$\% confidence intervals were
determined by fixing the model to the best-fit parameters and then
subsequently stepping one parameter at a time in both directions until $C =
C_{best} + 1$ is reached (while the other free parameters were kept
fixed). The upper and lower bound errors were then averaged to define a
symmetric error. Note that using $C_{best} + 1$ to determine the $68$\%
confidence intervals is only strictly correct in the case that there is one 
parameter of interest. In the case of the fitting performed here, this 
requires that the position and amplitude parameters are essentially 
independent (i.e. that the cross-correlation terms of the error matrix 
are negligible). This has been found through simulations to be an acceptable 
approximation in the present case (see also the discussion of the astrometric 
corrections in Sect.~\ref{astcorr}).

Four camera-specific X-ray colours, known as hardness ratios
(HR1\,--\,HR4), were obtained for each camera by combining corrected count
rates from energy bands $n$ and $n+1$:
\[ {\rm{HR}}n = (R_{n+1} - R_n) / (R_{n+1} + R_{n}) \]
where $R_n$ and $R_{n+1}$ are the corrected count rates in energy bands $n$
and $n+1$ ( $n=1-4$). Count rates, and therefore hardness ratios, are
camera dependent. In addition, they depend on the filter used for the
observation, especially for HR1. Note that HR1 is also a strong function of
Galactic absorption, $N_{\rm H}$. This needs to be taken into account when
comparing hardness ratios for different sources and cameras. It should be
stressed that a large fraction of the hardness ratios were calculated from
marginal or non-detections in at least one of the energy bands.
Consequently, individual hardness ratios should only be deemed reliable if
the source is detected in both energy bands, otherwise they have to be
treated as upper or lower limits. Similarly, the errors on the hardness
ratios will be affected for band-limited count rates in the Poisson regime
(Park et al. 2006).

\subsubsection{Extended-source parameterisation}  \label{extdsrc}

One of the enhancements incorporated in the 2XMM processing that was not
available in 1XMM was information about the potential spatial extent of
sources and, where detected, a measure of that extent.

The source extent characterisation was realised by fitting a convolution of
the instrumental PSF and an extent model to each input source. The extent
model was a $\beta$-model of the form
 \[ f(x,y)=\left(1+\frac{(x-x_0)^2+(y-y_0)^2}{r_{\mathrm c}^2}\right)^{-3\beta+1/2} , \]
where $\beta$ was fixed at the canonical value $\beta=2/3$ for the surface
brightness distribution of clusters of galaxies (Jones \& Forman 1984; but
see Sect.~\ref{extddiscussion} for a discussion of problems arising from
this assumption). The core radius, $r_{\mathrm c}$, the 
`extent' parameter of a source, was fitted with a constraint that $r_{\mathrm c} <
80\arcsec$.  Cases with $r_{\mathrm c} \le 6\arcsec$ were considered to be
consistent with a point source and $r_{\mathrm c}$ was reset to zero.

An extent likelihood based on the C-statistic and the best-fit point source
model as null hypothesis was calculated in an analogous way to that used in
the detection likelihood described in Sect.~\ref{det_emldetect}.  The
extent likelihood $L_{\rm ext}$ is related to the probability $P$ that the
detected source is spuriously extended due to Poissonian fluctuation (i.e.,
the source is point-like) by
 \[ L_\mathrm{ext} = - \ln(P)\, . \]
A source was classified as extended if $r_{\mathrm c} > 6\arcsec$ and if
the extended model improved the likelihood with respect to the point source
fit such that it exceeded a threshold of $L_\mathrm{ext,min}=4$.

Since source extent can be spuriously detected by the confusion of two or
more point sources, a second fitting stage tested whether a model including
a second source further improved the fit. If the second stage found an
improvement over the single-source fit, the result could be two point
sources or a combination of one point source and one extended source. Note,
however, that the previously fitted fainter sources
(Sect.~\ref{det_emldetect}) are not re-computed in such cases.

\subsection{Astrometric corrections}  \label{astcorr} 

The positions of X-ray sources were determined during the maximum
likelihood fitting of the source. These positions were placed into an
astrometric frame determined from the XMM-Newton on-board Attitude \& Orbit
Control Subsystem (AOCS) which uses XMM-Newton's two star trackers and its
``fine sun sensors". The overall accuracy of the XMM-Newton astrometric
frame (i.e., the difference between the XMM-Newton frame and the celestial
reference frame) is typically a few arcseconds although a few
observations suffer rather poorer accuracy.

As the mean positions of bright X-ray sources can be determined to a
statistical precision of $\ll 1\arcsec$ in the XMM-Newton images, and
typical sources to a precision of $1\arcsec -2\arcsec$, it is clearly
worthwhile to improve the astrometric precision of the positions. This was
done on an observation by observation basis by cross-correlating the source
list with the USNO B1.0 catalogue (Monet et al.\ 2003). This approach
depends on the assumptions, usually valid, that a significant number of
XMM-Newton detections will have an optical counterpart in the USNO
catalogue and that the number of random (false) matches is low. The
algorithm used a grid of trial position offsets (in RA and Dec) and
rotations between the XMM-Newton frame and the true celestial frame (as
defined by the USNO objects) and determined the optimum combination of
offset and rotation values which maximised a likelihood statistic related
to the X-ray/optical object separations.

To determine whether the offset/rotation parameters so determined
represented an acceptable solution, an empirically determined condition was
used. This was based on a comparison of the likelihood statistic determined
from the analysis with that calculated for purely coincidental
X-ray/optical matches in a given observation, i.e., if there were no true
counterparts.

In practice this approach worked very well at high Galactic latitudes,
resulting in a high success rate (74\% of fields with $|b|\ge20\degr$),
whilst at low Galactic latitudes (and other regions of high object density)
the success rate was much lower (33\% of fields with $|b|<20\degr$). The
typical derived RA, Dec offsets were a few arcseconds, and a few tenths of
a degree in field rotation, values consistent with the expected accuracy of
the nominal XMM-Newton astrometric frame as noted above.

The 2XMM catalogue contains equatorial RA and Dec coordinates with the
above determined astrometric corrections applied and corresponding
coordinates which are not corrected. Where the refined astrometric solution
was not accepted, the corrected and uncorrected coordinates are identical.

The catalogue also reports the estimated residual 
component of the position errors, $\sigma_{\rm sys}$.\footnote{In the catalogue and associated documentation
we refer to this as a `systematic' error. This nomenclature is somewhat misleading as the true nature of this component of the 
positional errors is
far from clear.\label{sysfoot}} 
 This has the value
$0\farcs35$ for all detections in a field for which an acceptable
astrometric correction was found and $1\farcs0$ otherwise. The values of
$\sigma_{\rm sys}$ in the catalogue are a new determination of the
residual error component based on further analysis undertaken after the
initial compilation of the catalogue was completed. The details of this
analysis are given in Sect.~\ref{astprop}. Higher initial values of
$\sigma_{\rm sys}$ ($0\farcs5$ and $1\farcs5$, respectively) were used in
earlier stages of the catalogue creation, for example in the external
catalogue cross-correlation (see Sect.~\ref{extcatcross}).

\subsection{Flux computation} \label{flux_computation}

The fluxes, $F_i$, given in the 2XMM catalogue have been obtained for each
energy band, $i$, as
\[
F_i={R_i  / f_i}
\]
where $R_i$ is the corrected source count rate and $f_i$ is the energy
conversion factor (ECF) in units of ${\rm
10^{11}\,ct\,cm^2\,erg^{-1}}$. The ECFs depend on camera, filter, data
mode, and source spectrum. Since the dependence on data mode is low (1--2\%), ECF values were calculated only for the full window mode which is
the most frequently used (cf.\ Table~\ref{obssumtab}).  To compute the ECF
values, a broad-band source spectrum was assumed, characterised by a power
law spectral model with photon index $\Gamma=1.7$ and observed X-ray
absorption $N_{\rm H}= 3 \times10^{20}\,{\rm cm}^{-2}$. As shown in
Sect.~\ref{xcol} (cf.\ Fig.~\ref{hrfig}), this model provides a reasonable
representation of the emission of the bulk of the sources in 2XMM. A single
model cannot, of course, provide the correct flux conversion for different
intrinsic spectra, and the effect of varying the shape of the assumed
power-law spectrum on the measured fluxes has been investigated. For
example, for $\Delta\Gamma=\pm0.3$ the fluxes can change by $\sim\!6$\% and
$\sim\!8$\% in bands~1 and~5, respectively. The effect is much less
($<2$\%) for bands~2\,--\,4 (i.e., between 0.5\,keV and 4.5\,keV). Very
soft or very hard spectra will, of course, produce much greater changes in
the conversion factor, particularly in the softest and hardest energy
bands.

Note that the fluxes given in 2XMM have {\it not} been corrected for
Galactic absorption along the line of sight. The ECF values used in the
2XMM catalogue are shown in Table~\ref{ecftab}.

\begin{table}[t]
\normalsize
\caption{Energy conversion factors used to compute 2XMM catalogue fluxes
(in units of ${\rm 10^{11}\,ct\,cm^2\,erg^{-1}}$). }
\label{ecftab}
\small
\begin{tabular}{lllll}
\hline \hline
Camera & Band & Thin & Medium & Thick \\
\hline       
pn    &	1   &	8.95403   &	7.82028   &	4.71096  \\
      &  2  &	8.09027   &	7.83782   &	6.02015  \\
      &  3  &	5.88255   &	5.78272   &	5.00419  \\
      &  4  &	1.92805   &	1.90529   &	1.80647  \\
      &  5  &	0.555226  &	0.554529  &	0.547205  \\
      &  9  &	4.53836   &	4.43953   &	3.74772  \\
MOS1  &	1   &	1.80399   &	1.60150   &	1.06500  \\
      &  2  &	1.88017   &	1.82853   &	1.48465  \\
      &  3  &	2.05034   &	2.01594   &	1.79446  \\
      &  4  &	0.746128  &	0.737800  &	0.707822  \\
      &  5  &	0.143340  &	0.143131  &	0.141213  \\
      &  9  &	1.42040   &	1.39361   &	1.23264  \\
MOS2  &	1   &	1.81179   &	1.60670   &	1.06620  \\
      &  2  &	1.88369   &	1.83088   &	1.48818  \\
      &  3  &	2.05117   &	2.01594   &	1.79530  \\
      &  4  &	0.750569  &	0.741687  &	0.711708  \\
      &  5  &	0.150769  &	0.150560  &	0.148537  \\
      &  9  &	1.42326   &	1.39647   &	1.23524  \\
\hline
\end{tabular}
\normalsize		
\end{table}

Publicly available response matrices (RMFs) were used in the computation of
the ECFs\footnote{EPIC RMFs are available at \newline
http://xmm.vilspa.esa.es/external/xmm\_sw\_cal/calib/epic\_files.shtml}.
For the pn they were on-axis matrices for single-only events for band~1 and
for single-plus-double events\footnote{Single-only events = pattern 0,
single-plus-double events = patterns 1\,--\,4.}  for bands~2\,--\,5 ({\it
epn\_ff20\_sY9\_v6.7.rmf}, {\it epn\_ff20\_sdY9\_v6.7.rmf}, respectively).
For the MOS cameras there has been a significant change in the low energy
redistribution characteristics with time, especially for sources close to
the optical axis. In addition, during XMM-Newton revolution 534 the
temperatures of both MOS focal plane CCDs were reduced (from -100C to
-120C), resulting in an improved spectral response thereafter (mainly in
the energy resolution). To account for these effects, epoch-dependent RMFs
were produced. However, in the computation of MOS ECFs time averaged RMFs
were used (for revolution 534). To be consistent with the event selection
used to create MOS X-ray images, the standard MOS1 and MOS2 on-axis RMFs
for patterns 0\,--\,12 were used ({\it m1\_534\_im\_pall.rmf}, {\it
m2\_534\_im\_pall.rmf}).

Note that for the computation of the ECFs, the effective areas used in the
spectral fitting were calculated without the corrections already applied to
the source count rates (i.e., instrumental effects including vignetting and
bad-pixel corrections, see Sect.~\ref{exposure_maps}), as well as for the
PSF enclosed-energy fraction.

\section{EPIC source-specific product generation    } \label{ssp}

The 2XMM processing pipeline was configured to automatically extract
source-specific products, i.e., individual time-series (including
variability measures) and spectra for the brighter detections. Sources were
selected when the following extraction criteria were satisfied: 1) they had
$\ge 500$ total-band EPIC counts\footnote{Where the source was only
observed with one or two cameras the equivalent EPIC counts were calculated
for the absent camera(s) using the pn to MOS count ratio $3.5 : 1$,
representative of the typical source count ratios.}, 2) the detector
coverage of the source, weighted by the PSF for the respective camera, was
$\ge 0.5$, and 3) the total-band detection likelihood for the respective
camera was $\ge 15$. The decision whether to extract products for a source
was based solely on it meeting these extraction criteria in the (merged)
exposures used in source detection (Sect.~\ref{sourcesearch}).  However,
products for qualifying sources were subsequently extracted for {\it all}
exposures (i.e., imaging event lists) of an observation that adhered to the
general exposure selection criteria given in Sect.~\ref{expsel} (i.e.,
items~1\,--\,7).

Table~\ref{evseltab} shows the event selection criteria for the extraction
of the source products. Instrumental GTIs (stored in the event list) are
always applied, while GTIs for masking out high background flaring (see
Sect.~\ref{images}) were only applied to spectra and the variability
tests. Source data were extracted from a circular region of radius $r =
28\arcsec$, centred on the detected source position, while the background
extraction region was a co-centred annulus with $60\arcsec \le r \le
180\arcsec$.  Circular apertures of radius $r=60\arcsec$ were masked from
the background region for any contaminating detection with a likelihood
$>15$ for that camera. These values represent a compromise choice for data
extraction by avoiding the additional complexity required to implement a
variable extraction radius optimised for each source. Note that the use of
an aperture-photometry background subtraction procedure here differs from
the use of the background maps applied at the detection stage.

\begin{table*}[ht]
\begin{minipage}[ht]{\columnwidth}
\normalsize
\caption{Event selection for source products.}
\label{evseltab}
\small
\centering
\renewcommand{\footnoterule}{} 
\begin{tabular}{lcc}
\hline \hline
 & pn & MOS \\
\hline
PATTERN\footnote{column in the event lists}: & $\le 4$ & $\le 12$ \\
FLAG$^a$ for spectra: & FLAG = 0 & (FLAG \& 0xfffffeff) = 0 \\
FLAG$^a$ for time-series: & (FLAG \& 0xfffffef) = 0 & (FLAG \& 0x766ba000) = 0 \\
energy range: & $0.2^b-12$\,keV & $0.2\footnote{the range
  $0.2-0.35\,$keV is set to bad in the spectra}-12\,$keV \\
GTIs for spectra: & instrumental and background flare GTIs & instrumental
      and background flare GTIs \\
GTIs for time-series: & instrumental GTIs & instrumental GTIs \\
GTIs for variability test: & merged instrumental and background flare GTIs
      & merged instrumental and background flare GTIs \\
\hline
\end{tabular}
\end{minipage}
\normalsize		
\end{table*}

\subsection{Spectra } \label{spec}

\begin{figure}[bt] 
\centering 
\resizebox{\hsize}{!}{\includegraphics[angle=0]{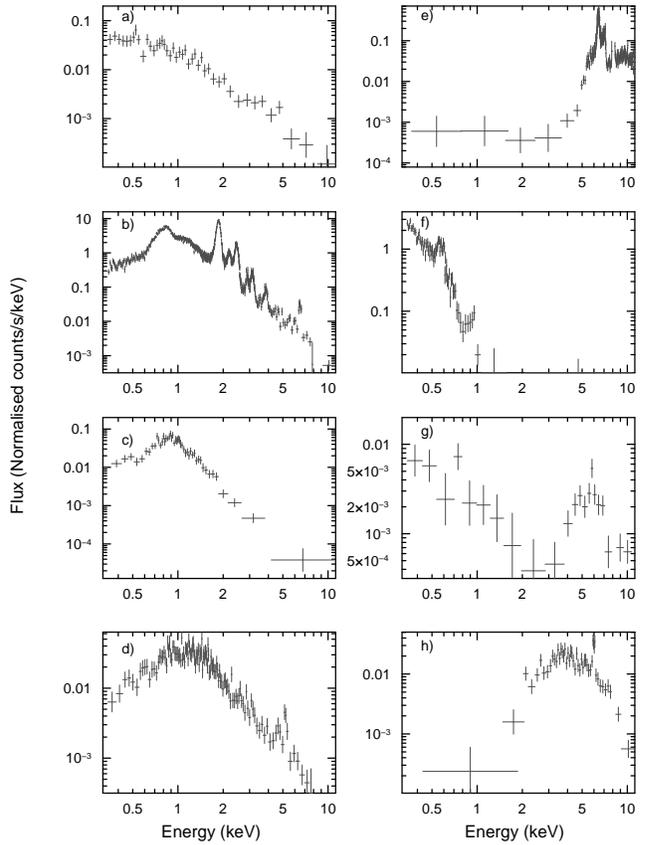}} 
\caption{Examples of auto-extracted 2XMM spectra. Sources are serendipitous objects
and spectra are taken from the EPIC pn unless otherwise stated. Panels: a)
a typical extragalactic source (Seyfert I galaxy); b) line-rich spectrum of
a localised region in the Tycho supernova remnant (target); c) MOS2
spectrum of a stellar coronal source (target; H {\protect\tiny II} 1384,
Briggs \& Pye 2003), described by two-component thermal spectrum; d)
spectrum of the hot intra-cluster gas in a galaxy cluster at $z=0.29$
(Kotov, Trudolyubov \& Vestrand 2006); e) heavily absorbed, hard X-ray
spectrum of the Galactic binary IGR~J16318-4848 (target; Ibarra et
al. 2007); f) spectrum of a super-soft source with oxygen line emission at
$\sim\!0.57$~keV; g) a relatively faint source showing a two-component
spectrum; h) source with power-law spectrum strongly attenuated at low
energies and with a notable red-shifted iron line feature around 6~keV.
}
\label{fig:example_spectra}
\end{figure}

For each source meeting the extraction criteria, the pipeline created the
following spectrum-related products: 1) a source+background spectrum
(grouped to 20 ct/spectral-bin) and a corresponding background-subtracted
XSPEC (Dorman \& Arnaud, 2001) generated plot; 2) a background spectrum; 3)
an auxiliary response file (ARF). Energies below $0.35$\,keV are considered
to be unreliable for the MOS due to low sensitivity and for the pn due to
the low-energy noise (in particular at the edges of the detector) and, as
such, were marked as `bad' in XSPEC terminology. Data around the Cu
fluorescence line for the pn ($7.875\,{\rm keV} \le E \le 8.225$\,keV) were
also marked `bad'. The publicly available `canned'\footnote{Pre-computed
for the instrument, mode, event pattern selection and approximate detector
location of the source.} RMF associated with each spectrum is conveyed by a
header keyword. Some examples of the diversity of source spectra contained
amongst the source-specific spectral products are shown in
Fig.~\ref{fig:example_spectra}.

\subsection{Time-series } \label{TS}

Light curves for a given source were created with a common bin-width (per
observation) that was an integer multiple of 10 seconds (minimum width 10
seconds), determined by the requirement to have at least 18 ct/bin for pn
and at least 5 ct/bin for MOS for the exposures used in source
detection. All light curves of a given source within an XMM-Newton
observation are referenced to a common epoch for ease of comparison.

The light curves themselves can include data taken during periods of
background flaring because background subtraction usually successfully
removes its effects. However, in testing for potential variability, to
minimise the risk of false variability triggers, only data bins that lay
wholly inside both instrument GTIs {\it and} GTIs reflecting periods of
non-flaring background were used.

Two simple variability tests were applied to the separate light curves: 1)
a Fast Fourier Transform and 2) a $\chi^2$-test against a null
hypothesis of constancy. While other approaches, e.g., the
Kolmogorov-Smirnov test, maximum-likelihood methods, and Bayesian methods
are potentially more sensitive, the $\chi^2$-test was chosen here as being
a simple, robust indicator of variability. The fundamental formula for
$\chi^2$ is
\[
\chi^2 = \sum_i \frac{(y_i - Y_i)^2}{\sigma_i^2} \, ,
\]
where $y_i$ is the $i$th data value, $Y_i$ the model at this point, and
$\sigma_i$ the uncertainty. In the present case, the model $Y_i$, which
incorporates the null hypothesis that the source flux is constant over
time, is constructed as follows:
\begin{equation}
\label{Y1}
  Y_i = f_{\mathrm{src},i} \ A_\mathrm{src} \ \Delta t \ [\phi_\mathrm{src}
  + \phi_{\mathrm{bkg},i}] \, ,
\end{equation}
where $f_i$ are exposure values, $A$ is the collecting area, $\Delta t$ is
the time-series bin duration, and $\phi$ is a (bin-averaged) `flux' in
counts per unit time per unit area.

The problem now is that \emph{a priori} the expectation values
$\phi_{\mathrm{bkg},i}$ for the background time-series is not known -- they
must be estimated, with as low an uncertainty as possible, by forming a
background time-series in an (ideally) fairly large area which is
sufficiently far from the source to avoid cross-contamination. Also, the
average source flux $\phi_\mathrm{src}$ is not known, which must also be
estimated from the (necessarily noisy) data at hand. After some algebra it
can be shown that the best estimate for $Y_i$ is given by
\begin{equation}
\label{Yprime}
  Y^\prime_i = \frac{f_{\mathrm{src},i}}{\Sigma_j f_{\mathrm{src},j}}
  \sum_{j=1}^{N}\left( y_j - \frac{A_\mathrm{src}}{A_\mathrm{bkg}}
  \frac{f_{\mathrm{src},j}}{f_{\mathrm{bkg},j}} \, b_j \right) +
  \frac{A_\mathrm{src}}{A_\mathrm{bkg}}
  \frac{f_{\mathrm{src},i}}{f_{\mathrm{bkg},i}} \, b_i \, ,
\end{equation}
where $b_i$ are the measured background counts.
The first term of equation 3 represents a constant, unweighted time-average 
of the background-subtracted source counts, derived from the whole light 
curve, while the second term reflects the background expected in the 
source aperture for time-bin, $i$.

The $\sigma$ values in the $\chi^2$ sum present a problem. In the Pearson
formula appropriate to Poissonian data, $\sigma^2_i$ is set to $Y_i$. If we
simply substitute $Y^\prime_i$ for $Y_i$ here, the resulting $\chi^2$
values are found via Monte Carlo trials to be somewhat too large. This is
because the use of the random background variate $b_i$ in Eq.~\ref{Yprime}
introduces extra variance into the numerators of the sum. A formula for
$\sigma$ which takes this into account is
\[
  \sigma^2_i = Y^\prime_i + \left( \frac{A_\mathrm{src}}{A_\mathrm{bkg}} \,
  \frac{f_{\mathrm{src},i}}{f_{\mathrm{bkg},i}} \right)^2 b_i \, .
\]

\begin{figure}[tb] 
\centering 
\resizebox{\hsize}{!}{\includegraphics[angle=0]{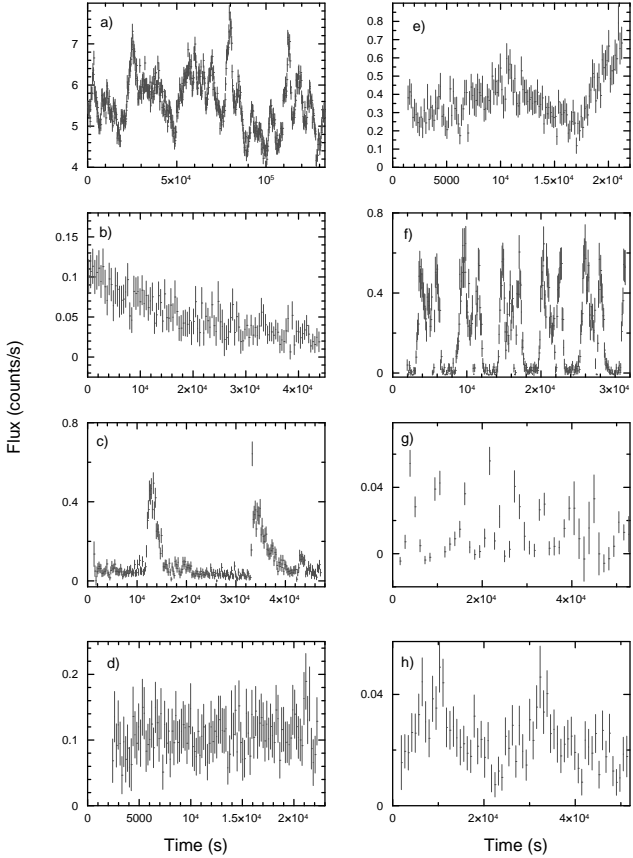}} 
\caption{Example auto-extracted 2XMM time-series. Sources are serendipitous objects and
the data are taken from the pn unless otherwise stated. Panels: a) MOS1
data for Markarian 335 (Seyfert I -- target); b) MOS1 data showing the
decay curve of GRB 050326 (target); c) X-ray flares from a previously
unknown coronally active star; d) time-series of the emission from a
relatively faint cluster of galaxies, showing no significant variability
(target); e) time-series of the obscured Galactic binary IGR~16318-4848
(target; Ibarra et al. 2007); f) previously unknown AM~Her binary showing
several phase-stable periodic features (Vogel et al.\ 2007); g) highly
variable AND periodic object, likely to be a cataclysmic or X-ray binary
(Farrell et al.\ 2008) -- the binning results in poor sampling of the
intrinsic periodic behaviour; h) source showing clear variability but not
flagged as variable in the catalogue (the probability of variability falls
below the threshold of $10^{-5}$). These last two cases highlight the
sensitivity of the variability characterisation on the time bin size.
} 
\label{fig:example_lcs}
\end{figure}

For each exposure used, the pipeline generated a background-subtracted
source time-series and the corresponding background time-series (corrected
for exposure, cosmic rays, and dead time), together with the graphical
representations of the data and of its power spectrum. The
$\chi^2$-statistics and probabilities are conveyed by header keywords. Some
example total-band time-series from these products that highlight the range
of source variability present in the 2XMM catalogue are shown in
Fig.~\ref{fig:example_lcs}.

\subsection{Limitations of the automatic extraction } \label{ssplimit}

As with any automated extraction procedure, a few source
products suffer from problems such as low photon statistics, low numbers
of bins, background subtraction problems, and contamination.

Spectra with few bins can arise for very soft sources where the total-band
counts meet the extraction criteria but the bulk of the flux occurs below
the $0.35$\,keV cut-off (Sect.~\ref{spec}). This can also occur if the
extraction is for an exposure with a shorter exposure time than those used
in the detection stage, especially if the detection was already close to
the extraction threshold. Similarly, background over-estimation in the
exposure (or underestimation in the original detection exposure) can result
in fewer source counts compared to those determined during the detection
stage, yielding poorer statistics and low bin numbers for the time-series
and spectra. This can occur when spatial gradients across the background
region are imperfectly characterised, e.g., where the source lies near
strong instrumental features such as OOT events, where there are marked
steps in the count-rate levels between adjacent noisy and non-noisy CCDs,
or where contaminant source exclusions are biased to one side of a
background region that overlaps the wings of a very bright source or bright
extended emission. In many cases the automatic (Sect.~\ref{autoflag}) as
well as manual flag settings (Sect.~\ref{manflag}) indicate whether source
products are likely to be reliable.

Contamination of the source extraction region (e.g., by another source, OOT
events, or single reflections) can also cause problems if the contamination
is brighter than or of comparable brightness to the extracted source. The
nearest-neighbour column can act as an initial alert in such cases -- 19\%
of the catalogue sources with spectra have neighbouring detections (of any
brightness) within $28\arcsec$ (i.e., the extraction radius).

The extraction process and exposure corrections are optimised for point
sources. Absolute fluxes in source-specific products of extended sources,
therefore, may not be reliable. However, relative measures such as
variability and spectral line detection should still be indicative.

\subsection{Known processing problems } \label{sspproblem}

A few products are affected by known processing problems:

(i) When the usable background region is very small, the background area
calculation becomes imprecise and results in an inaccurate
background-subtracted source spectrum. This can occur with bright sources
in MOS W2 and W4 partial window modes where most of the background region
lies outside the $110\arcsec \times 110\arcsec$ window or in crowded areas
where the source-free area is markedly reduced. In the former case the
source is usually bright enough that background subtraction has negligible
impact and so does not need to be performed.

(ii) Attitude GTIs were not included in the extraction criteria, and
occasionally the source was significantly displaced with respect to the
aperture as defined by the detection image (in extreme cases, off the
detector). This will affect the calculation of count rates in the spectra
and the variability measurements for the time-series.

(iii) Occasionally the light curve exposure correction failed (i.e., no
time-series were produced) or light curves were inadequately corrected for strong background
variations across CCDs (which can cause spurious variability
detection). The latter cases are confined to very bright extended sources
and are mostly associated with spurious detections.

(iv) Neither spectra nor time-series are corrected for pileup (nor are the
source count rates in the catalogue). Due to the difficulties in detecting
and quantifying pileup no attempt has been made to flag this effect.

\section{External catalogue cross-correlation   } \label{extcatcross}

As part of the XMM-Newton pipeline, the Astronomical Catalogue Data
Subsystem (ACDS) generated products holding information on the immediate
surrounding of each EPIC source and on the known astrophysical content of
the EPIC FOV, highlighting the possible non-detection of formerly known
bright X-ray sources as well as indicating the presence of particularly
important astrophysical objects in the area covered by the XMM-Newton
observation. 

In addition to Simbad\footnote{The SIMBAD Astronomical Database (Wenger et
al.\ 2000).} and NED\footnote{The NASA/IPAC Extragalactic Database.}, 202
archival catalogues and article tables were queried from
Vizier\footnote{The VizieR Service at CDS (Ochsenbein et al. 2000).}. They
were selected on the basis of their assumed high probability to contain the
actual counterpart of the X-ray source. Basically all large area ``high
density" astronomical catalogues were considered, namely the SDSS-DR3
(Abazajian et al. 2005) , USNO-A2.0 (Monet et al. 1998), USNO-B1.0 (Monet
et al. 2003), GSC 2.2 (STScI 2001), and APM-North (McMahon et al. 2000)
catalogues in the optical, the IRAS (Joint Science WG 1988; Moshir et
al. 1990), 2MASS (Cutri et al. 2003), and DENIS (DENIS consortium 2005)
catalogues in the infrared, the NVSS (Condon et al. 1998), WISH (de Breuck
et al. 2002), and FIRST (Becker et al. 1997) catalogues at radio
wavelengths, and the main X-ray catalogues produced by Einstein (2E; Harris
et al. 1994), ROSAT: RASS bright and faint source lists (Voges et al.\
1999, 2000), RBS (Schwope et al.\ 2000), HRI (ROSAT Team 2000), PSPC (ROSAT
2000), and WGACAT (White et al. 2000) catalogues of pointed observations),
and XMM-Newton (1XMM; XMM-SSC, 2003). Also included
were large lists of homogeneous objects (e.g., catalogues of bright stars,
cataclysmic variables, LMXBs, Be stars, galaxies, etc.). The full list of
archival catalogues queried is included as one of the pipeline products.

The XMM-Newton detections were cross-correlated with the archival entries
taking into account positional errors in both the EPIC and the archival
entries. The list of possible counterparts did not provide additional
information on the relative merits of the cross-correlation or on the
probability that the given archival entry was found by chance in the error
circle of the X-ray source.

The cross-correlation was based on the dimensionless variable:
\[r^2 = \frac{\Delta\alpha^2}{\sigma_{\alpha}^2} + \frac{\Delta\delta^2}{\sigma_{\delta}^2}\]
with $\sigma_{\alpha}^2 = \sigma_{\alpha_{x}}^2 + \sigma_{\alpha_{o}}^2$
and $\sigma_{\delta}^2 = \sigma_{\delta_{x}}^2 + \sigma_{\delta_{o}}^2$,
where $\sigma_{\alpha_{x}}$ and $\sigma_{\delta_{x}}$ are the standard
deviations in RA and Dec of the X-ray source position and
$\sigma_{\alpha_{o}}$ and $\sigma_{\delta_{o}}$ the corresponding errors on
the position of the archival catalogued object. The error on the X-ray
position is the quadratic sum of the statistical error with the additional
error which depends on the effectiveness of the astrometric correction
(cf.\ Sect.~\ref{astcorr}). Positional errors of the archival entries were
either read from the respective catalogue or fixed according to guidance in
the relevant catalogue literature. In all cases, the significance of the
error was rescaled to the $1\sigma$-level.

The probability density distribution of position differences between the
X-ray source and its catalogue counterpart due to measurement errors is a
Rayleigh distribution.  Hence, the probability of finding the X-ray source
at a distance between $r$ and $r+\delta r$ from its archival counterpart
is:
\[\delta p(r|id) = r \cdot {\rm e}^{(-r^2/2)}\, \delta r , \]
with a cumulative distribution function:
\[\int_{0}^{r} \delta p(r|id) = (1-{\rm e}^{(-r^2/2)}) . \]
Thus, lists of counterparts with $r(\sigma)< 2.146, 3.035$, and 3.439 are
90\%, 99\%, and 99.73\% complete, respectively.  Computing the actual
reliability of the identification requires a careful calibration of the
density of catalogue sources and of the likelihood ratio method applied; in
the near future, such reliabilities will be provided for candidates found
in the main archival catalogues. However, in the absence of such
information at the pipeline level, it was decided, for completeness to list
all possible identifications having positions consistent with that of the
X-ray source at the 99.73\% confidence level, corresponding to $3 \sigma$.

The ACDS results are presented in several interconnected HTML files
(together with copies in FITS format). Graphical products are 1) a plot
with the position of all quoted archival entries on the EPIC merged image,
2) an overlay of the position of the X-ray sources detected in the EPIC
camera and contours of the EPIC merged image on a ROSAT all-sky survey
image, and 3) finding charts based on sky pixel data provided by the CDS
Aladin image server (Bonnarel et al.\ 2000).

\section{Quality evaluation}   \label{qual_eval}

As part of the quality assurance for the data processing, a number of
procedures, both automated and manual, were performed on many of the data
products to take note of intrinsic problems with the data as well as to
detect software issues. Particular emphasis was given to potential problems
with the source detection and characterisation, and quality flags were set
accordingly.

\subsection{Visual screening of data products}  \label{screening}

The overall visual screening included data products from all three
instrument groups (EPIC, RGS, OM) as well as those from the external
catalogue cross-correlation (cf.\ Sect.~\ref{extcatcross}). Only products
that could be conveniently assessed were inspected using a dedicated
screening script, that is, most HTML pages, all PNG images and all PDF
plots (as representatives of data from the FITS files), all EPIC FITS
images and maps (including source-location overlays), and the mosaiced OM
FITS images with source overlay.  For each observation a screening report
with standardised comments was created, recording data and processing
problems (see, for example, Sect.~\ref{sspproblem}), and made available via
the XSA.

As a result of the visual screening, two otherwise eligible observations
(obtained for experimental mode tests) were excluded from the catalogue
since the tested mode was not properly supported by the processing system
and the source parametrisation was considered to be unreliable.

\subsection{Potential source detection problems}   \label{srcdetproblems}

Intrinsic features of the XMM-Newton instrumentation combined with some
shortcomings of the detection process have given rise to detections that
are obviously spurious\footnote{Spurious detections caused by the
background noise (as characterised by their likelihood) are not discussed
in this section, see Sects.~\ref{sourcesearch},
\ref{simulations}.}. Possible causes range from bright pixels and segments
to OOT events (in the case of pileup), RGA scattered light, single
reflections from the mirrors, and optical loading (cf.\ Appendix~\ref{newapp}
and Fig.~\ref{examplesfieldfig}\,a). In cases where the spatial background
varied rapidly (e.g., PSF `spikes', filamentary extended emission, edges of
noisy CCDs), the spline background map may deviate from the true
background. This could potentially have given rise to spurious source
detections and could also have affected the measured parameters (including
time-series and spectra) of real sources.

Extended sources were particularly difficult to detect and parametrise due
to their (often) filamentary or non-symmetric structure as well as the
maximum allowed extent in fitting ($80\arcsec$, Sect.~\ref{extdsrc}). This
often led to multiple detections of a large or irregular extended emission
region. On the other hand, multiple point sources (e.g., in a crowded
field) might also be detected as extended (due to computational
restrictions no attempt was made to distinguish more than two
overlapping/confused point sources). See Sect.~\ref{extddiscussion} and
Fig.~\ref{examplesextdfig} for a discussion of extended sources and some
examples.

\subsection{Automated quality-warning flags for detections}  \label{autoflag}

Some of the source detection problems could be identified and 
quantified so that the processing software could set automated quality warning
flags in the source lists.  For each detection, four sets of flags (one per
camera plus a summary set covering all cameras), each containing twelve
entries, were written into the observation source list.  Nine of the flags
in each set were populated based on other key quantities available in the
same source list.  The meaning of these flags is summarised in
Table~\ref{autoflagtab}.  The default value of every flag was False; when a
flag was set it means it has been changed to True.  For each detection,
Flags 2\,--\,7 were set in a common fashion across all four sets.  Flags 1,
8, and 9 are camera-specific, but any set to True were also reflected in
the summary set.

\begin{table*}[t]
\normalsize
\caption{Description of the automated (Flags 1\,--\,9) and manual (Flags
  11\,--\,12) quality warning flags.
}
\label{autoflagtab}
\small
\tabcolsep 1mm
\begin{tabular}{lp{8cm}p{9cm}}
\hline \hline
Flag & Description & Definition for flag to be set True (cf.\ Notes) \\
\hline
1 & Low detector coverage & $m_{camera} <  0.5$ \\
2 & Near other source & $r \le 65 \cdot \sqrt{R_{epic}}$\, AND $r_{min} =
10\arcsec$ AND  $r_{max} = 400\arcsec$ \\
3 & Within extended emission & $r \le 3 \cdot E$\, AND $r_{max} = 200\arcsec$ \\
4 & Possible spurious extended detection near bright source & Detection is
extended AND Flag~2 is set
    AND $c_{epic} \ge 1000$ \\
5 & Possible spurious extended detection within extended emission & Detection is
extended AND $r \le
    160\arcsec$ AND fraction of rate compared with causing source is $\le 0.4$ \\
6 & Possible spurious extended detection due to unusual large single-band
    detection likelihood & Detection is
extended AND fraction of detection likelihood per camera and
    band compared with the sum of all is $ \ge 0.9$ \\
7 & Possible spurious extended detection & At least one of the flags 4, 5, 6 is set \\
8 & On bright MOS1 corner or bright low-gain pn column  & Source position
    is located on one of the affected pixels \\
9 & Near bright MOS1 corner & Source position within $ r_f = 60\arcsec $ of a bright corner pixel \\
10 &  Not used &  \\
11 &  Within region where spurious detections occur  &  Set manually \\
12 &  Bright (`originating') point source in region where spurious detections
      occur & Set manually \\
\hline
\end{tabular}
\newline
{\bf Notes:} $m$ is the detector coverage of the detection weighted by the
PSF; $r$ is a radial distance in arcseconds from the `originating' source
within which all detections receive this flag; $R_{epic}$ is the EPIC
source count rate in ct/s of the `originating' source; $E$ is the extent
parameter (core radius) of the `causing source' in arcseconds (cf.\
Sect.~\ref{extdsrc}); $c_{epic}$ is the EPIC source counts of the `originating'
source; $r_f$ is the radius used for source PSF fitting in arcseconds.
\normalsize		
\end{table*}

The criteria used to set the flags were determined largely empirically from
tests on appropriate sample data-sets (cf.\ Fig.~\ref{examplesfieldfig}\,b
for some examples). Flags set to True should be understood mainly as a
warning: they identify possible problematic issues for a detection such as
proximity to a bright source, a location within an extended source
emission, insufficient detector coverage of the PSF of the detection, and
known pixels or clustering of pixels that tend to be intrinsically bright
at low energies. In all these cases the parameters of a real source may be
compromised and there is a possibility that the source is spurious.

Extended sources near bright sources and within larger extended emission
are most likely to be spurious and have been flagged as such.  In addition,
extended detections triggered by hot pixels or bright columns can be
identified since their likelihood in one band (of one camera) is
disproportionally higher than in the other bands and cameras.  However, no
attempt has been made to flag spurious extended detections in the general
case, that is, in areas where the background changes considerably on a
small spatial scale and the spline maps cannot adequately represent this.
At the same time, no point sources have been specifically flagged as
spurious (see Sect.~\ref{manflag} regarding manual flagging) though
they are often caused by the same features as the spurious extended
detections. The spatial density of real point sources is, in general, much
higher than for extended sources and the reliability of such a `spurious'
flag would be low. Instead, Flags~2, 3, and~9 can be used as a warning that
such a source could be spurious.

\subsection{Manual flag settings for detections} \label{manflag}

In addition to the automated quality flags, a more rigorous visual
screening of the source detection was performed for the EPIC fields to be
used in the catalogue. The outcome of this process was reflected in two
flags (11 and 12) as described below and summarised in
Table~\ref{autoflagtab}.

Images of each field, with source overlay, were inspected visually and
areas with likely spurious detections  were recorded (as
ds9-regions; Joye \& Mandel 2003). Such regions could be regular (circle,
ellipse, box) or irregular (polygon); in cases where only a single
detection was apparently spurious a small circle of $10\arcsec$ radius was
used, centred on this detection. It should be stressed that these regions,
except for the latter case, could include both suspected spurious and real
detections. In many cases (especially at fainter fluxes) it was impossible
to visually distinguish between a real source and a spurious detection that
was caused by artefacts on the detector or by insufficient background
subtraction. In addition, the effect of such features on the parameters of
a nearby real source has not been investigated in detail.  For example,
single reflections or the RGA scattered-light features were not included in
the background maps and may therefore have affected the source parameters.
On the other hand, as the source parameters are derived by the fitting
process in order of decreasing source brightness, the parameters of fainter
sources take the PSF of nearby bright sources into account
(Sect.~\ref{sourcesearch}).

The ds9-regions were converted to EPIC image masks where the bad areas have
the value zero and the rest of the field has the value one. These masks are
available as catalogue products (Sect.~\ref{cataccess}); they can be
combined with the camera detection masks to study, for example, the sky
coverage.

The masks were used to flag sources within the masked areas with
Flag~11. In many cases, the so-called `originating' source (a bright point
source, cf.\ Flag~2, or a large or irregular extended source, cf.\ Flag~3)
was located within the masked region. Though the brightest source was
fitted before the fainter ones, the contribution of the faint sources to
the fit of the bright source is considered to be negligible
(Sect.~\ref{det_emldetect}). Hence, the `originating' point source was
identified by setting its Flag~12\footnote{Note that Flag~12 was not set
when the source appeared to be split into two, cf.\ Sect.~\ref{extdsrc}, or
when a close-by fainter detection appeared to be of comparable brightness.}
to distinguish it from the other detections with Flag~11 in that particular
ds9-region, the parameters of which may be affected by the presence of the
indicated bright source due to imperfections in the PSF used.  In the case
of bright extended sources, however, the situation was different: the
extent parameter was obviously affected by nearby spurious detections, and
consequently the brightness was underestimated. Flag~12 was therefore only
set for point sources.

\subsection{Quality summary flag} \label{qualcolumns}

For easier use of the quality-flag information, the catalogue gives a
summary flag which combines the flags described above (11 per camera per
detection) to give a single, overall quality indication for each detection.
Its five possible values are as follows (in order of increasing severity):
\begin{enumerate}

\item[0:] There are no indications of problems for this detection; none of
the flags [$1-12$] for the three cameras [pn,M1,M2] are set to True.  This
value can be used to obtain the cleanest possible samples (but possibly at
the expense of omitting some otherwise acceptable detections). (71\% of all
detections.)

\item[1:] The source parameters are considered to be possibly compromised;
at least one of the warning flags [1,2,3,9] for any of the cameras
[pn,M1,M2] is True. This value can be used to accept detections for further
potential use, but they should be subjected to careful scrutiny dependent
on the specific application. (9\% of all detections.)

\item[2:] The detection may be spurious but was not recognised as such
during visual inspection; at least one of the automated `spurious
detection' flags [7,8] for any of the cameras [pn,M1,M2] is True but the
manual flag [11] is False. This value can be used to accept detections for
further potential use, but they should be subjected to careful scrutiny
dependent on the specific application. (1\% of all detections.)

\item[3:] The detection lies in a region where spurious detections occur
but which could not be dealt with in an automated way; the manual flag [11]
is True but the automated `spurious detection' flags [7,8] of all the
cameras [pn,M1,M2] are False. Detections with this value should be used
only after very careful scrutiny, as they may well be spurious, {\it
unless} flag~12 is True, in which case the detection (and possibly its
parameters) may well be valid, as it is likely to be a strong source. (15\%
of all detections, where Flag~12 was set for 600 detections.)

\item[4:] The detection lies in a region where spurious detections occur
and is flagged as likely spurious; the manual flag [11] is True and any of
the automated `spurious detection' flags [7,8] for any of the cameras
[pn,M1,M2] is also True. It is recommended that detections with this value
should not normally be used. (4\% of all detections.)

\end{enumerate}
Flag~12 was not included in the summary flag, selecting by Flag~12 as well can provide a clean as well as a
more complete sample, as
noted above, since this flag is usually given to reasonably bright point
sources.

The screening flags also offer a means of avoiding source-specific data products  with possible problems, noting that of all detections with products  a significant fraction have  summary flag $\ge 3$ indicating potential issues with the spectra and/or time series.

\subsection{Overall observation classification} \label{obsclass}

The summary flag assigned to each detection in the catalogue provides an overall classification of each detection
included in the catalogue. On the other hand, since about half of all observations in the
catalogue are little affected by artefacts and background subtraction
problems, an {\it observation classification} offers the possibility of
selecting good quality {\it fields} rather than good quality detections. This
classification is based on the fraction of area masked out in the flag mask
(Sect.~\ref{manflag}) as compared to the total area used in the source
detection (from the combined EPIC detection mask) for that observation. Six
classes of observations were identified. They are listed in
Table~\ref{obsclasstab} together with the percentage of observations
affected, the fractional area, and the approximate size of the
excluded region (note that the flag mask may comprise several regions in
various shapes).

\begin{table*}[tb]
\normalsize
\caption{Observation class definitions. 
}
\label{obsclasstab}
\small
\centering
\begin{tabular}{ccll}
\hline \hline
class & \%age of 2XMM obs & definition & comment \\
\hline
0 & 38\% & bad area = 0\%                & no region has been identified for flagging \\
1 & 12\% & 0\% $<$ bad area $<$ 0.1\% 	 &  $\la 3$ single detections \\
2 & 10\% & 0.1\% $\le$ bad area $<$ 1\%  & circular region with  $40\arcsec \la r \la 60\arcsec$ \\
3 & 25\% & 1\% $\le$ bad area $<$ 10\% 	 & circular region with $60\arcsec \la r \la 200\arcsec$ \\
4 & 10\% & 10\% $\le$ bad area $<$ 100\% &  circular region with $r \ga 200\arcsec$ \\
5 & \phantom{0}5\% & bad area = 100\%    & the whole field is flagged as bad
\\
\hline
\end{tabular}
\normalsize		
\end{table*}

\section{Catalogue compilation  } \label{catcompile}

The 2XMM catalogue is a catalogue of detections. As such, every row in the
2XMM catalogue represents a single detection of an object from a separate
XMM-Newton observation. The construction of the 2XMM catalogue consists of
two main steps. The first involves the aggregation of the data of
individual detections from the separate observation source lists into a
single list of detected objects, adding additional information about each
detection and meta-data relating to the observation in which the detection
was made.  The second step consists of cross-matching detections,
identifying resulting unique celestial objects and combining or averaging
key quantities from the detections into corresponding unique-source values.
Ultimately, the ensemble of data for both detections and unique sources
becomes the catalogue.

The primary source of data for the catalogue was the set of 3491 EPIC
summary source list files from the maximum-likelihood source-fitting
processes (Sect.~\ref{det_emldetect}). Additional information incorporated
into the catalogue for each detection includes the detection background
levels, the variability information (from the EPIC source time-series
files; see below) and the detection flags from the automatic flagging
augmented by the manual data screening process (see Sects.~\ref{manflag},
and~\ref{qualcolumns}). Ancillary information added to the catalogue
entries also includes various observation meta-data parameters (e.g.,
observation ID, filters and modes used) and the observation classification
determined as part of the data screening process (Sect.~\ref{obsclass}). In
the final catalogue table each detection is also assigned a unique
detection number.

The measured and derived parameters of the detections taken from the
pipeline product files are reflected in the 2XMM catalogue by a number of
columns described in Appendix~\ref{col_id} -- \ref{col_var}. For the
variability information for detections (Appendix~\ref{col_var}), the
variability identifier was set to True for a detection if at least one of
the time-series for this detection (derived from all appropriate exposures)
had a $\chi^2$-probability $\le\!10^{-5}$ based on the null hypothesis that
the source was constant (cf.\ Sect.~\ref{TS}). The probability threshold
was chosen to yield less than one false trigger over the entire set of
time-series. Where the flag was set, the camera and exposure ID with the
lowest $\chi^2$-probability were also provided for convenience.  No
assessment of potential variability has been made between observations for
those sources detected more than once.

\subsection{Unique celestial sources } \label{detmatching}

XMM-Newton observations can yield multiple detections of the same object on
the sky where a particular field is the subject of repeat pointings or
because of partial overlaps from dedicated mosaic observations or
fortuitous overlaps from unrelated pointings. As such, the catalogue
production process also sought to identify and collate data for all
detections pertaining to unique sources on the sky, providing a
unique-source indexing system within the catalogue. In parallel, the
catalogue provides a number of derived quantities relating to the unique
sources computed from the constituent detections.

To identify unique sources from multiple detections, reliable estimates of
the position error, $\sigma_{\rm pos}$, of each detection are
essential. The best estimate of the position error was found to be
\begin{equation}
\sigma_{\rm pos}= \sqrt{\sigma^2_{\rm sys} + \sigma^2_{\rm stat}} \, ,
\label{poserreq}
\end{equation}
where $\sigma_{\rm sys}$ is the additional error (Sect.~\ref{astcorr}, see also footnote~\ref{sysfoot}) and
$\sigma_{\rm stat}$ is the statistical centroid uncertainty measured from
the source-fitting stage (Sect.~\ref{det_emldetect}).

Two detections from different observations with respective position errors
of $\sigma_{\rm 1}$ and $\sigma_{\rm 2}$ were assumed to be potentially
associated with the same celestial source if their separation is
\[ r_{\rm sep} < 3(\sigma_{\rm1} + \sigma_{\rm 2}) \, ,\]
with $7\arcsec$ as an upper-limit. The $7\arcsec$ limit to position offsets
in the matching process was determined empirically as the best value to
prevent spurious matches (dominated by a few weak extended
sources with large position errors) without having a significant effect on
the number of genuine matches. A match was, however, rejected if $r_{\rm
sep} > 0.9 d_{\rm 1}$ or $r_{\rm sep} > 0.9 d_{\rm 2}$ where $d_{\rm 1}$
and $d_{\rm 2}$ are the distances from the detection to its nearest
neighbour in the same observation. This latter provision means that no two
distinct sources from the same image should be matched.  No quality flag
information was used in the matching process.

Using these constraints, the detection table was cross-correlated with
itself to find all possible pairs of detections having error-circle
overlaps. Some detections were found to have as many as 31 such overlaps,
since a few areas of sky were observed this many times (generally
calibration observations). Resolving this list into a set of unique
celestial sources required some experimentation because of potential
ambiguity in a few crowded or complex fields.  The extreme scenarios were
1) to assume a set of detections was associated with a unique source only
if they all overlapped each other -- this was considered too conservative;
2) to assume that a set of detections constituted a unique source if each
member overlapped at least one other member -- this was deemed overly
generous, i.e., it would have included a few pairs of detections whose
mutual separations would be incompatible with coming from a single
source. The algorithm adopted gave priority to those detections with the
highest number of overlaps (because they were likely to be near the true
source centre) and, this number being equal, to count-rate agreement. The
list of overlapping detections was therefore sorted in descending order of
the number of overlaps and the EPIC total-band count rate and then
processed in that order. Each detection was associated with all its
overlapping detections, except those which had already been removed from
the list by having been associated with another (better connected or
stronger) detection. In the final catalogue the number of detections which
might have been associated with a source different from the one actually
assigned to them, given a different order of processing, was about one
hundred, which was significantly lower than the figure from various
alternative algorithms. These ambiguous detections were almost all from
observations which the screening process flagged as unreliable, suggesting
that further refinements to the algorithm would have been of little
practical value.

The algorithm adopted for the identification of unique sources appears to
be reliable in the great majority of cases, but there are known to be a few
confused areas where the results are likely to be imperfect. The most
common cause is where real diffuse or bright objects give rise to
(generally spurious) additional detections which happen to approximately
coincide spatially in different observations. In most cases it is likely
that the sources will have received a manual flag. Incorrect matching can
also potentially occur where centroiding is adversely affected by pileup or
optical loading, where one or more contributing observations have
significant attitude errors which could not be astrometrically rectified
(Sect.~\ref{astcorr}), or where a real source is located close to another
detection associated with an artefact such as residual OOT events from a
strongly piled-up source elsewhere in the image. Where pileup or artefacts
are involved, affected sources may have been assigned automatic or manual
flags anyway. It should be emphasised, however, that flag information is
not used in the source matching process. Based on the extensive visual
inspection, incorrect detection matching is believed to be extremely rare
($<\!200$ detections affected). Inevitably, in a few cases, the matching
process fails to match some detections that belong together.

A number of quantities for unique sources are included in the 2XMM
catalogue, based on error-weighted merging of the constituent detection
values (see Appendix~\ref{col_unsrc}). The IAU name of each unique source
was constructed from its coordinates. Note that an individual detection is
completely specified by its IAU name {\it and} its detection identifier.
The unique-source data were augmented with five quantities that were not
based on error-weighted merging: 1) the unique-source detection likelihood
was set to the highest EPIC total-band detection likelihood, i.e., it
reflects the strongest constituent detection of a unique source. 2) A
unique-source extent likelihood was computed as the simple average of the
corresponding EPIC detection values. 3) The reduced $\chi^2$-probability
for the variability of a unique source was taken as the lowest of the
detection values, indicative of the detection with the highest likelihood
of being variable, where variability information was available. 4) Where
variability information existed for any of the constituent detections, a
unique-source variability identifier was set to True if any were True and
to False if none were True. Where no variability information was available,
the unique-source flag was set to Undefined. 5) A unique-source summary
flag took the maximum of the detection summary flag values
(Sect.~\ref{qualcolumns}), i.e., reflecting the worst-case flag from any of
the detections of the source.

The 2XMM catalogue was also cross-correlated against the 1XMM and 2XMMp
catalogues during the construction process. For each unique 2XMM source,
the most probable matching 1XMM counterpart and 2XMMp counterpart were
identified and listed in the 2XMM catalogue. The matching algorithm
employed was similar to the one described for identifying unique sources
but the maximum positional offset between the new catalogue and the older
ones was set at $3\arcsec$. This was a rather conservative value but since
a number of sources in 1XMM, especially, have positional errors greater  than
this, it ensures that there are very few incorrect matches or ambiguous
cases.

This resulted in $\sim\!88$\% of all 2XMMp sources having a match with 2XMM
sources. Apart from those lying outside the $3\arcsec$ matching circle,
non-matched sources are found to be either spurious, at the detection
limit, or the observation was not included in 2XMM. Comparison with 1XMM is
not straight forward due to the differences in the detection scheme (e.g.,
the source detection in 1XMM was done per camera) and likelihood cutoffs.
Note, though, that 1XMM comprises only 585 of the 2XMM fields.

\section{Catalogue characterisation \& results }  \label{catchar}

\subsection{Overall properties}  \label{properties}

The catalogue contains 246\,897 detections drawn from 3491 public
XMM-Newton observations (Fig.~\ref{skymapfig}). These detections relate to
191\,870 unique sources. Of these, 27\,522 X-ray sources were observed more
than once; some were observed up to 31 times in total due to the fact that
many sky regions are covered by more than one observation. Of the 246\,897
X-ray detections, 20837 are classified as extended. Table~\ref{dettab}
shows  the number of
detections and unique sources per camera and energy band (split into point sources and extended sources); a 
likelihood threshold $L\ge 10$ has been applied but no selection of detection flags has been made.

The catalogue contains detections down to an EPIC likelihood of 6. Around 90\% of the detections have $L>8$ and
$\sim\!82$\% have $L>10$. Simulations demonstrate that the false detection rate for typical high Galactic latitude fields is $\sim [2, 1, 0.5]$\% for
detections with $L>[6, 8, 10]$ respectively (Sect.~\ref{simulations}). We note that the source detection in 2XMM  has
a low degree of incompleteness 
$L\lesssim 10$. This arises from the fact that the first stage of the source detection
(Sect.~\ref{boxmapmode}) requires that each detection have $L\ge 5$. As this first stage of the processing is
relatively crude, the incompleteness primarily arises from this preselection of low significance detections. 

The 2XMM catalogue is intended to be a catalogue of serendipitous
sources. The observations from which it has been compiled, however, are of
course pointed observations which typically contain one or more target
objects chosen by the original observers, so the catalogue contains a small
fraction of targets which are by definition not serendipitous.
Appendix~\ref{newtargbit} provides details of the target identification and
classification. From this analysis we find that around 2/3 of the intended
targets are unambiguously identified in their XMM-Newton observations but,
allowing for multiple detections, only $\sim\!1400$ targets are plausibly
associated with 2XMM catalogue sources. This means that $<\!1$\% of 2XMM
sources are the target of the observation, although in a few observations
(e.g., nearby galaxies) the number of sources {\it associated} with the
target can clearly be much greater.

More generally the fields from which the 2XMM catalogue is compiled may
also not be representative of the overall X-ray sky. The classification of
the XMM-Newton observations (Appendix~\ref{newtargbit} and
Table~\ref{targettab}) is relevant to avoiding potential selection bias in
the use of the catalogue.

\begin{table}[tb]
\caption{
Numbers of detections with likelihood $L\ge10$ in the 2XMM catalogue.}
\label{dettab}
\small
\centering
\tabcolsep 1mm
\begin{tabular}{lc@{\extracolsep{1mm}}cccc}
\hline
  \multicolumn{1}{c}{Camera} &
  \multicolumn{1}{c}{Energy } &
  \multicolumn{1}{c}{Point} &
  \multicolumn{1}{c}{Ext'd } &
  \multicolumn{1}{c}{Unique point } &
  \multicolumn{1}{c}{Unique ext'd } \\
  \multicolumn{1}{c}{} &
  \multicolumn{1}{c}{band (keV)} &
  \multicolumn{1}{c}{source} &
  \multicolumn{1}{c}{source} &
  \multicolumn{1}{c}{source} &
  \multicolumn{1}{c}{source} \\
\hline
  pn & $0.2-\phantom{0}0.5$ & 38074 & 4319 & 30811 & 3843\\
  pn & $0.5-\phantom{0}1.0$ & 63248 & 7457 & 50639 & 6714\\
  pn & $1.0-\phantom{0}2.0$ & 68197 & 6217 & 55035 & 5555\\
  pn & $2.0-\phantom{0}4.5$ & 37511 & 3604 & 30702 & 3167\\
  pn & $4.5-12.0$ & 11144 & 1586 & \phantom{0}8682 & 1337\\
  M1 & $0.2-\phantom{0}0.5$ & 20841 & 3392 & 15887 & 2958\\
  M1 & $0.5-\phantom{0}1.0$ & 40965 & 6734 & 30998 & 5892\\
  M1 & $1.0-\phantom{0}2.0$ & 52569 & 6754 & 40062 & 5882\\
  M1 & $2.0-\phantom{0}4.5$ & 34230 & 4452 & 26710 & 3858\\
  M1 & $4.5-12.0$ & \phantom{0}7818 & 1825 & \phantom{0}5776 & 1547\\
  M2 & $0.2-\phantom{0}0.5$ & 20626 & 3485 & 15718 & 3012\\
  M2 & $0.5-\phantom{0}1.0$ & 42488 & 7045 & 32055 & 6149\\
  M2 & $1.0-\phantom{0}2.0$  & 56060 & 6997 & 42624 & 6107\\
  M2 & $2.0-\phantom{0}4.5$ & 36760 & 4703 & 28538 & 4080\\
  M2 & $4.5-12.0$ & \phantom{0}8546 & 2008 & \phantom{0}6265 & 1716\\
\hline\end{tabular}
\end{table}

\subsection{Sky coverage and survey sensitivity }  \label{skycov} 

To compute the effective sky coverage, the sky was notionally covered by a
grid of pixels using the HEALPix projection (Gorski et al. 2005). Adequate
resolution was obtained using pixels $\sim\!13\arcsec$ across. For each
observation included in 2XMM the exposure times were computed for each
HEALPix pixel taking into account the exposure map for each observation
(i.e., the actual coverage taking into account observing mode, CCD gaps,
telescope vignetting, etc.). From this analysis we find that in total the
catalogue fields cover a sky area of more than 500\,deg$^2$. The
non-overlapping sky area is $\sim\!360$\,deg$^2$ ($\sim\!1$\% of the sky).

\begin{figure}[t]
\resizebox{\hsize}{!}{\includegraphics{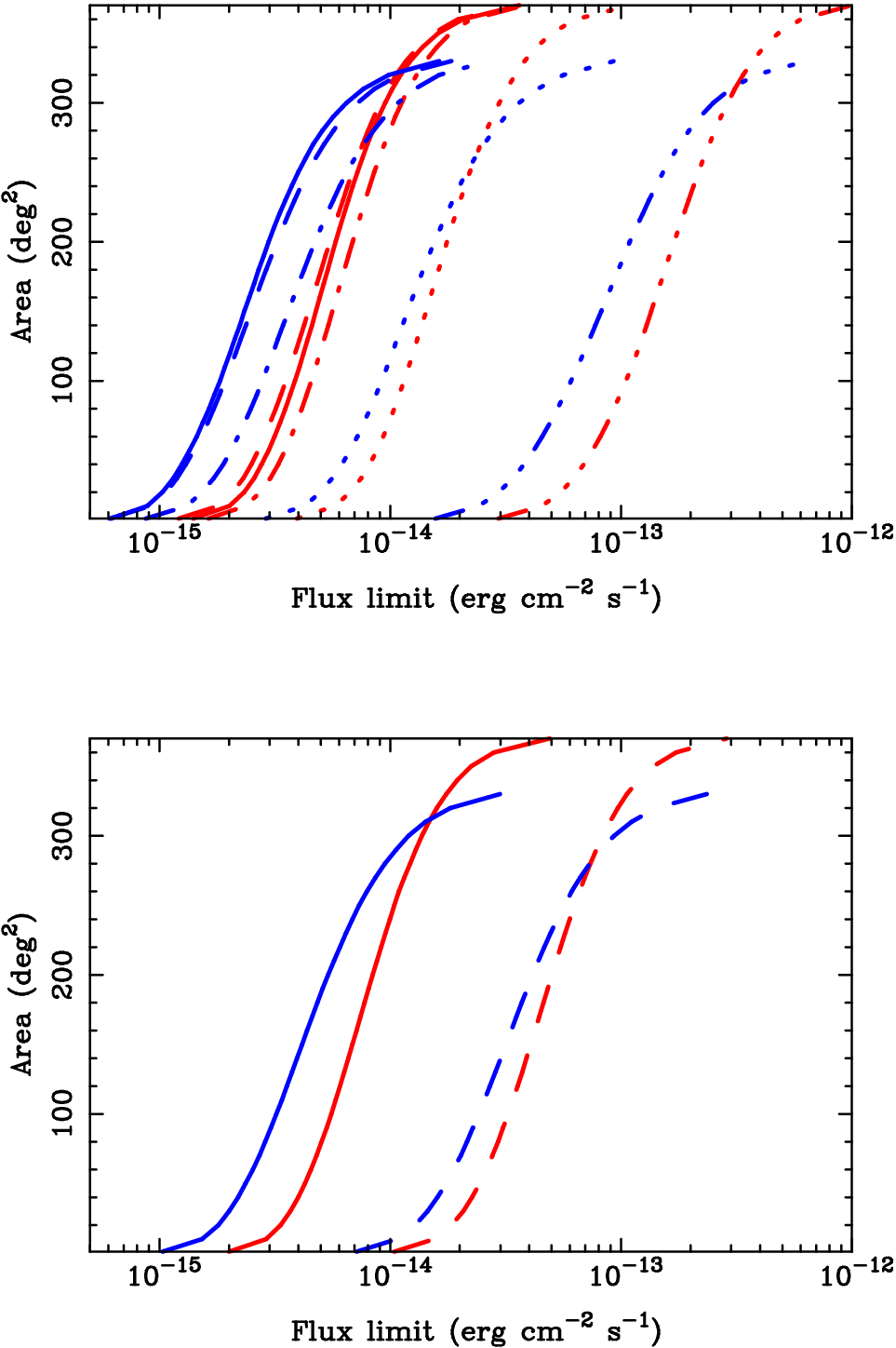}}
\caption[]{Sky area as a function of flux limit for the 2XMM catalogue
computed for sources with a detection likelihood limit $L\ge10$ in the
respective energy band. Red curves are for MOS2; blue curves are for
pn. (MOS1 is not shown but is very similar to MOS 2). \\ {\sl Top panel}:
Energy bands 1, 2, 3, 4, \& 5 for each camera are shown with solid,
long-dash, dash-dot, dotted, \& dot-dot-dot-dashed line styles,
respectively.\\ {\sl Bottom panel}: Energy bands 6 \& 7 for each camera are
shown with solid \& long-dash styles, respectively. 
}
\label{fluxareafig} 
\end{figure}

The sensitivity of the 2XMM survey catalogue was estimated empirically
using the method of Carrera et al.\ (2007). The algorithm presented in
their Appendix~A was used to compute sensitivity maps for each instrument
and energy band, using data from the exposure maps and background maps from
each observation. Using a grid of HEALPix pixels in a similar way to that
outlined above, the limiting flux of the {\it most sensitive} observation
of each part of the sky was estimated.  Figure~\ref{fluxareafig} shows the
sky area against limiting flux for each EPIC camera and energy band
separately. This analysis provides a relatively robust estimate of the
total sky area of the 2XMM catalogue for each of the three EPIC cameras,
although it does not take into account those sky regions which are
effectively useless for serendipitous source detection due to the presence
of bright objects or certain instrumental artefacts (see discussion in
Sect.~\ref{obssel} and Fig.~\ref{examplesfieldfig}b and c). \footnote{Figure~\ref{fluxareafig} also does not take into account the effects of Poisson noise which produces a  probability distribution for source detectability about the sensitivity limit. These effects are only important at the low count limit, i.e. essentially only at faint fluxes,  cf. Georgakakis et al. (2008). }
These area-flux
plots computed for $L>10$ show that the effective sky coverage for the MOS2
camera is $\sim\!370$\,deg$^2$ (for the MOS1 camera it is
$\sim\!360$\,deg$^2$ due to the loss of one of the MOS1 CCDs in March
2005), whilst for the pn camera the area is $\sim\!330$\,deg$^2$, due
primarily to reduced or zero imaging sky area provided by some of the pn
observing modes. The limiting fluxes vary between camera and energy
band. For the pn camera which provides the highest sensitivity, the minimum
detectable fluxes in the soft ($0.5-2$ keV), hard ($2-12$ keV) and hardest
($4.5-12$ keV) bands at 10\% sky coverage are $\sim$~[2, 15, 35]~$\times
10^{-15}\rm\,erg\,cm^{-2}\,s^{-1}$, respectively. The fluxes for $>\!90$\%
sky coverage (i.e., close to complete coverage) in these bands are $\sim$
[1, 9, 25]~$\times 10^{-14}\rm\,erg\,cm^{-2}\,s^{-1}$ respectively.

\subsection{Flux and count distributions  }\label{fluxdist}

\begin{figure}[t] 
\resizebox{\hsize}{!}{\includegraphics{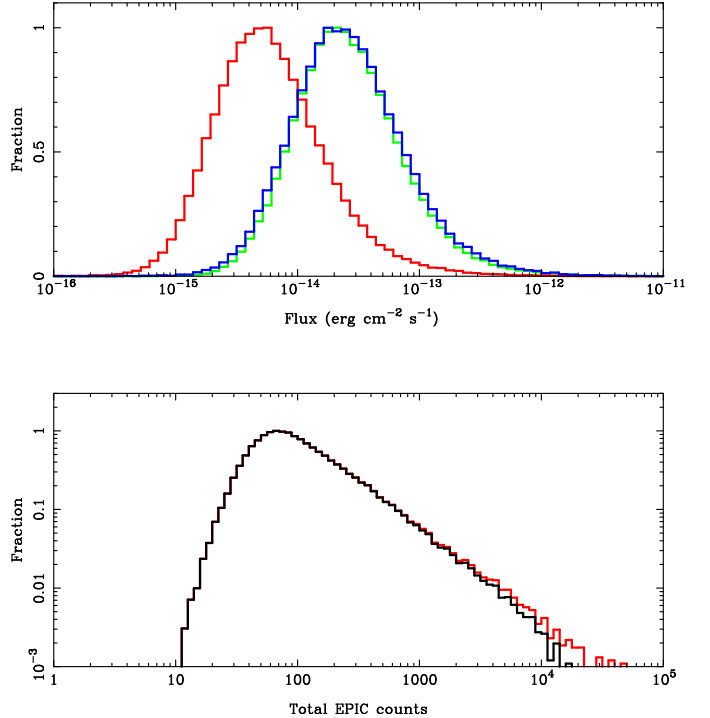}} 
\caption[]{Top: Distribution of point source fluxes for the 2XMM catalogue
in the soft (red), hard (blue), and total band (green) energy bands. The
targets of the individual XMM-Newton observations are excluded from these
distributions (see Sect.~\ref{formtargetid}). Detections selected for these
distributions have likelihood $L\ge 10$ in the relevant bands. Only sources
with summary flag 0 are included. Bottom: distribution of total EPIC counts
for the same sample of 2XMM detections.The red histogram shows the
distribution if the XMM-Newton targets are included.
}
\label{flux_dist}  
\end{figure}

The distribution of fluxes for the 2XMM catalogue detections is shown in
Fig.~\ref{flux_dist}. This figure illustrates that the typical soft-band
flux for the catalogue sources is
$\sim\!5\times10^{-15}\rm\,erg\,cm^{-2}\,s^{-1}$ and is
$\sim\!2\times10^{-14}\rm\,erg\,cm^{-2}\,s^{-1}$ in the hard and total
bands. These values correspond quite closely to the fluxes of the sources
which dominate the cosmic X-ray background (where the slope of the
extragalactic source counts breaks), demonstrating the importance of 2XMM
in providing large samples at these fluxes.
 
Also shown in Fig.~\ref{flux_dist} is the distribution of total counts in
the combined EPIC images for the same sample of 2XMM detections. As
expected the distribution is dominated by low count sources, with the peak
lying at $\sim\!70$ counts. This plot also illustrates the effect of the
targets of the XMM-Newton fields themselves which only contribute
significantly, not surprisingly, above $\sim\!2000$ EPIC counts.

We note that it would be possible to combine the survey sensitivity curves
discussed in Sect.~\ref{skycov} and the flux distributions discussed here
to construct the source counts (i.e., the $\log N - \log S$ relationship)
for the 2XMM catalogue. In practice, however, the results of this exercise
would have limited value due to the large uncertainties in the correct
area-sensitivity corrections for the substantial number of fields included
in 2XMM which contain, for example, bright objects or are subject to
problematic instrumental effects. A separate paper, Mateos et al.\ (2008),
presents the $\log N - \log S$ relationship and results for a carefully
selected subset of the 2XMM fields at high Galactic latitudes.

\subsection{False detection rate \& likelihood calibration } \label{simulations}

The significance of the source detection in the 2XMM catalogue is
characterised by the maximum likelihood parameter for the detection, $L$
(cf.\ Sect.~\ref{det_emldetect}). Although the detection likelihood
values are formally defined in terms of the probability of the detection
occurring by chance, the complexity of the data processing implies that the
computed likelihoods need to be carefully assessed.
To investigate the calibration of the likelihood values and  the
expected false detection rate, we thus carried out realistic Monte-Carlo
simulations of the 2XMM catalogue source detection and parameterisation
process.  The simulations performed were chosen to represent typical
high-latitude fields without bright sources  
or extended X-ray emission  apart from the unresolved cosmic X-ray
background.  The simulations include a particle background component and a
distribution of X-ray point sources with uniform spectral shape drawn from
a representative extragalactic $\log N - \log S$ relationship (eg. Hasinger et al., 2001).
 The source spectrum assumed is a power law characterised by $\Gamma=1.7$ with a Galactic
column density $N_H=3\times 10^{20}\rm\ cm^{-2}$. 

The simulation creates images (and exposure maps etc.)  in the five
standard energy bands using the appropriate calibration information (i.e.,
energy- and position-dependent PSFs, vignetting, detection efficiency,
etc.). The simulated data are then processed with exactly the same steps
used in the actual 2XMM pipeline (Sect.~\ref{processing}) and the derived
source parameters, such as likelihoods, were compared with the input (i.e.,
simulated) parameters.
\begin{figure}[hbt]
\centering
\resizebox{\hsize}{!}{\includegraphics[angle=270]{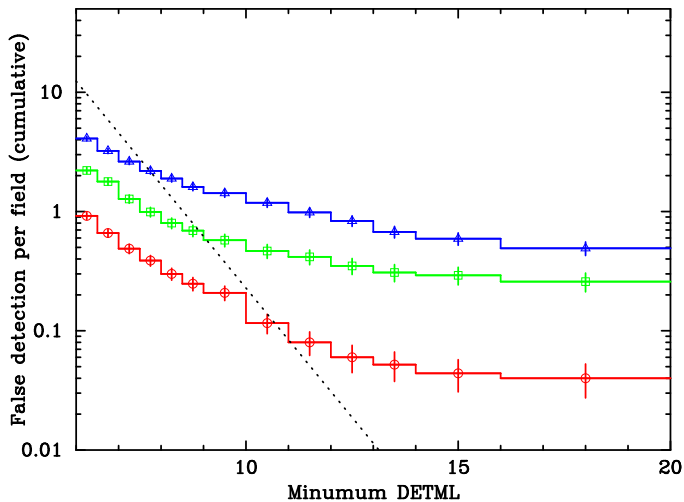}} 
\caption[]{
The number of false detections per field estimated via simulations for
typical high Galactic latitude fields as a function of $L_{\rm min}$ for various
exposure times. The red circles show the results for exposures
of 12\,ks for MOS and 8\,ks for pn ($\sim 70$\% of the median values), whereas the green squares and blue triangles
show those with the exposures of 3 and 10 times higher
respectively. The dotted line represents the theoretical false detection
number assuming 5,000 independent detection cells per field (see text).
}
\label{fig:cumulated-false-detection}
\end{figure}

Figure~\ref{fig:cumulated-false-detection} shows the number of false
detections per field derived from the simulations as a function of the
minimum $L$ for three different exposure times: 12\,ks for MOS and
8\,ks for pn, corresponding to around 70\% of the median exposure, and
three and ten times higher exposure values. Also shown is the expected
false detection number $n$ for an assumed $N_c=5,000$ independent
detection cells per field, calculated simply as $n=N_c\ .\
exp(-L)$. The value of $N_c$ of course depends on the effective
'beam-size' for EPIC observations. The value $N_c=5,000$ we adopt is
based on the area of the search box ($20\arcsec \times 20\arcsec$,
Sect.~\ref{boxlocalmode}), corrected downwards to take into account
the degradation and change of shape of the PSF off-axis. This value is
a factor $\sim 4$ times less than would be derived from assuming
the beam-size is of the order the PSF width (eg. 15\arcsec\ HEW),
highlighting that this is a poorly defined quantity.

The results shown in Fig.~\ref{fig:cumulated-false-detection}
demonstrate: (i) the number of false detections per field is low even
for $L\ge 6$ ; (ii) the dependence of the number of false detections
on $L$ is much flatter than simple expectations; (iii) the number of
false detections depends on the exposure time.

For typical observations included in the catalogue (represented by the
red curve in Figure~\ref{fig:cumulated-false-detection}), the number
of false detections is $\sim [1,0.3, 0.1]$ per field at likelihood
limits of $L\ge[6, 8, 10]$ respectively. These values increase to $\sim
[4, 2, 1.5]$ for the longest exposure time represented in
Figure~\ref{fig:cumulated-false-detection}. For each of the three
exposure times adopted, we also compared the numbers of false
detections with the average number of sources detected in
corresponding exposures of typical XMM-Newton high Galactic latitude
fields, i.e. $\sim [60, 100, 200]$ sources per field, to derive false
detection {\it rates}. We find that these rates have only a low
dependence on the exposure time, ie. the false detection {\it rate} is
approximately constant at $\sim [2, 1, 0.5]$\% for likelihood limits
$L\ge [6, 8, 10]$ over the range of exposures investigated.

Our simulation results can be compared with the analysis presented by  Brunner et al. (2008), carried out in the
context of the very deep XMM-Newton  observation of the Lockman Hole. Their simulations are for a detection approach
similar to that presented here and their results are also broadly similar (cf. their Fig.4 which shows a qualitatively
similar dependence of false number with likelihood), albeit they are presented for different energy bands. The
number of false detections in their simulations is higher, but of course corresponds to an observation with an
exposure time $\sim 100$ times longer. Brunner et al. comment that the significant difference between the simulation 
results and simple expectations primarily originates in the multi-step detection procedure (which introduces two
effective detection thresholds) and the 
simultaneous multi-band fitting of source positions and fluxes, both of which result in a
reduction of the effective number of independent trials. The fact that the number of false 
detections depends on the exposure time is not in line with simple expectations, but is probably a
reflection of a combination of Eddington bias and source confusion effects. The much larger than expected false detection numbers at $L>12$ may arise from a too stringent matching criterion between the input and output sources in the simulations. Other similar studies of the false detection rate in XMM-Newton observations
include 
Loaring et al. (2005) for the relatively deep XMM-Newton 13$^{\rm H}$ field  and 
Cappelluti et al. (2007) for the COSMOS field. Both studies determined false detection rates which are 
somewhat higher than our estimates for 2XMM, but these can be reconciled with detailed differences in the
assumptions made in these studies.

We also investigated the sensitivity of the false detection number to 
the background and to the assumed spectral shape. The largest differences are an increase by a factor $\sim 2$ at the lowest
likelihoods ($L<8$) for background conditions 3 times higher than typical. Assuming much softer or harder spectral shapes
produces a similar increase in the false detection number, again restricted to the lowest likelihood bins. 

In addition to the false detection rate and calibration of the likelihood values, these simulations also provide a means to address the issue of catalogue completeness, ie. the effects of Poisson noise which produces a  probability distribution for source detectability about the sensitivity limit. This study is beyond the scope of the current paper, but we note that completeness corrections relating to these source detection biases are expected to be small except at the lowest fluxes, cf. Georgakakis et al. (2008).

The simulation work also allows us to address the astrometric performance
of the processing. Comparison of the input and output positions shows that:
(i) there is no measurable average offset; (ii) the distribution of
position offsets closely follows the expected statistical form (cf.\
Sect.~\ref{astprop}), validating the statistical position error
estimates. This distribution does, however, show offsets that are
statistically too large for simulated sources with position errors
$\leq\!0.5\arcsec$. The origin of this effect is unclear, although it may
be related to the discrete sampling of the PSF representation 
in the XMM-Newton calibration data. 

Full details of the evaluation of the 2XMM catalogue with the simulations
will be presented elsewhere (Sakano et al., in preparation).

\subsection{Astrometric properties } \label{astprop} 
 
In order to investigate the overall astrometric accuracy of the 2XMM
catalogue, in particular the extent to which the position error estimates
correctly describe the true positional uncertainty, we tested the catalogue
positions against the Sloan Digital Sky Survey (SDSS) DR5 Quasar Catalog
(Schneider et al. 2007) which contains 77\,429 objects classified as
quasars by their SDSS optical spectra. The sky density of the Sloan quasars
is $\sim\!10$ per square degree, and their positional accuracy is better
than $0\farcs1$, making this an excellent astrometric reference set. This
approach has the advantage that XMM-Newton is expected to detect a large
fraction of all Sloan quasars in X-rays (especially at the bright magnitude
limit for SDSS spectroscopy) and thus, {\it a priori}, it seems safe to
assume that essentially all positional matches are actually real
associations and that the SDSS provides the true celestial position of the
object.

To carry out the analysis, the 2XMM catalogue was cross-correlated with the
DR5 Quasar Catalog, keeping all matches within $20\arcsec$. This produced
around 1600 matches, corresponding to 1121 unique 2XMM sources. The total
sky area for the matches (out to $20\arcsec$ radius) was
$\sim\!0.2$\,deg$^2$.  Given the sky density of Sloan quasars this
translates to $\sim\!2$ false matches overall, or $\sim\!0.5$ false matches
if we use just the inner $10\arcsec$ of the distributions. We can thus be
confident that the false match rate is negligible for this
investigation. This is the real advantage of using Sloan quasars over other
comparison catalogues.

For the astrometry evaluation a subset of these matches was used with
detection likelihood $L\ge 8$, summary flag 0, off-axis angle
$<\!13\arcmin$, and excluding extended sources. These selections reduce the
total number of detection matches to 1007 (corresponding to 656 unique
sources).

\begin{figure}[t]
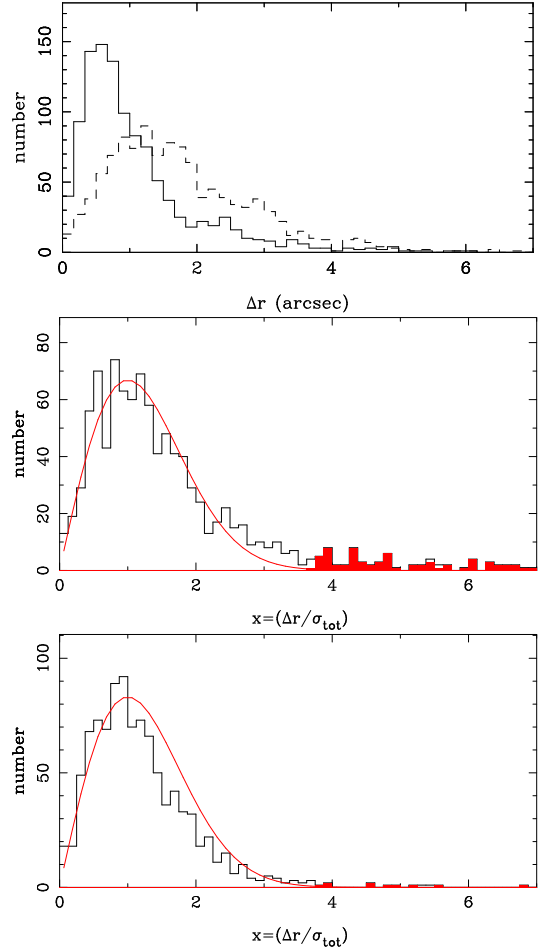
 
\begin{center}
\resizebox{7cm}{!}{\includegraphics[angle=270]{astrom_f1.ps}}
\resizebox{7cm}{!}{\includegraphics[angle=270]{astrom_f2.ps}}
\resizebox{7cm}{!}{\includegraphics[angle=270]{astrom_f3.ps}}
\caption[]{ Top: X-ray/optical position separation for each match for the
corrected (solid histogram) and uncorrected (dashed histogram) XMM-Newton
coordinates.  Centre: Distribution of separation sigma ($x$) for
$\sigma_\mathrm{sys}=0$. Bottom: Distribution of separation sigma ($x$) for
$\sigma_\mathrm{sys}=0\farcs35$. For the centre and bottom plots --
histogram: separation sigmas; filled histogram: outliers with $x >3.7$ (and
$\Delta r<10\arcsec$); smooth curve: expected distribution $N(x)$
normalised to fit the peak of the distribution.
}
\label{astrom}
\end{center}
\end{figure}

Figure~\ref{astrom} shows the distribution of the X-ray/optical position
separations for each match for both the corrected and uncorrected 2XMM
coordinates. As can be seen, the uncorrected separations peak at
$\sim\!1\farcs5$ and show a broad distribution out to $4\arcsec-5\arcsec$,
whereas the corrected separations peak at $<\!1\arcsec$ and show a narrower
spread. This result of course reflects the overall success of the
astrometric rectification carried out as part of the processing
(Sect.~\ref{astcorr}).

To make a more detailed comparison of the observed and expected
distributions, we consider the separations normalised by the position
errors. If we define $x=\Delta r/\sigma_\mathrm{pos}$ where $\Delta r$ is
the angular separation and $\sigma_\mathrm{pos}$ is the total position
error, the expected distribution function $N(x)$ takes the form
\[
N(x)dx \propto x \ {\rm e}^{-x^2/2}dx \, . 
\]
Thus comparing the empirical $N(x)$ distribution with the expected form
provides a means to determine the correct $\sigma_\mathrm{pos}$ value.  We
expect $\sigma_\mathrm{pos}$ to have two components:
$\sigma_\mathrm{stat}$, the statistical error already determined in the
maximum likelihood fit (Sect.~\ref{det_emldetect}) and a possible
additional, residual component, $\sigma_\mathrm{sys}$ (see comment on nomenclature: footnote~\ref{sysfoot}), to take into account any
residual errors in the position determination and correction process, cf.\
Sect.~\ref{astcorr}. Although it is not completely obvious how
$\sigma_\mathrm{stat}$ and $\sigma_\mathrm{sys}$ should be combined because
the nature of the residual error is formally not known, the analysis
reported here assumes eq.~(\ref{poserreq}), cf.\ Sect.~\ref{detmatching}
(other assumptions such as the linear combination of the errors provide
worse fits to the distributions).

Figure~\ref{astrom} (centre) shows the distribution, for corrected
XMM-Newton coordinates only, of the X-ray/optical position separation
sigmas (i.e., $x=\Delta r/\sigma_\mathrm{pos}$) for the matched detection
sample assuming $\sigma_\mathrm{sys}=0$. Although the observed distribution
is reasonably close to the expected form at low $x$-values, there is a long
tail of outliers at $x >3.7$ amounting to $\sim\!8$\% of the total sample,
whereas we would expect $<\!0.1$\% to lie at $x >3.7$. More detailed
investigation of these outliers shows that they are dominated by sources
with low $\sigma_\mathrm{stat}$-values (mostly $<\!0\farcs5$), clearly
indicating the need for an additional component, $\sigma_\mathrm{sys}$, of
the order $0\farcs5$.

We investigated a range of possible values of $\sigma_\mathrm{sys}$ and
found that $\sigma_\mathrm{sys}=0\farcs35$ provides the best overall fit
between the observed and expected distributions, as is shown in
Fig.~\ref{astrom} (bottom). For this choice of $\sigma_\mathrm{sys}$ there
are still more outliers at large $x$-values than expected if the position
errors were perfectly described, but we find that at least some of these
can be explained on astrophysical grounds (e.g., source confusion, lensed
objects), so regard our choice as the best overall value to represent the
global additional error estimate for the catalogue.

A detailed comparison between the observed and expected distributions
(Fig.~\ref{astrom}) shows that there is a deficit of points at low
$x$-values and indeed this is true for any $\sigma_\mathrm{sys}>0$. This
indicates that the true value of the statistical position error,
$\sigma_\mathrm{stat}$, is slightly overestimated by the fitting routine.
Attempts to model this effect with a simple rescaling of the
$\sigma_\mathrm{stat}$-value were not successful. We note that the typical
error estimate of the rectification of the XMM coordinates is
$\sim\!0\farcs6$ with a spread from $0\farcs3$ to $>\!1 \arcsec$. This
suggests that most of the additional error component needed is related to
the rectification residuals, with other effects being at a lower
level. An obvious alternative approach is thus to use the explicit values
of the errors determined by the rectification algorithm for
$\sigma_\mathrm{sys}$ (which thus vary from field to field and indeed from
source to source if the error in the field rotation is taken into account)
instead of the empirically determined -- and fixed value -- described
above. Overall this approach gives similar results, but gives $x$-values
which are systematically significantly too low, implying the
uncertainties derived by the rectification algorithm may also be
significantly over-estimated (by up to 50\%). We conclude that using a
fixed value of the additional error provides the best empirical description
of the data. On this basis the value $\sigma_\mathrm{sys}=0\farcs35$ was
adopted for the 2XMM catalogue. The total position error given in the
catalogue, $\sigma_\mathrm{pos}$, combines the statistical and additional
errors in quadrature, see eq.~(\ref{poserreq}). We note that the effect
described here may be identical to that discovered through the simulation
work described in Sect.~\ref{simulations}. If this is the case it would
imply that the residual errors associated with the rectification must
indeed be rather lower than the formal estimated values overall.

We repeated the analysis described above for the {\it uncorrected} XMM
coordinates to determine the $\sigma_\mathrm{sys}$-value appropriate to
those XMM-Newton fields for which astrometric rectification was not
possible (see Sect.~\ref{astcorr}). For the uncorrected XMM-Newton
coordinates we determine a good fit between the observed and expected
distributions for $\sigma_\mathrm{sys}=1\farcs0$. This value is adopted in
the catalogue for sources in those fields for which astrometric
rectification was not possible.

For completeness we looked for possible correlations between outliers and
the obvious XMM-Newton detection parameters (e.g., detection likelihoods,
off-axis angle). Rather surprisingly no clear correlations were found,
except with off-axis angle where it was noted that detections at very high
off-axis values ($>\!15\arcmin$) were somewhat more likely to have
statistically too large separations. By no means all high off-axis
detections are affected in this way, however. Essentially this means that
the statistical position error estimates are robust over a very wide range
of detection parameters and a single additional error component provides a
very adequate representation of the data.  Finally we note that properties
of the 2XMM/Sloan DR5 Quasar sample are reasonably representative of the
whole 2XMM catalogue. There is a bias towards higher X-ray fluxes and thus
lower statistical position errors, but a significant number of lower flux
objects are included and the full range of total counts and likelihoods is
sampled.

\subsection{Photometric properties  }

We have evaluated the flux cross-calibration of the XMM-Newton EPIC cameras
based on the calibration used to compute 2XMM fluxes (see
Sect.~\ref{flux_computation}). To do this we performed a statistical
analysis, comparing the fluxes between cameras for sources common to both,
selected from the entire FOV. The parameter used to quantify the difference
in flux was defined as $(S_i-S_j)/S_j$, where $S_i$ and $S_j$ are the
fluxes of the sources in each pair of cameras ($i$,$j$).
 
To minimise the impact of other effects, we performed the following
filtering on the comparison samples:
\begin{enumerate}

\item We used only point sources, as the uncertainties in the measured flux
for extended sources are much higher.

\item We used only sources having at least 250 counts in the energy band
and for each camera. This requirement was used to avoid Eddington bias
effects (an increase in the measured flux due to statistical fluctuations).

\item We did not use sources with a 2\,--\,12~keV flux ${\rm \ga
6\times10^{-12}\,erg\,cm^{-2}\,s^{-1}}$ as these objects suffer from pileup
and therefore their measured flux is underestimated.

\end{enumerate}

The distributions obtained were fitted with a Gaussian profile, which in
all cases provided a good representation of the data. The best-fit mean
values obtained from each distribution are listed in Table~\ref{flux_comp}.

\begin{table*}[ht]
\caption{Summary of the statistical comparison of the 2XMM fluxes from the
EPIC cameras.}
\label{flux_comp}      
\small
\begin{center}
\begin{tabular}{c r r r r r r}
\hline\hline
Energy band &  pn\,--\,M1 & ${\rm N_{pn-M1}}$ & pn\,--\,M2 & ${\rm
  N_{pn-M2}}$ & M2\,--\,M1 & ${\rm N_{M2-M1}}$ \\ 
\ [keV] & [\%] & & [\%] & & [\%] & \\
(1) & (2) & (3) & (4) & (5) & (6) & (7)\\
\hline
$0.2-\phantom{0}0.5$ &  4.9$\pm$1.2 &  785 &  8.4$\pm$0.9 &  771 & -0.9$\pm$0.6 &  987 \\
$0.5-\phantom{0}1.0$ & -2.4$\pm$0.3 & 1906 & -2.7$\pm$0.2 & 1957 &  1.0$\pm$0.3 & 2384 \\
$1.0-\phantom{0}2.0$ & -7.6$\pm$0.3 & 2394 & -8.6$\pm$0.3 & 2461 &  0.6$\pm$0.2 & 2932 \\
$2.0-\phantom{0}4.5$ & -6.1$\pm$0.3 & 1311 & -5.4$\pm$0.3 & 1342 & -0.8$\pm$0.2 & 1552 \\
$4.5-12.0$           &-12.4$\pm$0.7 &  387 & -9.5$\pm$0.6 &  408 & -3.2$\pm$0.4 &  441 \\
\hline
\end{tabular}
\end{center}
(1): Energy band definition in keV. 
(2) and (3): difference (in \%) in the measured flux in pn and M1 
and number of sources used in the comparison. 
(4) and (5): same as Cols (2) and (3) but for pn and M2.
(6) and (7): same as Cols (2) and (3) but for M2 and M1.
\normalsize		
\end{table*}

There is an excellent agreement in the measured fluxes between the two MOS
cameras, better that 5\% in all 2XMM energy bands. The agreement between
pn--MOS fluxes is also good, better than 10\% at energies below 4.5\,keV
and $\sim\!10-12$\% above 4.5\,keV. These flux differences are in broad
agreement with the results of Stuhlinger et al. (2008) who find a small
excess, $5-10$\%, of the MOS cameras with respect to pn, using a sample of
very bright on-axis sources. A more detailed analysis will be presented in
Mateos et al. (2008).

\subsection{X-ray hardness (colour) distributions } \label{xcol}

\begin{figure*}[!ht]
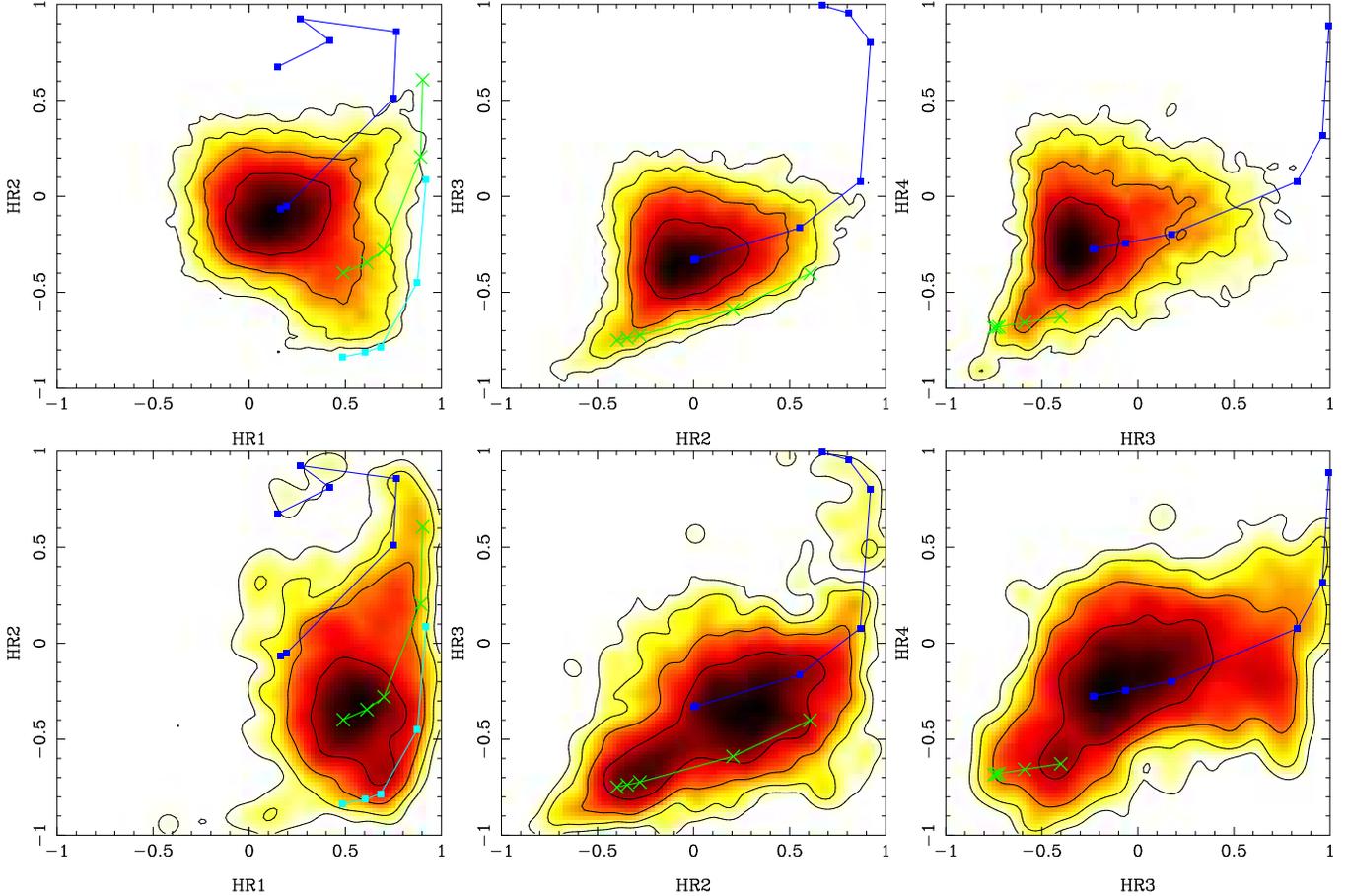

\includegraphics[width=6cm,angle=270]{cat_hr12_hixx.ps}
\includegraphics[width=6cm,angle=270]{cat_hr23_hixx.ps}
\includegraphics[width=6cm,angle=270]{cat_hr34_hixx.ps}\\
\includegraphics[width=6cm,angle=270]{cat_hr12_loxx.ps}
\includegraphics[width=6cm,angle=270]{cat_hr23_loxx.ps}
\includegraphics[width=6cm,angle=270]{cat_hr34_loxx.ps}
\caption[]{Top row: EPIC pn X-ray hardness ratio density plots for high
Galactic latitude ($|b|>10^\circ$) 2XMM sample. Bottom row: X-ray hardness
ratio density plots for low Galactic latitude ($|b|<10^\circ$) 2XMM
sample. Density is displayed on a logarithmic scale with a dynamic range of
100. The spectral tracks overlaid are for (i) power-law spectra with
$\Gamma=1.9, 1.7, 1.4$ (blue) for the left, middle, and right panels,
respectively; (ii) thermal spectrum (APEC model) with $kT=0.3$ keV (cyan;
HR1-HR2 plot only); (iii) a composite thermal model with three components
with $kT=[0.3, 1, 3]$ keV (green). In each case hardness values are shown
for $N_{\rm H}=[0.03, 0.4, 1, 5, 10, 50]\times 10^{22}\rm\ cm^{-2}$
(power-law model) and $N_{\rm H}=[0.01, 0.05, 0.1, 0.5, 1 ]\times
10^{22}\rm\ cm^{-2}$ (thermal models). For each spectral track the
left-most point marked corresponds to the lowest $N_{\rm H}$ value, ie. $N_{\rm H}$ increases towardsthe top right.
}
\label{hrfig} 
\end{figure*}

For each 2XMM source there are four X-ray hardness ratios (X-ray `colours')
which provide a crude representation of the X-ray spectrum (cf.\
Sect.~\ref{det_emldetect} for hardness ratio definition). In
Fig.~\ref{hrfig} we show the hardness ratio density plots for 2XMM
catalogue sources at high and low Galactic latitudes. These plots are for
the pn camera hardness ratios only, as they typically are better
constrained. Density plots are constructed for sources which have detection
likelihood $L>8$ in the energy bands comprising each pair of hardness
ratios: this means that the subsample included in each plot differs and
there is an inevitable bias towards softer sources for the HR1-HR2 plot and
to harder sources for the HR3-HR4 plot. Imposing the same likelihood
threshold for {\it all} bands would produce a bias towards higher flux
sources and in fact would restrict this exercise to relatively small
samples from the whole catalogue. We also exclude sources with summary flag
4; a more severe restriction on the flag produces relatively small changes
to the overall distributions. Overlaid on these hardness ratio density
plots are spectral tracks for representative simple power law and thermal
spectral models with a range of absorbing column densities.

These density plots provide an excellent statistical characterisation of
the spectral properties of the catalogue sources, thus potentially
providing constraints on the overall X-ray population. Although a detailed
analysis is beyond the scope of the present paper, we comment here on how
these match simple expectations about the underlying source populations.

For the high latitude regions of the sky, the density plot is dominated by
sources with power-law spectra and column densities $N_{\rm H}\lesssim
10^{22}\rm\ cm^{-2}$, as expected from the dominant population of AGN. The
fraction of AGN in 2XMM with $N_{\rm H} > 10^{22}\rm\ cm^{-2}$ can be seen
from these plots to be quite low. The high latitude plots also show an
extension to much softer hardness ratios. The main contributors to this are
likely to be coronally active stars and non-active galaxies (see comment
below about the thermal spectra). Due to the bias towards softer (harder)
sources in the HR1-HR2 (HR3-HR4) plots noted above, the power-law tracks
overlaid have different indices to approximately match the observed density
distributions.

At low Galactic latitudes, in contrast, the plots show a more complex
structure (albeit the sample sizes are smaller). The overall low latitude
density pattern is consistent with a large population of coronally active
stars (particularly evident in the HR1-HR2 plot) with relatively soft
thermal spectra together with a significant population of much more
absorbed objects: background AGN together with distant accreting binaries
in the Galactic plane (e.g., Hands et al. 2004). Sources with very
low-temperature thermal spectra (i.e., $kT \sim 0.3$ keV) are only evident
as a small component in the HR1-HR2 plot. We note that the density peak in
the low latitude density plots is {\it not} consistent with what is
expected for a distribution of single-temperature thermal spectra with a
range of intrinsic temperatures. Instead the peak is much better matched by
a {\sl multi-temperature} spectrum which we have here characterised
empirically as a composite three-component model with $kT=[0.3, 1, 3]$ keV
with equal weighting (emission measure) of the three components. This
finding is broadly consistent with the spectral properties of X-ray
selected active star samples (e.g., Lopez-Santiago et al. 2007 and
references therein) in which such objects typically are best-fit with
two-temperature models with $kT\sim [0.3, 1]$~keV. The fact that our
hardness density plots are better characterised with the {\sl ad hoc}
addition of a third higher temperature component clearly points to a harder
component being present in a significant number of the objects contributing
to the hardness density plots.

\subsection{Variability characterisation} \label{varevaluation}

\begin{figure*}[ht]
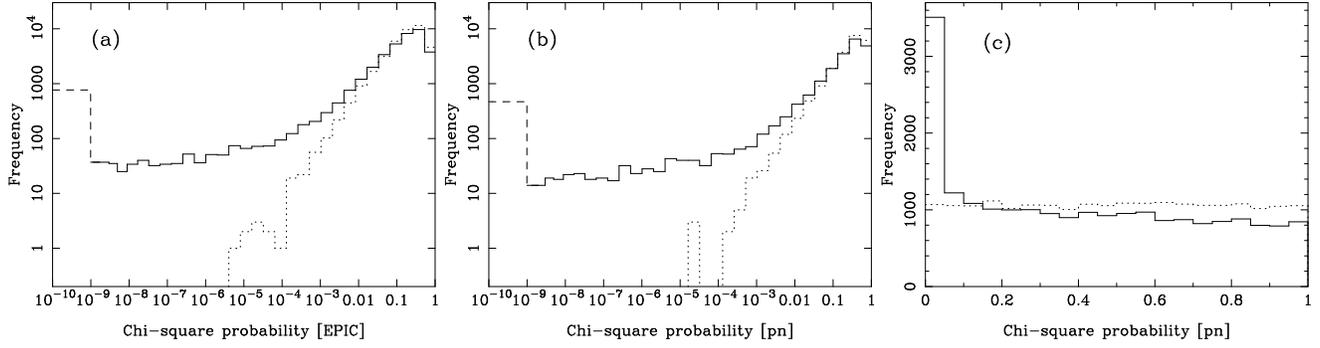

\begin{center}
\includegraphics[width=4.5cm,angle=270]{varplot-a-ep-log.ps}
\includegraphics[width=4.5cm,angle=270]{varplot-b-pn-log.ps}
\includegraphics[width=4.5cm,angle=270]{varplot-c-pn-lin.ps} 
\end{center}
\caption[]{(a) Frequency distribution of $P(\chi^2)_{\rm EPIC}$, with log
  scales on both axes: solid line -- observed; dotted line -- simulation
  for random noise, taking into account that there is not always a complete
  set of 3 camera values for each detection. (b) Frequency distribution of
  $P(\chi^2)_{\rm pn}$, with log scales on both axes: solid line --
  observed; dotted line -- simulation for random noise. (c) As (b) but with
  linear scales on both axes.
}
\label{varfig01} 
\end{figure*}

In the whole 2XMM catalogue there are 2307 detections indicated as variable
(cf.\ Sect.~\ref{catcompile}), which relate to 2001 unique
sources. Evaluation of the frequency distributions of the
$\chi^2$-probability, $P(\chi^2)$, from the time-series analysis reveals no
significant systematic effects and shows the expected behaviour for the
parts of the distributions dominated by random noise. For example, the
frequency distribution of $P(\chi^2)$, as shown in Figs.~\ref{varfig01}\,b)
and c) for the pn (the distributions for MOS1 and MOS2 are very similar),
is almost constant per unit interval of probability down to low
probabilities ($\la 0.1$). Obviously, a non-variable set of time-series
would have this property across the whole probability range $0.0 - 1.0$.

Figure~\ref{varfig01}\,a) shows the observed frequency distribution of
$P(\chi^2)_{\rm EPIC}$ compared with a simulated distribution for a
non-variable set of time-series. As there are many detections with less
than the full set of [pn, M1, M2] time-series, it was necessary to
reproduce this incompleteness in the simulation. The numbers of detections
with 3, 2, 1, or 0 $P(\chi^2)$-values are: 14917, 11330, 11917, 156,
respectively. The simulation was conducted by generating three vectors
representing pn, M1, M2, with each element containing a uniform, random
number in the range $0.0 - 1.0$. For each element, a check was performed to
see if there was a valid $P(\chi^2)$-value for the associated, real camera
data; if not, the random value was set to NULL (so that the correct `run'
of valid values was mimicked in the simulations). These values simulate the
expected distribution of [pn, M1, M2]-probabilities for the case of no real
variability (see Fig.~\ref{varfig01}\,a). As expected, the resulting
distributions are `flat' (on a linear scale), as discussed above. A fourth
vector was then computed with the minimum simulated $P(\chi^2)$, i.e., a
simulated set of $P(\chi^2)_{\rm EPIC} =$ min($P(\chi^2)_{\rm pn},
P(\chi^2)_{\rm M1}, P(\chi^2)_{\rm M2}$) over all available values for each
detection.

Visual inspection of samples of time-series flagged as not variable,
indicated a number of cases and types of variability that were likely to
have been `missed' by the 2XMM variability test, implying that the
catalogue is conservative in this respect. These included relatively
short-duration increases or decreases, and low-level trends/ramps.

We have compared the fraction of variable sources (or detections) to all
sources (or detections) having time-series as a function of various other
parameters of the catalogue. As a function of flux (specifically EPIC
total-band flux), we find this fraction to be $\sim\!25\%, 10$\%, and 5\%
for fluxes $\ga 10^{-10},\ \sim 10^{-11}$, and $\la 10^{-12}$,
respectively. This is broadly as expected as the ability to detect
variability falls towards lower fluxes.

We have also carried out an initial evaluation of the variable 2XMM sources
using secure positional matches with objects in the Simbad database. From
this study we estimate that, for serendipitous (i.e., non-target) sources,
$\sim\!40$\% are `normal' (i.e., non-degenerate) stars, $\sim\!5$\% are
X-ray binaries, $\sim\!3$\% are cataclysmic variables and $\sim\!5$\% are
AGNs, plus lower percentages of objects such as GRBs. Of order 45\% could
not be identified from Simbad. The above figures relate primarily to the
$\sim\!1000$ sources with quality summary flag values 0\,--\,2. Although
this is not a definitive study as the completeness of Simbad for different
object types is highly non-uniform, it does nevertheless provide
confirmation of the utility of the catalogue variability characterisation
to select known types of variable objects efficiently.

\subsection{Extended sources}  \label{extddiscussion}

\begin{figure*}[ht!]
\resizebox{\hsize}{!}{\includegraphics{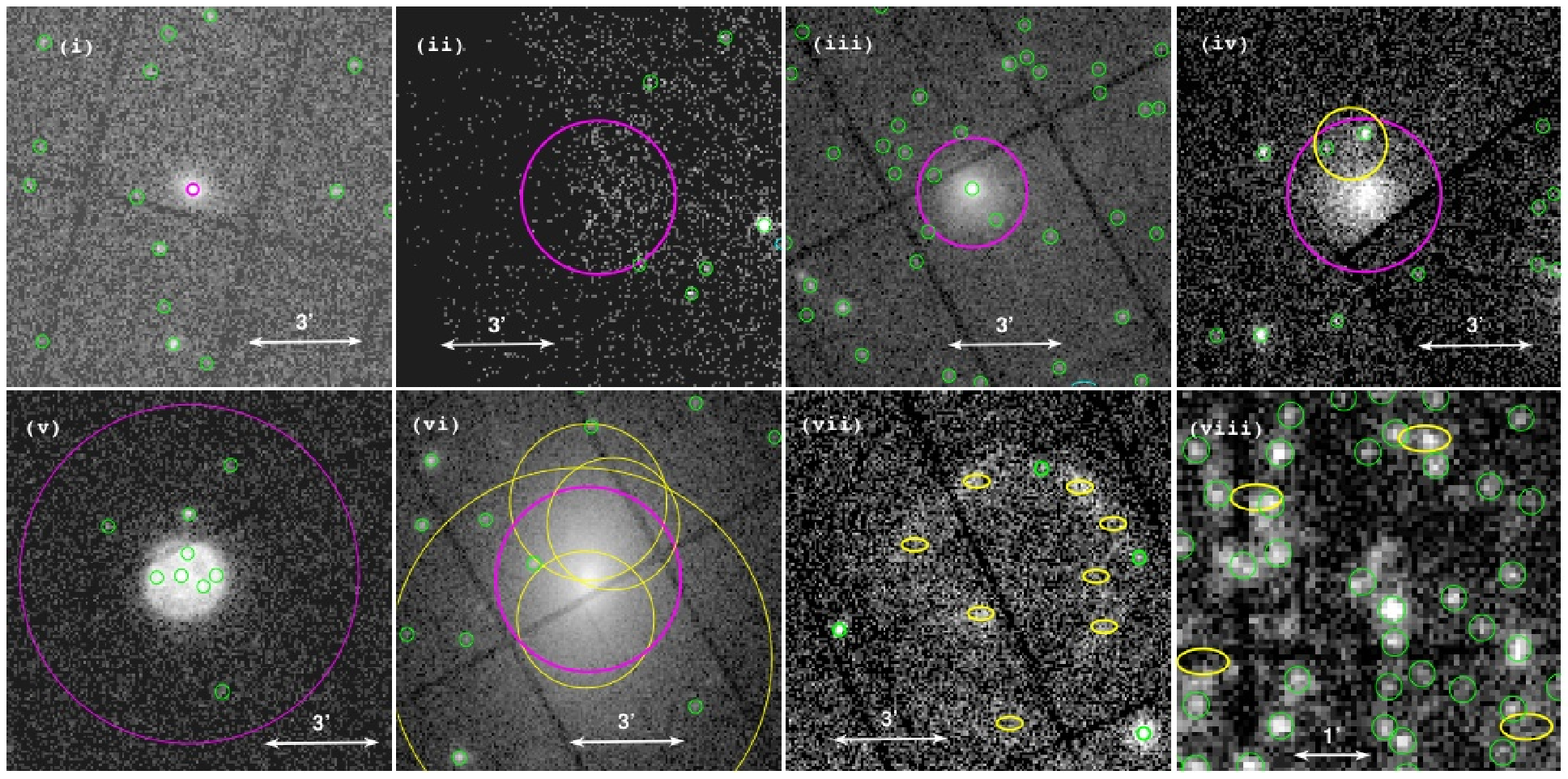}} 
\caption[]{Examples of extended source detections. Green circles mark point
  source detections. In panels (i) -- (vi) the magenta and yellow circles
  mark real and spurious extended detections respectively, plotted with
  their fitted extent (i.e., core radius, see Sect.~\ref{extdsrc}). In
  panels (vii) \& (viii) the yellow ellipses indicate the position of
  spurious extended source detections.\\
{\sl Top row}: (i) a compact extended source with a small core radius; (ii)
  a large, low surface brightness extended source at the edge of the FOV
  with low likelihood but high flux (see Fig.~\ref{extsrc}); (iii) an
  object with a point-like core detected both as a point source as well as
  an extended source; (iv) a clearly extended source with a spurious
  detection nearby (yellow circle) which is smaller and fainter (by a
  factor of 45) and which therefore does not significantly affect the
  parameters of the real source.\\
{\sl Bottom row}: (v) a SNR in the LMC where intrinsic structure is
  detected as point sources (note that the core radius is not
  representative as the extended emission does not follow the $\beta$-model
  fitted); (vi) a bright extended source with multiple spurious detections
  around the centre: the core radii of these spurious detections are
  comparable to or greater than the extent of the real source and will thus
  significantly affect the parameters of the real source (note that the
  maximum core radius allowed in the fitting is $80\arcsec$); (vii) a faint
  filamentary structure broken up into several extended detections where
  the parameters have little meaning (due to the circularly symmetric
  nature of the fit); (viii) a crowded region where several point sources
  are detected as extended due to source confusion (the algorithm is
  restricted to fitting at most two confused sources simultaneously).
  }
\label{examplesextdfig}
\end{figure*}

The 2XMM catalogue contains more than 20\,000 entries of extended
detections. The reliable detection and parameterisation of extended sources
is significantly more demanding than for point-like sources because there
are many more degrees of freedom in the parameter space. The relatively
simple analysis approach used in the creation of the catalogue
(Sect.~\ref{extdsrc}) means that the catalogue contains a significant
number of extended object detections that are either spurious or at least
uncertain (cf.\ Sects.~\ref{srcdetproblems} and~\ref{autoflag}). The most
common causes of problems with extended sources are summarised below and
illustrated in Fig.~\ref{examplesextdfig}.

\begin{description}

\item{\bf Spurious detections near bright point sources:} These are mostly
due to inaccuracies of the PSF models, leading to inaccurate modelling of
the internal background by the source fitting routine.

\item{\bf Confusion of point sources:} Pairs or multiples of point sources
can be detected as one extended source since only up to two point sources
are modelled simultaneously by the fitting algorithm.

\item{\bf Insufficient background subtraction:} Some spatial variations of
the intrinsic background are poorly modelled by the spline map. In regions
where the background is underestimated, spurious detections of extended
sources are possible. (In many cases the extent parameter of these sources
is at or near the maximum of the allowed range, $80\arcsec$.)

\item{\bf Multiple detections of extended sources:} The surface brightness
distribution of extended sources is generally more complex than the fitted
$\beta$-model. This can lead to additional detections in the wings of
extended sources. The most extreme cases are observations of complex,
bright extended sources (e.g., Galactic supernova remnants), leading to the
detection of numerous extended sources in one field. Also, extended
emission following the fitted $\beta$-model, but with an extent greater than
the maximum allowed in the fit, tends to be broken up into multiple
detections.
 
\item{\bf Instrumental artefacts:} OOT events of piled-up sources, single
reflections arcs, and scattered light from the RGA (cf.\
Fig.~\ref{examplesfieldfig}\,c) can cause both point-like and extended
spurious detections.

\end{description}  
 
The catalogue contains extensive detection flags (Sect.~\ref{qual_eval})
which can be used to produce much cleaner extended-source samples, albeit
at the expense of removing some genuine extended objects. (This is the case
as the flagging scheme only provides warnings about generic problems with
the analysis or the data rather than a specific assessment of the reality
of each detection.) In particular the automated quality Flags 4, 5, and 6
(see Table~\ref{autoflagtab}) are set to warn about possible spurious
detections of extended sources. The combined Flag~7 for extended sources is
set if one of the Flags $4-6$ is set. This flag is set for 9\,882 out of
20\,837 detections, indicating a potential spurious fraction of about
$50$\%. However, the rate of spurious detections is distributed very
unevenly over the catalogue observations as is discussed below.

\begin{figure}[ht]
\resizebox{\hsize}{!}{\includegraphics[angle=270]{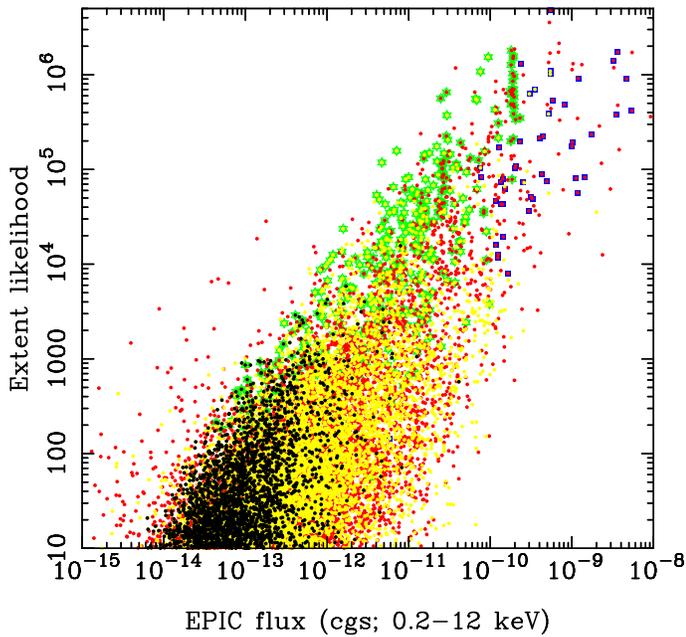}} 
\caption[]{Distribution of extent likelihood as a function of total-band
EPIC flux for the extended source detections in the 2XMM catalogue. Red
dots are potentially spurious detections with Flag 7=T, yellow dots are
detections with Flag 7=F, black dots are the `best' sample detections with
summary flag $<\!2$. Green stars mark the targets of the XMM-Newton
observations classified as extended object types and blue squares targets
which are object types classified as point-like. The vertical
concentrations of target points at flux $\sim\!3\times 10^{-11}$ and
$\sim\!2\times 10^{-10}\rm\,erg\,cm^{-2}\,s^{-1}$ are real, being due to
multiple detections of two different SNRs used as XMM-Newton calibration
targets.
}
\label{extsrc} 
\end{figure}

Figure~\ref{extsrc} illustrates some of the main features of the extended
source detections in the catalogue. The plot shows that there is, as
expected, an overall correlation of extent likelihood with EPIC flux. The
considerable scatter in the plot has three origins: (i) the observations
from which the detections are drawn have a considerable range of exposure
times and background values; (ii) source extent: sources with larger
spatial extent have lower likelihoods at the same integrated flux; (iii)
the presence of significant numbers of spurious detections. The detections
with Flag~7 set show, as expected, a broader distribution than those
without this flag, and a much broader distribution than for the detections
with `best' summary flags (i.e., summary flag $<\!2$). This is, of course,
due to the fact that spurious detections will often have implausible
likelihoods for the fitted flux or correspond to very large source extent
which is rare in genuine detections.

Based on the sample with `best' summary flags it is clear that there are
very few reliable extended source detections with extent likelihood above
$\sim\!1000$ or flux above $\sim\!4\times
10^{-13}\rm\,erg\,cm^{-2}\,s^{-2}$, highlighting the problems that the
detection algorithm has with bright objects.\footnote{We also note that
this is what is expected from the source counts of clusters of galaxies
which are expected to dominate the extended detections, at least at high
Galactic latitudes.} Indeed the majority of reliable extended objects in
this region of the diagram are the XMM-Newton targets themselves (but note
that many of these have Flag~7 set which would otherwise indicate
potentially spurious detections). At the highest fluxes a large fraction of
the detections relate to very bright point-like targets that are
incorrectly parameterised as being extended due the deficiencies of the
fitting algorithm noted above.

We have investigated a small subset of the extended detections at high
Galactic latitudes covered by SDSS DR6 (excluding targets). We selected
detections with extent likelihood $>\!100$ and no warning flags set (i.e.,
summary flag 0) and evaluated their validity by examining the X-ray images
visually and by searching for matches with catalogued objects. We find that
less than 5\% of these may be spurious extended source detections, around
40\% are clearly associated with catalogued clusters or groups of galaxies
and a few percent are associated with single nearby galaxies.  For a
further $\sim\!30$\% of the detections we find convincing evidence of a
previously uncatalogued cluster or group of galaxies at the X-ray source
location from visual inspection of the SDSS DR6 images.  These results
demonstrate that the overall reliability of the `best' extended source
sample is high, at least at higher likelihoods, and that, as expected, the
extended source sample is dominated by groups and clusters of galaxies. We
have not carried out a similar exercise systematically at low Galactic
latitudes but checks of selected detections demonstrate the expected
associations with SNRs, HII regions, and discrete extended features in the
Galactic Centre region.

\section{Availability of the catalogue and catalogue products } \label{cataccess}

The 2XMM catalogue table itself is essentially a flat file with 246\,897
rows and 297 columns (described in Appendix~\ref{catcolumns}). Access to
the catalogue file in various formats (FITS and comma-separated-variable
[CSV]) is available from the XMM-SSC catalogues web-page: {\sl
http://xmmssc-www.star.le.ac.uk/Catalogue/}.  This XMM-SSC web-page is the
primary location for information about the 2XMM catalogue. It provides
links to the other hosting sites and the documentation for the
catalogue. It also provides a `slimline', reduced volume version of the
2XMM catalogue, which is based on the 191\,870 unique sources and contains
just 39 columns. The columns in this version are restricted to just the
merged source quantities, together with the 1XMM and 2XMMp
cross-correlation counterparts.

Ancillary tables to the catalogue also available from the XMM-SSC web-page
include the table of observations incorporated in the catalogue
(Appendix~\ref{osc}) and the target identification and classification table
(Appendix~\ref{newtargbit}).

Associated with the 2XMM catalogue itself is an extensive range of data
products such as the EPIC images from each observation and the spectra and
time-series data described in Sect.~\ref{ssp}. These products are accessible, 
along with the catalogue itself, from ESA's XMM Science Archive
(XSA\footnote{http://xmm.esac.esa.int/xsa/}), the
LEDAS\footnote{http://www.ledas.ac.uk/xmm/2xmmlink.html} (LEicester
Database and Archive Service) system and
are being made available through the Virtual Observatory via LEDAS using
AstroGrid\footnote{http://www.astrogrid.org} infrastructure.

LEDAS also provides access to a single HTML summary page for each detected
source in the catalogue. These summary pages provide the key detection
parameters and parameters of the corresponding unique source, links to
other detections of the same source, thumbnail X-ray images and graphical
summaries of the X-ray time-series and spectral data where these exist.

The results of the external catalogue cross-correlation carried out for the
2XMM catalogue (Sect.~\ref{extcatcross}) are available as data products
within the XSA and LEDAS or through a dedicated on-line database system
hosted by the Observatoire de Astrophysique,
Strasbourg\footnote{http://amwdb.u-strasbg.fr/2xmm/home}.

\section{Summary} \label{conclusion}

We have presented the 2XMM catalogue, described how the catalogue was
produced and discussed the main characteristics of the
catalogue. Table~\ref{sumtable} provides a summary of its main properties,
bringing together information presented elsewhere in this paper.

\begin{table}[bt]
\caption{Summary of 2XMM catalogue characteristics}
\label{sumtable}
\small
\centering
\tabcolsep 1.5mm
\begin{tabular}{llrl}
\hline
\multicolumn{2}{l}{Energy range (keV; set by EPIC cameras)} & 0.2\,--\,12.0 \\
\hline
Observations & {\sl total} & 3491 \\
             & pn data     & 2674 \\
	     & MOS1 data   & 3384 \\
	     & MOS2 data   & 3394 \\
\hline
Time interval   & \multicolumn{2}{r}{Feb 2000\,--\,Mar 2007} \\
\hline
Detections & {\sl total}            & 246897 \\
           & {\sl  total} L$\ge 10$ & 201275 \\
           & {\sl  total} sum flag 0& 199359 \\
           & point-like             & 226060 \\
	   & extended               & 20837 \\
           & with products          & 38320 \\
\hline
Unique sources & {\sl total}   & 191870 \\
	       & point-like    & 173066 \\
	       & extended      & 18804 \\
\hline
Sky area \ (deg$^2$ ) & {\sl total}\,$^{a}$ & $\sim\!560$ \\
            	    & net MOS1/2          & $\sim\!355$ \\
	            & net pn              & $\sim\!330$ \\
\hline
Median exposure time & MOS1/2 & $\sim\!16000$ s\\
\ \ \ \ \ \ (per observation)    & pn     & $\sim\!12500$ s\\
\hline
Flux limit (pn) at  $\sim 10$\% sky & 0.5\,--\,2.0 keV   & $\sim\!2$ \\
\ coverage \  ($10^{-15} \rm\ erg\ cm^{-2}\ s^{-1}$ )  & 2.0\,--\,12 keV    & $\sim\!15$ \\	
                    & 4.5\,--\,12 keV  & $\sim\!35$ \\
\hline
Flux limit (pn) at  $\sim 90$\% sky & 0.5\,--\,2.0 keV  & $\sim\!10$ \\
\ coverage  \  ($10^{-15} \rm\ erg\ cm^{-2}\ s^{-1}$ )     & 2.0\,--\,12 keV   & $\sim\!90$ \\	
                 & 4.5\,--\,12 keV & $\sim\!250$ \\
\hline
 Astrometric accuracy \ (1$\sigma$)& typical    & 1.5\arcsec \\
\ \ \ \ \ \   & best$^{b}$ & 0.35\arcsec \\
\hline
Photometric accuracy & MOS1/2 comparison & $\le 5$\% \\
		     & pn/MOS comparison    & $\le 10$\% \\ 
\hline
\end{tabular}
\begin{list}{}{}
\item [$^{\mathrm{a}}$]overlaps included
\item [$^{\mathrm{b}}$]limited by systematics
\end{list}
\normalsize		

\end{table}

2XMM is the largest X-ray source catalogue ever produced, containing almost
twice as many discrete sources as either the ROSAT survey or ROSAT pointed
catalogues. The catalogue complements deeper Chandra and XMM-Newton small
area surveys, and probes a large sky area at the flux limit where the bulk
of the objects that contribute to the X-ray background lie.  The catalogue
has very considerable potential a detailed account of which lies outside the scope of this
paper. In particular the catalogue provides a rich resource for generating
sizeable, well-defined samples for specific studies, utilising the fact that
X-ray selection is a highly efficient (arguably the most efficient) way of
selecting certain types of object, notably active galaxies (AGN), clusters
of galaxies, interacting compact binaries and active stellar coronae. The
large sky area covered by the serendipitous survey, or equivalently the
large size of the catalogue, also means that 2XMM is a major resource for
exploring the variety of the X-ray source population and identifying rare
source types. Although the 2XMM catalogue alone provides a powerful way of
studying the X-ray source population, matching the X-ray data with, e.g.,
optical catalogues can offer an even more effective way to generate considerable
samples of particular object types.  Projects that exploit some of these
characteristics are already underway.

Finally we note that, since the XMM-Newton spacecraft and instruments
remain in good operational health, we can anticipate a substantial growth
in the pool of serendipitous X-ray sources detected, increasing at a rate
of $\sim$35\,000 sources/year. With this backdrop, further XMM-Newton
catalogue releases are planned at regular intervals. The first such
incremental release is planned for August 2008.

\begin{acknowledgements}

We gratefully acknowledge the contributions to this project made by our
colleagues at the XMM-Newton Science Operations Centre at ESA's European
Space Astronomy Centre (ESAC) in Spain. We thank Steve Sembay for useful
comments and the CDS team for their active contribution and
support.

M. Ceballos, F. Carrera and X. Barcons acknowledge financial support by the
Spanish Ministerio de Educacion y Ciencia under projects ESP2003-00812 and
ESP2006-13608-C02-01.  The French teams are grateful to CNES for supporting
this activity.  In Germany the XMM-Newton project is supported by the
Bundesministerium f\"{u}r Wirtschaft und Technologie/Deutsches Zentrum
f\"{u}r Luft und Raumfahrt e.V. (DLR) and the Max Planck Society. Part of
this work was supported by the DLR project numbers 50 OX 0201, 50 OX 0001,
and 50 OG 0502.  The Italian team acknowledges financial support from the
Agenzia Spaziale Italiana (ASI), the Ministero dell' Istruzione,
Universit\`a e Ricerca (MIUR) and the Istituto Nazionale di Astrofisica
(INAF) over the last years; they are currently supported by the grant
PRIN-MIUR 2006-02-5203 and by the ASI grants n.I/088/06/0 and n.I/023/05/0.
UK authors thank STFC for financial support.

This research has made use of the NASA/IPAC Extragalactic Database (NED),
which is operated by the Jet Propulsion Laboratory, California Institute of
Technology, under contract with the National Aeronautics and Space
Administration.  This research has also made use of the SIMBAD database, of
the VizieR catalogue access tool, and of Aladin, operated at CDS,
Strasbourg, France, and of the Digitized Sky Surveys (produced at the Space
Telescope Science Institute under U.S. Government grant NAG W-2166).

\end{acknowledgements}

\pagebreak

\appendix
\section{Summary of XMM-Newton and EPIC camera  terminology}\label{newapp}

\begin{description}

\item [\bf On-axis position]: the telescope optical axes, defined by the geometry of each of the three
X-ray mirror modules,
is not coincident with 
geometrical centres of the EPIC detectors. The target of the observation is
preferentially placed close to, but slightly offset from, the optical axis.

\item [\bf Point-Spread-Function (PSF)]: the telescope optics spread X-ray
photons from a point source into a centrally-peaked distribution which is
oversampled by the EPIC cameras. The PSF is energy-dependent and becomes
broader with increasing (off-axis) angle from the telescope optical axis
but also suffers a distortion which elongates the profile in the azimuthal
direction.

\item [\bf  Event patterns]: an X-ray photon incident in a given CCD location
causes charge deposition in several surrounding CCD pixels, often not
symmetrically distributed around the central pixel. Several distinct charge
distributions (patterns) are recognised as real events by the on-board
processing electronics for the MOS cameras, whilst this processing takes
place on the ground for the pn camera.

\item [\bf Out-Of-Time (OOT) events]: EPIC camera exposures are composed of many
short-duration frames during which the recorded events are rapidly read out
and processed by the on-board electronics. The total time between frames
(frame-time) depends on the observing mode but is a maximum of 73\,ms for
the pn and 2.6\,s for the MOS. The cameras are shutterless and record data
during the readout (`out-of-time') as well as the processing (`imaging')
period, leading to a faint trail of the `out-of-time' events along the
readout direction which becomes obvious for bright sources (see
Fig.~\ref{examplesfieldfig}c). The percentage of OOT events is a function
of the ratio of the frame readout time to the frame integration time for a
given mode. The highest percentage of OOT events at 6.3\% is for the pn
full frame mode, while it never exceeds 0.5\% for the MOS.

\item [\bf Pileup]: for bright sources, pixels in the core of the PSF can receive
multiple X-ray photons during an integration frame. The on-board processing
electronics cannot recognise them as distinct events within that frame and
either treats them as a single event with higher energy or rejects them
entirely if the resultant pixel pattern of the combined event lies outside
the pre-defined X-ray pattern library. As a result the recorded counts are
lower in the core of the source profile, producing a flattening or even
depression of the source profile (see Fig.~\ref{examplesfieldfig}c). In
addition, it has an impact on the spectral profile (i.e., a hardening of
the spectrum).

\item [\bf Optical loading]: The EPIC cameras (more so the MOS detectors) are
also sensitive to optical photons so that optically bright objects generate
recordable events in EPIC images (Lumb 2000, and references therein). The
level of contamination depends on the filters used and the optical
brightness of the object. In most observations the filter used is
conservatively selected to minimise this effect. Note that in the case of
the pn some apparently very soft sources are affected by optical loading.

\item [\bf RGA scattered light]: scattering of incident X-rays by the RGAs in the
two telescope modules that feed the MOS cameras causes a diffuse bright
narrow band in the X-ray images which is detectable for bright X-ray
sources (see Fig.~\ref{examplesfieldfig}c).

\item [\bf Good-Time-Interval (GTI)]: data from EPIC camera frames can be
accepted or rejected according to the state of various housekeeping and
science parameters, e.g., spacecraft attitude stability and particle
background level. The `GTI's are the time periods during which the
parameter(s) being monitored are within the acceptable thresholds.

\end{description}

A more detailed description of the instruments can be found in the on-line
version of the XMM User Handbook (Ehle et al.\ 2007) and on the ESAC
documentation web-pages for
calibration\footnote{http://xmm2.esac.esa.int/external/xmm\_sw\_cal/calib/documenta\-tion/index.shtml}.

\section{Observation summary table}\label{osc}

\onllongtabL{1}{
\include{obs_tab_short}
}

Table~\ref{obstab} presents the observations and exposures included in 2XMM
and is available on-line at \aap as well as at the XMM\_SSC catalogue
web-page (cf.\ Sect.~\ref{cataccess}). The columns in this table are as
follows.

{\it Column 1:} satellite revolution number (consecutive in time).

{\it Column 2:} observation number (10 digit ID).

{\it Column 3:} ODF version number.

{\it Column 4 and 5:} nominal field Right Ascension and declination (J2000)
in degrees.

{\it Column 6:} target name (20 characters).

{\it Column 7:} Quality classification of the whole observation based on
the area flagged as bad in the manual flagging process as compared to the
whole detection area, see Sect~\ref{manflag}. 0 means nothing has been
flagged; 1 indicates that 0\% $<$ area $<$ 0.1\% of the total detection
mask has been flagged; 2 indicates that 0.1\% $\le$ area $<$ 1\% has been
flagged; 3 indicates that 1\% $\le$ area $<$ 10\% has been flagged; 4
indicates that 10\% $\le$ area $<$ 100\% has been flagged; and 5 means that
the whole field was flagged as bad.

{\it Column 8:} number of detections in this field.

{\it Column 9:} number of detections in this field that have not received
manual Flag~11 and are considered to be `good'.

{\it Column 10:} number of the pn exposures merged for the source detection
(cf.\ Sect.~\ref{expsel}).

{\it Column 11:} filter of the pn exposures: Tn1 stands for Thin1, Tn2 for
Thin2, Med for Medium, and Tck for Thick.

{\it Column 12:} observing mode (cf.\ Table~\ref{modestab}) of the pn
exposures.

{\it Column 13:} total exposure time of the pn exposures in seconds.

{\it Column 14\,--\,17:} same as columns~10\,--\,13 but for MOS1.

{\it Column 18\,--\,21:} same as columns~10\,--\,13 but for MOS2.

\section{Target identification and classification procedures} \label{newtargbit}

In the following are described the procedures adopted to identify and
classify the targets of each XMM-Newton observation included in the 2XMM
catalogue. The results of this exercise are available on-line at \aap.

As any attempt to identify and classify a target is subjective and likely
to be incomplete (only the investigators of that observation know all the
details), two different approaches were chosen to give the user a choice
regarding detail and reliability: on the one hand some formal information
associated with an observation is provided; on the other hand, a manual
classification scheme tries to supply interpretation of sometimes ambiguous
target names and to directly identify associated 2XMM detections.

\subsection{Formal target identification}  \label{formtargetid}

There are three kinds of coordinates associated with each observation:
\begin{enumerate}

\item The median of the spacecraft attitude (`pointing direction',
independent of the instrument) usually points to approximately the same
position on the detectors and defines best the centre of the FOV (this is
given in Table~\ref{obstab}).

\item The proposal position refers to the position given by the observer;
this position is placed at a specified detector location which depends on
the prime instrument (EPIC or RGS) as indicated by the observer and which
avoids chip gaps, dead spots etc, unless an offset is indicated by the
investigator.

\item The XSA gives the coordinates of the prime instrument viewing
direction which are corrected for the star tracker mis-alignment.

\end{enumerate}

In most cases, the proposal position is the best representative of the
target object as chosen by the investigators. However, there are cases
where the actual target object is deliberately off-set from the proposal
position, or the proposal position is not very accurate. The latter can be
due to catalogue errors, positions with large uncertainties (e.g., gamma
ray sources), or an error by the observer. In cases where more than one
object is the target the proposal position can either be located on one of
the objects or between them. In a few cases, the image was not obtained at
the proposed position due to a slew failure or a `Target of Opportunity'
(ToO) observation that was not properly registered in the ODF.

The XSA coordinates are usually near the centre of the field and/or the
target but do not represent the target position as well as the proposal
position.

The target identification table (Appendix~\ref{targtable}) lists the
proposal and XSA positions together with the proposal category and proposal
program information as given in the XSA. The latter provide a coarse
classification of the target as determined by the observer. Note though,
that the proposal category of calibration observations are often
meaningless since they are often instrument related for which there is no
particular proposal category.

\subsection{Manual target identification}  \label{mantargetid}

In many ways the target name as given in the proposal gives a better
indication of the field content than the coordinates since a target can
comprise more than one object or it may be diffuse emission that can only
be detected in the spectra of background objects. In other words, if a
target name can be resolved by on-line data bases like Simbad and NED one
can easily derive more information about that object, e.g., object type,
other names, or references.

On the other hand, an XMM-Newton target name can be descriptive or refer to
a personal choice of the observer, it can be abbreviated, or additional
information is added. It was therefore necessary to `interpret' many of the
target names before Simbad could recognise them.

The target identification table lists therefore, next to the XMM-Newton
target name, the best estimate of the Simbad-recognisable name where
possible (usually very close to the given target name), together with the
Simbad coordinates\footnote{Note that Simbad frequently up-dates its
information and the coordinates given here may be out of date.} and Simbad
object type for classification purposes. In cases where Simbad gives more
than one object type, the one closest to the proposal category was
given. Where no Simbad name could be identified a NED identification may be
given instead, and where possible an estimated object type based on the
proposal information is given.

For the use of the catalogue, however, it is most helpful to know which and
how many sources are `targets' and therefore not serendipitous. The
observations are thus classified by their field content (i.e., target
classification; see Fig.~\ref{examplesfieldfig} for some examples), using
the following categorisation:
\begin{itemize}

\item a point or point-like source, that is, a single detection in the
catalogue (excluding spurious detections);

\item an extended source (the target can be the detection of the extended
emission as well as point sources associated with it, e.g., galaxies in a
cluster);

\item a field, that is, all detections are potential targets (e.g., distant
AGNs);

\item diffuse  emission;  the detections in such a field are
considered to be all serendipitous but the {\it location} of the field was
chosen specifically by the observer because of the presence of the -- often
large-scale -- diffuse emission;

\end{itemize}
or a combination of these. Occasionally the field is totally serendipitous
due to operational issues. For fields that could not be easily classified, the
content is `unknown'. The class of extended targets was further divided as
follows:
\begin{itemize}

\item small extended source (i.e., well within the FOV) with a radius of
$<3\arcmin$ (covering roughly 3\% of the full FOV),

\item large extended source with a radius of $>3\arcmin$ and often
extending beyond the FOV,

\item extended source of undetermined radius: these are either not
detected, not identifiable (more than one object fitted the description), or
offset and beyond the edge of the FOV.

\end{itemize}

In cases where one or two point sources are the target, the catalogue
detection IDs (for a match within $\sim\!10$\arcsec) are given as well. In
cases of extended targets a catalogue detection ID is only given if the
match is unambiguous and the centre of the extended emission well
represented by the XMM-Newton detection (the parameters of the detection,
however, may be unreliable). In a few cases a positive identification could
be achieved through another but deeper observation of the same target.

Because neither the formal nor the manual classification can be perfect in
every case, the table also lists, for quick reference, an indicator for the
positions (proposal or Simbad) which best represents the target (subject to
changes and improvements in Simbad). In some cases both positions were
deemed to be equally viable (e.g., in field observations or large offsets
of extended objects) and no preference is given in the table.

\subsection{Problem cases}  \label{problemtargetid}

Not all targets fit unambiguously into the field content classes. In a few
cases where no decision could be made the target was classified as
`unknown'. Otherwise the following guidelines were used.

\begin{description}

\item[\bf Galaxies:] A galaxy was classified as `point source' when the
  emission from the (active) nucleus was dominant. It was classified as `extended' 
  when either diffuse emission was
  apparent or if the galaxy was large enough for discrete X-ray sources in
  the galaxy to be resolved (in case of doubt a comparison was made with an
  optical image downloaded from the DSS\footnote{The STScI Digitized Sky
  Survey.}) or if the galaxy was detected as a single point source in the
  catalogue but it clearly consisted of several (unresolved) sources.

  In two cases, a `field' classification was preferred: observations of the
  M31 halo and offset pointings of M33. In both cases the galaxy is
  considerably larger than the FOV. Note that the observations of
  the centre of M31 (often called M31 core) are classified as `large
  extended' instead since the field includes diffuse emission.

\item [\bf Galaxy clusters:] Galaxy clusters usually show X-ray emission
  from the intracluster gas as well as emission from some of the galaxies
  within that cluster. Most
  galaxy clusters were classified as `large'. Exceptions are distant
  clusters which are significantly smaller than $r = 3\arcmin $ and where
  no point sources could be discerned within the diffuse emission.

\item[\bf Galaxy groups:] Galaxy groups have fewer members than galaxy
  clusters. In many cases there is no detectable intracluster emission and
  the X-ray images  show only emission from some of the members. In
  some cases there is a prominent galaxy in the centre with a large X-ray
  halo. Despite this diversity it was preferred to classify all groups in
  the same way as galaxy clusters, that is, as extended emission, mixed
  with point or other extended sources.

\item[\bf Extragalactic point sources:] In a few cases a bright X-ray
  source within a galaxy was the target (e.g., `super Eddington' sources);
  these were treated like AGNs, that is, if no galaxy emission could be
  discerned the target was classified as `point source', otherwise as
  `extended'.

\item[\bf Mixed targets:] Examples for mixed targets are a particular
  galaxy within a galaxy cluster or a Central Compact Object in a
  SNR. These were classified by the `larger' target, that is, in the
  examples given the class would be `extended', while the Simbad object
  type is likely to refer to the point(-like) source. There are a number of
  cases where such a connection was not obvious or could not be easily
  determined (e.g., a connection between a quasar and a galaxy cluster
  which may be hosting the quasar or simply be superimposed in the
  line-of-sight) and the class refers to the quoted object only. In case of
  a calibration observation the object is more likely to be chosen for its
  own properties and not for its possible connection/interaction with the
  environment.

\item[\bf Solar system objects:] There are a number of observations of
  planets or comets in our solar system. A special object type, `com' for `comet' and `plt' for `planet', is
  listed for these. The field classification depended on what was visible
  in the image, e.g., if there was visible (and detected) diffuse emission
  in case of a comet, or if a planet was observed long enough to produce a
  elongated trace on the image (the pipeline processing corrects for any
  attitude shift so that a fixed point in the sky is always at the same
  location in the image).

\end{description}

\subsection{Target classification}  \label{stattargetid}

There are 3491 fields in total in the 2XMM catalogue. For 3044 fields
(87\%) a Simbad name could be found, and in 53 cases (1.5\%) a NED
identification is given. Of the remaining 394 fields only 56 (1.6\%) do not
have an estimated object type.

About 10\% of the observations were obtained for calibration purposes, and
3\% are ToO observations. Table~\ref{pcategorytab} lists the distribution
of the proposal category for 2XMM observations, and
Table~\ref{fieldcontenttab} gives the same for the field content
classes. The ratio of point source to extended source to field observation
is roughly 5:3:1.

For best results on identifying target objects in the catalogue, it is
recommended to use both the field content class as well as the Simbad
object type.

\begin{table}[t]
\normalsize
\caption{Proposal category given by the XSA }
\label{pcategorytab}
\small
\begin{center}
\begin{tabular}{lp{5.5cm}r}
\hline \hline
Class & Description & Percentage \\
\hline
I   & Stars, White Dwarfs and Solar System & 16\% \\
II  & White Dwarf Binaries, Neutron Star Binaries, Cataclysmic Variables,
ULXs and Black Holes & 15\% \\
III & Supernovae, Supernova Remnants, Diffuse Emission, and Isolated
Neutron Stars & 14\% \\
IV  & Galaxies and Galactic Surveys & 9\% \\
V   & Groups of Galaxies, Clusters of Galaxies, and Superclusters & 14\% \\
VI  & Active Galactic Nuclei, Quasars, BL Lac Objects, and X-ray Background
& 23\% \\
VII & X-ray Background and Surveys & 8\% \\
\hline
\end{tabular} 		
\end{center}
\normalsize
\end{table}		

\begin{table}[h]
\normalsize
\caption{Target / field content classification }
\label{fieldcontenttab}
\small
\begin{center}
\begin{tabular}{cp{5.5cm}l}
\hline \hline
Class & Description & Percentage \\
\hline
p & point or point-like source & 50\% \\
s & small extended ($r < 3\arcmin $) & 10\%  \\
l & large extended ($r > 3\arcmin $) & 22\% \\
e & extended source of unknown extent &  \phantom{0}0.7\% \\
f & `field' (all detections are potential targets) &  12\% \\
x & `X-ray shadow experiment' and similar, that is, only the spectra of
     fore- and background objects are of interest (though the location of the
     field should be considered as `target') & \phantom{0}2.5\% \\
t & two clearly identified targets (e.g., a double star) & \phantom{0}0.4\% \\
n & there is no target associated with the field & \phantom{0}0.2\% \\
u & unknown target, i.e., the target could not be classified or is of unknown
nature  & \phantom{0}2\% \\

\hline
\end{tabular} 		
\end{center}
\normalsize		
\end{table}

\subsection{Target table}  \label{targtable}

\onllongtabL{2}{
\include{tab_target_short}
}

The columns in Table~\ref{targettab}, which is available on-line at \aap as
well as at the XMM-SSC catalogue web-page (cf.\ Sect.~\ref{cataccess}), are
as follows.

{\it Column 1:} satellite revolution number (consecutive in time).

{\it Column 2:} observation number (10 digit ID).

{\it Column 3:} a star indicates if there is a note for this observation or
for this proposal-ID (first 6 digits of an observation, referring to
several observations for this proposal) as detailed below.

{\it Column 4:} the source number per observation of the identified target
taken from the column SRC\_NUM in the catalogue.

{\it Column 5:} the detection ID of the identified target taken from the
column DETID in the catalogue.

{\it Column 6:} field classification as described in
Table~\ref{fieldcontenttab}.

{\it Column 7:} coordinate preference between proposal position and Simbad
position, depending on which defined the target better; in case of offset
positions (usually indicated in the field name from the proposal, Col.~12)
no preference is given.

{\it Column 8:} proposal category as taken from the XSA as described in
Table~\ref{pcategorytab} (note that some of the calibration observations
are not properly classified).

{\it Column 9:} proposal program as taken from the XSA: GO stands for Guest
Observer, Cal for Calibration, ToO for Targets of Opportunity, Cha for
Co-Chandra, ESO for Co-ESO, Trig for Triggered, and Large.

{\it Columns 10} and {\it 11:} Right Ascension and declination (J2000) in
degrees as given in the proposal (taken from the RA\_OBJ and DEC\_OBJ
keywords in the attitude time-series file).

{\it Column 12:} field name as given in the proposal (taken from the OBJECT
keyword in the calibration index file).

{\it Column 13} and {\it 14:} Right Ascension and declination (J2000) in
degrees as extracted from Simbad using the Simbad name given in Col.~16.

{\it Column 15:} object type as given by Simbad.  If no Simbad object is
given a type was estimated. Additional types not recognised by Simbad are:
XRN for X-ray reflection nebula, sfr for star forming region, plt for
planet, and com for comet.

{\it Column 16:} modified field name which Simbad recognises (and can be
used in a script), except for 53 cases that have a name recognised by NED
(indicated with `[ned]' after the name). Modifications include dropping
offset indicators, completing coordinates, and adjusting the prefix to a
recognised convention as described in Simbad's dictionary of nomenclature.

{\it Column 17} and {\it 18:} Right Ascension and declination (J2000) in
degrees as given in the XSA; they represent the prime instrument viewing
direction (median value) and are corrected for the star tracker
mis-alignment.

A list of observations (10 digits) or proposal-IDs (6 digits) in numerical
order with special remarks as indicated in Col.~3 of the table follows.
\begin{description}

\item[0002740101:] CFHT-Pl-12 appears to be the name of a CFHT plate, and
the proposal abstract suggests that this is a field observation.

\item[0002970401:] The coordinates of the proposal position and image do
not agree.  The Observation Log Browser web-page at ESAC refers to an
`earth limb test'. The field of the observation is therefore as a whole
serendipitous.

\item[0008820401:] The observation of HD\,168112 was replaced by a ToO
observation of GRB 020321 which, however, was not registered in the ODF.

\item[004534:] This is a double star but the X-ray detection is not at the
Simbad position, and the field classification is ambiguous.

\item[0075940101:] Simbad recognises the field name `30 Ari' but returns
two objects (30 Ari A and 30 Ari B). Due to the ambiguity no Simbad name and coordinates are given.

\item[0093550401:] This observation was intended to have Z And as a target but
due to an operational issue  a different position was observed. The field of the
observation is therefore as a whole serendipitous.

\item[0094360201:] There seems to be an error in the proposal
coordinates in the proposal; the field of the observation is therefore as a
whole serendipitous.

\item[0094380101:] The observation of 1ES1255+244 was replaced by a ToO
observation of GRB 011211 which, however, was not registered in the ODF.

\item[0094530401:] The observation has a large offset observation from 3C192.

\item[0106860101:] There is a source at the proposal position, however, it
is possibly only a spurious extended detection, and therefore no source ID
is given.

\item[010806:] The field name is AXAF Ultra Deep Field; this appears to be 
the same as the Hubble Ultra Deep
Field with very similar coordinates (53.1625, -27.7914).

\item[0109060201:] Ambiguous because target name is not precise enough.

\item[010986:] The target name is A 189 but the proposal abstract indicates
that NGC 533 group is the object. It is not obvious whether both are
target.

\item[0111520301:] This is a ToO observation of GRB 010220, the field name
as given in the proposal is wrong.

\item[0112200601:] Unclear whether the extended
emission around the pulsar is connected to it, the field classification is
therefore `unknown'.

\item[0112200701:] Unclear whether the extended
emission around the pulsar is connected to it, the field classification is
therefore `unknown'.

\item[0112201101:] pulsar is located in  SNR W44 (cf.\ proposal-ID
  008327), and extended emission is detected; the field classification is
  taken to be the same as for proposal-ID 008327.

\item[011226:] The target is a merging galaxy cluster, A399/A401.  There
are four observations in different offset positions. The Simbad column
lists for each observation the cluster that is nearer to the centre of the
FOV, where possible.

\item[011305:] The proposal abstract mentions clumpy sources in the
neighbourhood of pulsars, and the field classification is somewhat
ambiguous (with respect to actual detections).

\item[0135960101:] The proposal abstract describes the object as X-ray
reflection nebula. There is no Simbad type for that but it seems
appropriate to use.

\item[0141610601:] The Simbad position appears to be wrong (the coordinates
in the name were assumed to be B1950 and converted to J2000 coordinates).

\item[014363:] This is a double star but the X-ray detection is not at the
Simbad position, and the field classification is ambiguous.

\item[0149630301:] The proposal explains LMC1 to be a supergiant shell,
while Simbad knows only a symbiotic star named LMC1. Instead Simbad knows
the supergiant shell as LMC-SGS~1.

\item[0154750401:] Both the proposal position and the Simbad position are
offset from the identified source. The correct identification of this
source comes from other observations of the same object (proposal-ID
020100).

\item[0154750301:] Though the proposal position and Simbad position are not
centred on the source identification given, the identification seems
unambiguous (note that the Simbad position is not very precise which would
explain the offset).

\item[0201270101:] The Simbad position appears to be wrong (the coordinates
in the name were assumed to be B1950 and converted to J2000 coordinates).

\item[0202940201:] The declination is wrong, the field of the observation
is therefore as a whole serendipitous.

\item[0203540901:] From the field name and proposal abstract it is not
clear whether this is a field or point-source observation.

\item[0204010101:] The target is three point sources.

\item[020422:] The field name is a composite of several target names.

\item[020619:] According to the proposal abstract the target type is an
      X-ray compact source.

\item[021047:] This is an observation of a super-bubble; the field
classification is ambiguous (`x' or `l').

\item[0303670101:] The proposal abstract indicates that this is an
observation of two galaxy groups, the Simbad name is given for the first
name only.

\item[0304050101:] It is not  clear if this is a point source or a
small extended source.

\end{description}

\section{Catalogue columns} \label{catcolumns}

The catalogue contains 297 columns. Each detection was observed with up to
three cameras. For the source detection, the total energy range
($0.2-12$\,keV) was split into five sub-bands as well as the XID wide-band
($0.5-4.5$\,keV), see Table~\ref{EnergyBands}. As a result, some of the
source parameters (like count rates or fluxes) are given for each camera
and band as well as for the combined cameras (EPIC) and total band. The
column names reflect this by using a two-letter prefix to indicate the
camera [ca = EP,PN,M1,M2]; in case of parameters that refer to a unique
source rather than an individual detection (Sect.~\ref{detmatching}) the
prefix [SC] is used (it stands for `source'). Following the prefix comes an
energy band indicator where applicable (b = 1,2,3,4,5,8,9). Entries are
NULL when there is no detection with the respective camera (that is, the
detector coverage of the detection weighted by the PSF, MASKFRAC, $<
0.15$).

In the following, a description for each column is given. The name is given
in capital letters, the FITS data format in brackets, and the unit in
square brackets. If the column originates from a SAS task\footnote{The
documentation on SAS tasks are available through the public XMM-SAS
distribution from the ESAC web pages.}, the name of the task follows.

For easier reference the columns are grouped into seven sections.

\subsection{Identification of the detection}  \label{col_id}

Next to the various identifications, cross matches with the 1XMM and 2XMMp
catalogues are given here. There are 9 columns in this section.

{\it DETID} (J): A consecutive number which identifies each entry
(detection) in the catalogue.

{\it SRCID} (J): A unique number assigned to a group of catalogue entries
which are assumed to be the same source. To identify members of the same
group the distance in arcseconds between each pair of sources was compared
on the $3\sigma$-level of both positional errors. A maximum distance of
$7\arcsec$ was assumed, which was reduced to $0.9\, \cdot \,$DIST\_NN
(distance to the nearest neighbour) where necessary. See
Sect.~\ref{detmatching} for a more detailed description. The combined
parameters for the unique sources are described in Sect.~\ref{col_unsrc}.

{\it IAUNAME} (21A): The IAU name assigned to the unique SRCID.

{\it SRC\_NUM} (J), SAS task {\tt srcmatch}: The (decimal) source number in
the individual source list for this observation as determined during the
source fitting stage; in the hexadecimal system it identifies the
source-specific product files belonging to this detection.

{\it MATCH\_1XMM} (21A): The IAU name of the closest 1XMM source within
$r=3\arcsec$, cf.\ Sect.~\ref{detmatching}.

{\it SEP\_1XMM} (E) [arcsec]: The distance between this source and the
matched 1XMM source, MATCH\_1XMM.

{\it SRCID\_2XMMP} (J): The unique source ID of the closest 2XMMp source
within $r=3\arcsec$, cf.\ Sect.~\ref{detmatching}.

{\it MATCH\_2XMMP} (22A): The IAU name of the closest 2XMMp source, cf.\
Sect.~\ref{detmatching}.

{\it SEP\_2XMMP} (E) [arcsec]: The distance between this source and the
matched 2XMMp source, MATCH\_2XMMp.

\subsection{Details of the observation and exposures}  \label{col_obsexp}

There are 11 columns in this section which covers the meta-data of a
detection. Details on XMM-Newton filters and modes can be found in the XMM
User Handbook (Ehle et al. 2007).

{\it OBS\_ID} (10A): The XMM-Newton observation identification.

{\it REVOLUT} (4A) [orbit]: The XMM-Newton revolution number.

{\it MJD\_START} (D) [d]: Modified Julian Date (i.e., JD\,--\,2400000.5) of
the start of the observation.

{\it MJD\_STOP} (D) [d]: Modified Julian Date (i.e., JD\,--\,2400000.5) of
the end of the observation.

{\it OBS\_CLASS} (J): Quality classification of the whole observation based
on the area flagged as bad in the manual flagging process as compared to
the whole detection area, see Sect~\ref{manflag}. 0 means nothing has been
flagged; 1 indicates that 0\% $<$ area $<$ 0.1\% of the total detection
mask has been flagged; 2 indicates that 0.1\% $\le$ area $<$ 1\% has been
flagged; 3 indicates that 1\% $\le$ area $<$ 10\% has been flagged; 4
indicates that 10\% $\le$ area $<$ 100\% has been flagged; and 5 means that
the whole field was flagged as bad.

{\it PN\_FILTER} (6A): PN filter. The options are Thick, Medium, Thin1, and
Thin2, indicating the degree of the optical blocking desired.

{\it M1\_FILTER} (6A): M1 filter. The options are Thick, Medium, and Thin1,
indicating the degree of the optical blocking desired.

{\it M2\_FILTER} (6A): Same as M1\_FILTER but for M2.

{\it PN\_SUBMODE} (23A): PN observing mode. The options are full frame mode
with the full FOV exposed (in two sub-modes), and large window mode with
only parts of the FOV exposed (Sect.~\ref{obssel}).

{\it M1\_SUBMODE} (16A): M1 observing mode. The options are full frame mode
with the full FOV exposed, partial window mode with only parts of the
central CCD exposed (in different sub-modes), and timing mode where the
central CCD was not exposed (`Fast Uncompressed'), see Sect.~\ref{obssel}.

{\it M2\_SUBMODE} (16A): Same as M1\_SUBMODE but for M2.

\subsection{Coordinates}  \label{col_coord}

The catalogue lists rectified (`external') equatorial and Galactic
coordinates as well as uncorrected (`internal') equatorial coordinates. Two
independent error estimates are combined into a third error column. There
are 9 columns in this section.

{\it RA} (D) [deg], SAS task {\tt evalcorr}: Corrected Right Ascension of
the detection (J2000) after statistical correlation of the {\tt emldetect}
coordinates, RA\_UNC and DEC\_UNC, with the USNO~B1.0 optical source
catalogue. In case where the cross-correlation is determined to be
unreliable no correction is applied and this value is therefore the same as
RA\_UNC (Sect.~\ref{astcorr}).

{\it DEC} (D) [deg], SAS task {\tt evalcorr}: Corrected declination of the
detection (J2000) after statistical correlation of the {\tt emldetect}
coordinates, RA\_UNC and DEC\_UNC, with the USNO~B1.0 optical source
catalogue. In case where the cross-correlation is determined to be
unreliable no correction is applied and this value is therefore the same as
DEC\_UNC (Sect.~\ref{astcorr}).

{\it POSERR} (E) [arcsec]: Total position uncertainty calculated by
combining the statistical error, RADEC\_ERR, and the `systematic' error,
SYSERR, as follows:
\[{\rm POSERR} = \sqrt{{\rm RADEC\_ERR}^2 + {\rm SYSERR}^2}\, .\]

{\it LII} (D) [deg], SAS task {\tt evalcorr}: Galactic longitude of the
detection corresponding to the (corrected) coordinates RA and DEC.

{\it BII} (D) [deg], SAS task {\tt evalcorr}: Galactic latitude of the
detection corresponding to the (corrected) coordinates RA and DEC.

{\it RADEC\_ERR} (E) [arcsec], SAS task {\tt emldetect}: Statistical
$1\sigma$-error on the detection position (RA\_UNC and DEC\_UNC).

{\it SYSERR} (E) [arcsec]: The estimated `systematic' $1\sigma$-error on the
detection position. It is set to be $0\farcs35$ if the SAS task {\tt
eposcorr} resulted in a statistically reliable cross-correlation with the
USNO B1.0 optical catalogue, otherwise the error is $1\farcs0$
(Sect.~\ref{astcorr}).

{\it RA\_UNC} (D) [deg], SAS task {\tt emldetect}: Right Ascension of the
source (J2000) as determined by the SAS task {\tt emldetect} by fitting a
detection simultaneously in all cameras and energy bands
(Sect.~\ref{det_emldetect}).

{\it DEC\_UNC} (D) [deg], SAS task {\tt emldetect}: Declination of the
source (J2000) as determined by the SAS task {\tt emldetect} by fitting a
detection simultaneously in all cameras and energy bands
(Sect.~\ref{det_emldetect}).

\subsection{Detection parameters}  \label{col_srcpar}

This section lists 22s3 columns. The fitted and combined detection
parameters as well as auxiliary information are taken directly from the
source lists created by the SAS tasks {\tt emldetect} and {\tt srcmatch}.

Instead of listing {\it each} column, descriptions of the general parameter
(and their errors) are given followed by an indicator for which bands and
camera combinations this parameter is available. Most parameters were
determined by the SAS task {\tt emldetect} which is described in detail in
Sect.~\ref{sourcesearch}, while some others were derived by the SAS task
{\tt srcmatch}. XID-band parameters are derived in a separate {\tt
emldetect} run and are therefore single-band values which ensures a better
handling of the error values.

{\it ca\_b\_FLUX} and {\it ca\_b\_FLUX\_ERR}: (E) [erg/cm$^2$/s], SAS tasks
  {\tt emldetect, srcmatch}: Fluxes are given for all combinations of ca =
  [EP, PN, M1, M2] and b = [1, 2, 3, 4, 5, 8, 9]; they correspond to the
  flux in the entire PSF and do not need any further corrections for PSF
  losses.

For the individual cameras, single-band fluxes are calculated from the
respective band count rate using the filter- and camera-dependent energy
conversion factors given in Table~\ref{ecftab} and corrected for the dead
time due to the read-out phase. These can be 0.0 if the detection has no
counts. The errors are calculated from the respective band count rate error
using the respective energy conversion factors.

Total-band fluxes and errors for the individual cameras are the sum of the
fluxes and errors, respectively, from the bands 1\,--\,5.

The EPIC flux in each band is the mean of the band-specific detections in
all cameras weighted by the errors, with the error on the weighted mean
given by
\[{\rm EP\_b\_FLUX\_ERR} = \sqrt{1.0 / \sum{ 1 / {\rm ca\_b\_FLUX\_ERR^2} }}\,, \]
where ca = [PN,M1,M2].

{\it ca\_b\_RATE} and {\it ca\_b\_RATE\_ERR} (E) [count/s], SAS task {\tt
  emldetect}: Count rates and errors are given for all combinations of ca =
  [PN, M1, M2] and b = [1, 2, 3, 4, 5, 8, 9] as well for ca = [EP] and b =
  [8, 9].

The single-band count rate is the band-dependent source counts (see
ca\_b\_CTS) divided by the exposure map, which combines the mirror
vignetting, detector efficiency, bad pixels and CCD gaps, and an OOT-factor
(Out Of Time) depending on the PN modes. The source counts and with it the
count rates were implicitly background subtracted during the fitting
process. They correspond to the count rate in the entire PSF and do not
need any further corrections for PSF losses. Note that rates can be 0.0
(but not negative) if the source is too faint in the respective band to be
detectable.

Total-band count rate for each camera are calculated as the sum of the
count rates in the individual bands 1\,--\,5.

The EPIC rates are the sum of the camera-specific count rates in the
respective band.

{\it ca\_b\_CTS} and {\it ca\_b\_CTS\_ERR} (E) [count], SAS task {\tt
emldetect}: Source counts and errors are given for ca = [EP, PN, M1, M2]
and b = [8].

The single-band source counts (not given in the catalogue) are derived
under the total PSF (point spread function) and corrected for
background. The PSF is fitted on sub-images of $r=60\arcsec$ in each band,
which means that in most cases at least 90\% of the PSF (if covered by the
detector) was effectively used in the fit.

Combined band source counts for each camera are calculated as the sum of
the source counts in the individual bands 1\,--\,5.

The EPIC counts are the sum of the camera-specific counts.

The error is the statistical $1\sigma$-error on the total source counts of
the detection.

{\it ca\_b\_DET\_ML} (E), SAS task {\tt emldetect}: Maximum likelihoods are
  derived for all combinations of ca = [PN, M1, M2] and b = [1, 2, 3, 4, 5,
  8, 9] as well for ca = [EP] and b = [8, 9].

The single-band maximum likelihood values stand for the detection
likelihood of the source, $L = - \ln{P}$, where $P$ is the probability the
detection is spurious due to a Poissonian fluctuation. While the detection
likelihood of an extended source is computed in the same way, systematic
effects such as deviations between the real background and the model, have
a greater effect on extended sources and thus detection likelihoods of
extended sources are more uncertain.

To calculate the maximum likelihood values for the total band and EPIC the
sum of the individual likelihoods is normalised to two degrees of freedom
using the function
\[L = -\ln (1-P_{\Gamma}(\frac{\nu}{2},L')) \;\;\; 
     {\rm with} \;\;\;  L' = \sum_{i=1}^{N} L_i \;,\]
where $P_{\Gamma}$ is the incomplete Gamma function, $N$ is the number of
energy bands involved, $\nu$ is the number of degrees of freedom of the fit
($\nu = 3+N$, if the source extent is a fitted parameter, see
Sect.~\ref{extdsrc}, and $\nu = 2+N$ otherwise).

{\it EP\_EXTENT} and {\it EP\_EXTENT\_ERR} (E) [arcsec], SAS task {\tt
  emldetect}: The extent radius (i.e., core radius) and error of a source
  detected as extended is determined fitting a beta-model profile to the
  source PSF (Sect.~\ref{extdsrc}). Anything below $6\arcsec$ is considered
  to be a point source and the extent is re-set to zero. To avoid
  non-converging fitting an upper limit of $80\arcsec$ has been introduced.

{\it EP\_EXTENT\_ML} (E), SAS task {\tt emldetect}: The extent likelihood
  is the likelihood of the detection being extended as given by $L_{\rm
  ext} = - \ln{p}$, where $p$ is the probability of the extent occurring by
  chance.

{\it ca\_HRn} and {\it ca\_HRn\_ERR} (E), SAS tasks {\tt emldetect,
  srcmatch}: The hardness ratios are given for ca = [EP, PN, M1, M2] and n
  = [1, 2, 3, 4]. They are defined as the ratio between the count rates $R$
  in bands $n$ and $n+1$:
\[ {\rm HR}n\ = (R_{n+1} - R_{n}) / (R_{n+1} + R_{n}) \, . \]
In the case where the rate in one band is 0.0 (i.e., too faint to be
detected in this band) the hardness ratio will be $-1$ or $+1$ which is
only a lower or upper limit, respectively. In case where the rate in both
bands is zero, the hardness ratio is undefined (NULL).

Errors are the $1\sigma$-error on the hardness ratio.

EPIC hardness ratios are calculated by the SAS task {\tt srcmatch} and are
averaged over all three cameras [PN, M1, M2]. Note that no energy
conversion factor was used and that the EPIC hardness ratios are de facto
not hardness ratios but an equivalent parameter helpful to characterise the
hardness of a source.

{\it ca\_b\_EXP} (E) [s], SAS task {\tt emldetect}: The exposure map values
are given for combinations of ca = [PN, M1, M2] and b = [1, 2, 3, 4, 5].
They are the PSF-weighted mean of the area of the sub-images
($r=60\arcsec$) in the individual-band exposure maps (cf.\
Sect.~\ref{sourcesearch}).

{\it ca\_b\_BG} (E) [count/pixel], SAS task {\tt emldetect}: The background
map values are given for combinations of ca = [PN, M1, M2] and b = [1, 2,
3, 4, 5]; they are derived from the background maps at the given detection
position. Note that the source fitting routine uses the background map
itself rather than the single value given here. The value is (nearly) zero
if the detection position lies outside the FOV.

{\it ca\_b\_VIG} (E), SAS task {\tt emldetect}: The vignetting values are
given for combinations of ca = [PN, M1, M2] and b = [1, 2, 3, 4, 5]. They
are a function of energy band and off-axis angle. Note that the source
parametrisation uses the vignetted exposure maps instead.

{\it ca\_ONTIME} (E) [s]: The ontime values, given for ca = [PN, M1, M2],
are the total good exposure time (after GTI filtering) of the CCD where the
detection is positioned. Note that some source positions fall into CCD gaps
or outside of the detector and will have therefore a NULL given.

{\it ca\_OFFAX} (E) [arcmin], SAS task {\tt emldetect}: The off-axis
angles, given for ca = [PN, M1, M2], are the distance between the detection
position and the on-axis\footnote{This is the optical axis which is close
to but not the same as the geometrical centre of the detector.} position on
the respective detector; the off-axis angle for a camera can be greater than
$15\arcmin$ when the detection is located outside the FOV of that camera.

{\it ca\_MASKFRAC} (E), SAS task {\tt emldetect}: The maskfrac values,
given for ca = [PN, M1, M2], are the PSF weighted mean of the detector
coverage of the detection. It depends slightly on energy; only band 8
values are given here which are the minimum of the energy-dependent
maskfrac values. Sources which have less than 0.15 of their PSF covered by
the detector are considered as being not detected.

{\it EP\_DIST\_NN} (E) [arcsec], SAS task {\tt emldetect}: The distance to
the nearest neighbouring detection; note that there is an internal
threshold of $6\arcsec$ (before positional fitting) for splitting a source
into two.

\subsection{Detection flags}  \label{col_flags}

This section lists quality flags as well as flags for the presence of
time-series or spectra for a detection. There are 7 columns in this
section.

{\it SUM\_FLAG} (J): The summary flag of the source is derived from the
EPIC flag EP\_FLAG as explained in detail in Sect.~\ref{qualcolumns}.  They
are:

    0 = good,

    1 = source parameters may be affected,

    2 = possibly spurious,
 
    3 = located in an area where spurious detection may occur,

    4 = located in an area where spurious detection may occur and possibly
    spurious.

{\it EP\_FLAG} (12A), SAS task {\tt dpssflag}: EPIC flag that combines the
flags in each camera [PN\_FLAG, M1\_FLAG, M2\_FLAG], that is, a flag is set
in EP\_FLAG if at least one of the camera-dependent flags is set.

{\it PN\_FLAG} (12A), SAS task {\tt dpssflag}: PN flag made of the flags
[$1 - 12$] (counted from left to right) for the PN source detection. A flag
is set to True according to the conditions summarised in
Sect.~\ref{autoflag} for the automatic flags and Sect.~\ref{manflag} for
the manual flags. In case where the camera was not used in the source
detection a dash is given. In case a source was not detected by this camera
the flags are all set to False (default). Flag~[10] is not used.

{\it M1\_FLAG} (12A), SAS task {\tt dpssflag}: Same as PN\_FLAG but for M1.

{\it M2\_FLAG} (12A), SAS task {\tt dpssflag}: Same as PN\_FLAG but for M2.

{\it TSERIES} (L): The flag is set to True if this source has a time-series
made in at least one exposure (Sect.~\ref{ssp}).

{\it SPECTRA} (L): The flag is set to True if this source has a spectrum
made in at least one exposure (Sect.~\ref{ssp}).

\subsection{Variability information}  \label{col_var}

This section lists 7 columns with variability information for those
detections for which time-series were extracted.

{\it EP\_CHI2PROB} (E): The minimum value of the available camera
probabilities [PN\_CHI2PROB, M1\_CHI2PROB, M2\_CHI2PROB].

{\it PN\_CHI2PROB} (E), SAS task {\tt ekstest}: The $\chi^2$-probability
(based on the null hypothesis) that the source as detected by the PN camera
is constant. The Pearson's approximation to $\chi^2$ for Poissonian data
was used, in which the model is used as the estimator of its own variance
(Sect.~\ref{TS}). If more than one exposure (that is, time-series) is
available for this source the lowest value of probability was used.

{\it M1\_CHI2PROB} (E), SAS task {\tt ekstest}: Same as PN\_CHI2PROB but
for M1.

{\it M2\_CHI2PROB} (E), SAS task {\tt ekstest}: Same as PN\_CHI2PROB but
for M2.

{\it VAR\_FLAG} (L): The flag is set to True if this source was detected as
variable, that is, EPIC $\chi^2$-probability $< 10^{-5}$ (see
EP\_CHI2PROB).

{\it VAR\_EXP\_ID} (4A): If the source is detected as variable (that is, if
VAR\_FLAG is set to True), the exposure ID (`S' or `U' followed by a
three-digit number) of the exposure with the lowest $\chi^2$-probability
is given here.

{\it VAR\_INST\_ID} (2A): If the source is detected as variable (that is,
if VAR\_FLAG is set to True), the instrument ID [PN,M1,M2] of the exposure
given in VAR\_EXP\_ID is listed here.

\subsection{Unique source parameters}  \label{col_unsrc}

This section lists 31 columns with combined parameters for the unique
sources (using the prefix 'SC') together with the total number of
detections per source. For a detailed description on how the detections are
matched see Sect.~\ref{detmatching}.

{\it SC\_RA} (D) [deg]: The mean Right Ascension in degrees (J2000) of all
the detections of the source SRCID (see RA) weighted by the positional
errors POSERR.

{\it SC\_DEC} (D) [deg]: The mean declination in degrees (J2000) of all the
detections of the source SRCID (see DEC) weighted by the positional errors
POSERR.

{\it SC\_POSERR} (E) [arcsec]: The error of the weighted mean position
given in SC\_RA and SC\_DEC in arcseconds.

{\it SC\_EP\_b\_FLUX} and (E) [erg/cm$^2$/s]: The mean band~b flux of all
the detections of the source SRCID (see EP\_b\_FLUX) weighted by the errors
(EP\_b\_FLUX\_ERR), where b = [1,2,3,4,5,8,9].

{\it SC\_EP\_b\_FLUX\_ERR} (E) [erg/cm$^2$/s]: Error on the weighted mean
band~b flux in SC\_EP\_b\_FLUX, where b = [1,2,3,4,5,8,9].

{\it SC\_HRn} (E): The mean hardness ratio of the bands $n$ and $n+1$ of
all the detections of the source SRCID (see EP\_HRn) weighted by the errors
(see EP\_HRn\_ERR), where n = [1, 2, 3, 4].

{\it SC\_HRn\_ERR} (E): Error on the weighted mean hardness ratio in
SC\_HRn.

{\it SC\_DET\_ML} (E): The total-band detection likelihood of the source
SRCID is the maximum of the likelihoods of all detections of this source
(see EP\_8\_DET\_ML).

{\it SC\_EXT\_ML} (E): The total-band detection likelihood of the extended
source SRCID is the average of the extent likelihoods of all detections of
this source (see EP\_EXTENT\_ML).

{\it SC\_CHI2PROB} (E): The $\chi^2$-probability (based on the null
hypothesis) that the unique source SRCID as detected by any of the
observations is constant, that is, the minimum value of the EPIC
probabilities in each detection (see EP\_CHI2PROB) is given.

{\it SC\_VAR\_FLAG} (L): The variability flag for the unique source SRCID
is set to VAR\_FLAG of the most variable detection of this source.

{\it SC\_SUM\_FLAG} (J): The summary flag for the unique source SRCID is
taken to be the maximum flag of all detections of this source (see
SUM\_FLAG).

{\it N\_DETECTIONS} (J): The number of detections of the unique source
SRCID used to derive the combined values.

\end{document}